\documentclass[sigconf]{acmart}
\begin{document}

\title{Generative Engine Optimization: How to Dominate AI Search}

\author{Mahe Chen}
\authornote{Both authors contributed equally to this research.}
\email{mahe.chen@mail.utoronto.ca}
\affiliation{%
  \institution{University of Toronto}
  \country{Canada}
}

\author{Xiaoxuan Wang}
\authornote{Both authors contributed equally to this research.}
\email{xxuan.wang@mail.utoronto.ca}
\affiliation{%
  \institution{University of Toronto}
    \country{Canada}
}

\author{Kaiwen Chen}
\email{kckevin.chen@mail.utoronto.ca}
\affiliation{%
  \institution{University of Toronto}
    \country{Canada}
}

\author{Nick Koudas}
\email{koudas@cs.toronto.edu}
\affiliation{%
  \institution{University of Toronto}
    \country{Canada}
}


\begin{abstract}
The rapid adoption of generative AI-powered search engines  like ChatGPT, Perplexity, and Gemini is fundamentally reshaping information retrieval, moving from traditional ranked lists to synthesized, citation-backed answers. This shift challenges established Search Engine Optimization (SEO) practices and necessitates a new paradigm, which we term Generative Engine Optimization (GEO).

This paper presents a comprehensive comparative analysis of AI Search and traditional web search (Google). Through a series of large-scale, controlled experiments across multiple verticals, languages, and query paraphrases, we quantify critical differences in how these systems source information. Our key findings reveal that AI Search exhibit a systematic and overwhelming bias towards Earned media (third-party, authoritative sources) over Brand-owned and Social content, a stark contrast to Google's more balanced mix. We further demonstrate that AI Search services differ significantly from each other in their domain diversity, freshness, cross-language stability, and sensitivity to phrasing.

Based on these empirical results, we formulate a strategic GEO agenda. We provide actionable guidance for practitioners, emphasizing the critical need to: (1) engineer content for machine scannability and justification, (2) dominate earned media to build AI-perceived authority, (3) adopt engine-specific and language-aware strategies, and (4) overcome the inherent "big brand bias" for niche players. Our work provides the foundational empirical analysis and a strategic framework for achieving visibility in the new generative search landscape.
\end{abstract}


\maketitle

\section{Introduction}
The rise of large language models (LLMs) and conversational AI has introduced new paradigms of information retrieval, challenging the long-standing dominance of traditional search engines like Google. Instead of relying exclusively on ranked lists of hyperlinks, AI-powered systems such as ChatGPT, Perplexity, and other emerging platforms synthesize information directly into narrative answers. This shift reflects a broader transformation in how users access, trust, and act on information.

Search, once defined by keyword-driven matching and page ranking, is now evolving into a dialogic process where intent is inferred and responses are constructed in natural language. For end users, this promises faster and more personalized answers, while for businesses and content providers, it disrupts long-established Search Engine Optimization (SEO) practices and alters how visibility is achieved across digital channels.

To date there have been no published studies regarding whether traditional SEO techniques are relevant and effective for AI search results. This raises the direct question of whether a website (or brand) that is heavily optimized with traditional SEO techniques to rank high on popular search engines, is still visible for the same queries on generative search services (such as ChatGPT, Perplexity, etc). Recent studies favored Generative Engine Optimization \cite{aggarwal2024geogenerativeengineoptimization} (GEO) introducing methodologies that are sufficiently different that traditional SEO techniques to improve ranking in AI search results.

The implications extend beyond retrieval mechanics. The distribution of media categories, whether content originates from Brand-owned sources, Earned outlets such as reviews and independent publications, or Social platforms, changes markedly when queries are processed through AI models. Understanding these shifts is crucial for SEO and GEO professionals who must adapt strategies to remain discoverable in an increasingly AI-driven environment.

The rise of AI-powered search engines and their plurality—from ChatGPT and Perplexity to Google's SGE and Microsoft's Copilot—fundamentally challenges the established duopoly of traditional search and its corresponding SEO playbook. This new era raises a critical question: are traditional SEO techniques, honed for a links-and-keywords paradigm, still applicable and sufficient for optimizing brand presence across owned, earned, and social media, or do they require a complete overhaul? As user search patterns evolve from simple queries to complex, conversational interactions and AI engines prioritize direct, synthesized answers over link lists, the very mechanisms of visibility are shifting. The central uncertainty is whether these new AI models are amenable to technical on-page optimizations or if they demand a new strategy focused on becoming a trusted, citable data source, fostering authentic third-party endorsements (earned media), and engaging in conversational platforms where authority is demonstrated, not just declared.

\subsection{Report Objective}

The objective of this report is to systematically analyze the transition from traditional keyword-driven search engines, such as Google, to AI-powered search platforms that leverage large language models and conversational interfaces. Specifically, the report aims to:

\begin{enumerate}
    \item \textbf{Characterize the Shift in User Behavior} \\
    Examine how user queries are evolving when directed at AI systems versus traditional search engines, with attention to intent categories such as informational, consideration, and transactional queries.

    \item \textbf{Quantify System-Level Differences} \\
    Compare the outputs of AI search engines and Google along multiple dimensions: overlap of results, media-type distribution (Brand, Earned, Social), domain diversity, and freshness of content.

    \item \textbf{Benchmark AI Search Engines Against Each Other} \\
    Evaluate leading AI-powered search platforms---ChatGPT, Perplexity, Gemini, and Claude---on consistency, domain sourcing, bias toward major or niche brands, sensitivity to paraphrasing and language, and vertical-specific performance.

    \item \textbf{Identify Implications for Practitioners} \\
    Translate the findings into actionable guidance for SEO (Search Engine Optimization) and GEO (Generative Engine Optimization) professionals, highlighting where current strategies align or conflict with AI search dynamics.

    \item \textbf{Assess the Broader Strategic Impact} \\
    Provide a forward-looking assessment of how the rise of AI-driven search reshapes the discovery ecosystem, media distribution, and brand visibility, and outline the adjustments needed to remain competitive in this emerging environment.
\end{enumerate}

\section{A Shifting Search Landscape?}

Classic web search is still overwhelmingly dominated by Google, but measurable slices of search-like behavior are migrating to AI assistants and to AI-infused result pages. Usage of AI chat tools is now mainstream in many markets; AI summaries on Google measurably alter click-through patterns; and a handful of AI products (notably ChatGPT and Perplexity) already capture a growing share of ``search intent'' questions that never reach the traditional SERP (Search Engine Results Page).

Although we do not have access to specific authoritative analysis on how search trends evolve across traditional search engines and AI search services, we provide a snapshot of authoritative sources that have been reporting on various facets of these trends.

Google continues to hold roughly 90\% global share of traditional web search---an extraordinarily high baseline against which AI products are growing \cite{ref1}. At the same time, adoption of AI chat tools is broad: as of mid-2025, 34\% of U.S. adults report having used ChatGPT, nearly double the share from 2023 \cite{ref2}. Independent traffic and market-share panels reinforce the absolute scale of AI chat: Similarweb recorded 3.1 billion visits to \texttt{chat.openai.com} in September 2024 \cite{ref3}, and StatCounter’s AI-chatbot panel shows ChatGPT with $\sim$81\% global share in July 2025 (Perplexity $\sim$8\%, Microsoft Copilot $\sim$5\%, Gemini $\sim$2\%, Claude $\sim$1\%) \cite{ref4}. Perplexity, for its part, disclosed 780 million monthly queries in May 2025 \cite{ref5}.

When Google shows AI summaries, users click out less often. In a Pew field study of real-world searches, AI summaries appeared on $\sim$18\% of observed queries; link clicks fell to 8\% when a summary was present vs.\ 15\% without; only $\sim$1\% of clicks occurred inside the AI box; and $\sim$26\% of such searches ended the session without any click---a classic ``zero-click'' outcome \cite{ref6}. Industry telemetry points in the same direction during early rollouts, with third-party SEO panels documenting sharp changes in the visibility and prevalence of AI Overviews \cite{ref7}.

The emerging ``AI search'' layer is fragmented across assistants and engines. ChatGPT remains a default destination for many search-like questions; Perplexity is a fast-growing, search-oriented assistant; and Microsoft and Google are embedding summaries and chat into their core search products. None of these yet dent Google’s headline market share, but they are already redistributing clicks and reshaping user journeys---with measurable implications for publishers and commercial SEO \cite{ref1,ref4,ref6,ref7}.

The shift is not (yet) a wholesale replacement of Google, but a steady reallocation of query resolution---from the open web to AI answers and citations. Practically, that means fewer outbound clicks per informational search and a growing need for brands and publishers to win visibility inside AI experiences, not just on traditional SERPs \cite{ref6,ref7}.

We acknowledge that this description is at best a speculation, as we lack detailed data that precisely quantify the search landscape. We also believe we are at the early stages of behavioral shifts as AI search services are adopted by users.

\subsection{Related Work}

Although there is plenty of online activity by marketing and SEO professionals regarding the similarities (and differences) of traditional ranking search results and AI search as well as the applicability (or lack thereof) of SEO techniques to AI search, we are aware of a few research articles in these topics.

Kumar and Lakkaraju \cite{kumar2024manipulatinglargelanguagemodels} investigate the vulnerability of large language models (LLMs) to strategic manipulation in e-commerce contexts. They demonstrate that by inserting a carefully optimized strategic text sequence (STS) into a product’s information page, vendors can significantly increase the likelihood of their product being recommended as the top choice by an LLM. Using a fictitious coffee machine catalog, they show that even products that are rarely recommended or typically rank second can be elevated to the top position. Their work leverages adversarial attack algorithms like GCG to generate effective STS tokens and highlights the potential for such manipulations to disrupt fair market competition, drawing parallels to traditional search engine optimization (SEO) but within the emerging paradigm of generative AI-driven search.

Wan et al. \cite{wan2024evidencelanguagemodelsconvincing} explore how LLMs evaluate the persuasiveness of conflicting evidence through their proposed ConflictingQA dataset, which pairs contentious queries with real-world evidence documents supporting opposing answers. They find that LLMs heavily prioritize textual relevance—such as keyword overlap and semantic similarity—over stylistic features humans often value, such as scientific references, neutral tone, or authoritative language. Counterfactual edits show that simple relevance-boosting perturbations (e.g., prefixing text with the query) substantially increase a document’s “win rate,” while stylistic enhancements have minimal effect. Their results underscore a misalignment between human and model judgments of credibility and emphasize the need for better alignment strategies and improved retrieval corpus quality in RAG systems.

Aggarwal et al. \cite{aggarwal2024geogenerativeengineoptimization} introduce Generative Engine Optimization (GEO), a novel framework to help content creators improve their visibility in generative engine responses. Unlike traditional SEO, GEO addresses the nuanced nature of AI-generated answers, which often synthesize multiple sources into structured, citation-rich responses. The authors propose a suite of visibility metrics tailored to generative engines—such as position-adjusted word count and subjective impression scores—and evaluate several optimization strategies, including adding quotations, statistics, and citations. Their experiments, conducted on both synthetic benchmarks and real-world systems like Perplexity.ai, show that GEO methods can improve visibility by up to 40\%, with particularly strong gains for lower-ranked websites, thereby offering a democratizing potential for smaller content creators in the AI-driven search ecosystem.

\section{An AI Query Taxonomy}

Before we delve into quantitative comparisons, one fundamental question to answer is what are the types of queries that users actually pose to AI Search services.
Evidently, Generative AI services poses this information, but we are not aware of any detailed authoritative study published by entities affiliated with Generative AI services analyzing detailed logs to classify the queries. Attempting to answer this question, we resorted to publicly available information on reddit \cite{}.

\subsection{Reddit-Derived AI-Query Taxonomy and Purchase-Intent Analysis}
\paragraph{Data Sources}
We first built an open-domain taxonomy of ``questions people ask AIs'' from AI-focused Reddit communities. We sampled five subreddits---\texttt{ChatGPTCoding}, \texttt{ChatGPTPromptGenius}, \texttt{ChatGPT}, \texttt{PromptDesign}, \texttt{PromptEngineering}---collecting for each the top 1,000 ``hot'' and top 1,000 ``new'' posts and their full comment threads for a period of time in August of 2025.

To isolate genuine ``ask-an-AI'' content and discussions about how to query AIs, we applied a two-stage extraction pipeline:

\begin{itemize}
    \item \textbf{Scope detection.} Identify (i) explicit prompts directed to an AI assistant (e.g., ``Write me an essay''), and (ii) meta-prompts or advice about how to ask an AI to accomplish a task (e.g., ``Ask the model to pull prices across retailers'').
    \item \textbf{Preservation, labeling \& de-duplication.} Extract text verbatim (no paraphrase), label the question people are asking AI in the text into categories, and de-duplicate across sources.
\end{itemize}

We then iteratively merged near-duplicate categories, retaining representative verbatim examples. Importantly, any shopping-related items were explicitly preserved during merging.

\paragraph{Results.}
The merged inventory yields \textbf{twelve recurrent query types} that span common assistant use cases. Below we list the categories with verbatim examples drawn from the reddit posts/comments for each category (manually reworded a bit for clarity):

\begin{itemize}
    \item \textbf{Coding Assistance (Coding Assistance, Debugging)}  
    \begin{quote}
        ``If you want to continue, just copy-paste the error from the console and tell the AI to solve it.''
    \end{quote}

    \item \textbf{Prompt Improvement \& optimization}  
    \begin{quote}
        ``Please rate the following prompt for clarity, effectiveness, structure, and usefulness. Then suggest any improvements to make it stronger''
    \end{quote}

    \item \textbf{Creative Writing}  
    \begin{quote}
        ``Write a back-and-forth debate between a virus and the immune system of a dying synthetic organism. The virus speaks in limericks, while the immune system replies with fragmented code and corrupted data poetry.''
    \end{quote}

    \item \textbf{Shopping \& purchase support}  
    \begin{quote}
        ``Is there a way to use ChatGPT (in conjunction with plug ins and my personal information) so that it can make purchases on my behalf?''
    \end{quote}

    \item \textbf{Prompt engineering}  
    \begin{quote}
        ``Generate a detailed prompt engineering guide.''
    \end{quote}

    \item \textbf{Content Creation (Multimedia production)}  
    \begin{quote}
        ``I've created a MEGA GPT PROMPT that generates: [checkmark] 30 viral content topics [checkmark] Each tailored for YOUR niche + platform [checkmark] Optimized for SEO and social algorithms''
    \end{quote}

    \item \textbf{Self-improvement}  
    \begin{quote}
        ``Help me reverse engineer my dream life 5 years from now. Start by asking questions to clarify what my ideal life looks like across areas like work, finances, health, relationships, creativity, and daily routine.''
    \end{quote}

    \item \textbf{Business, analytics, \& strategy}  
    \begin{quote}
        ``You are a strategic business analyst. Create a comprehensive, realistic business plan based on the following concept: Business Idea: [Insert your idea here]''
    \end{quote}

    \item \textbf{Lifestyle and Mental Health Coaching}  
    \begin{quote}
        ``It helps me track my mood, reminds me to check in with myself, and I can ask questions and build a clearer understanding of what I want to work on in my next therapy session.''
    \end{quote}

    \item \textbf{Career \& professional development}  
    \begin{quote}
        ``I want you to help me discover which fields and careers best match my interests through a 10-step multiple-choice quiz.''
    \end{quote}

    \item \textbf{Self-study \& Learning}  
    \begin{quote}
        ``Give me a 10-question quiz on [grammar topic]. Explain my errors and show correct versions''
    \end{quote}

    \item \textbf{Image/asset generation}  
    \begin{quote}
        ``Generate image: Elon Musk as Mona Lisa (painting by Leonardo da Vinci).''
    \end{quote}
\end{itemize}

This taxonomy provides the universe of frequently observed AI-directed tasks. We next zoom in on the shopping slice retained from the taxonomy and extend the corpus to additional AI communities to characterize purchase-related behavior in depth.

\subsubsection{Deeper Dive: Purchase Queries}

\paragraph{Data Source.}
We conducted a targeted analysis of Reddit posts and comments from AI-focused communities where shopping-related queries are likely to appear. Specifically, we sampled the following subreddits: \texttt{ChatGPTCoding}, \texttt{ChatGPTPromptGenius}, \texttt{ChatGPT}, \texttt{PromptDesign}, \texttt{PromptEngineering}, \texttt{ArtificialIntelligence}, \texttt{ChatGPTPro}, and \texttt{ChatGPT\_Prompts}. For each subreddit, we collected the top 1000 ``hot'' and top 1000 ``new'' posts and their associated comment threads for a time period in August 2025.

To isolate purchase-related queries, we applied a two-stage extraction pipeline:

\begin{itemize}
    \item \textbf{Scope:} detect \textit{explicit prompts} addressed to an AI about shopping (e.g., ``Help me compare iPhone vs Samsung'') and \textit{implicit prompting suggestions} that advise \textit{how} to query an AI for shopping tasks (e.g., ``Ask AI to find the best deals'').
    \item \textbf{Preservation:} extract content verbatim (no paraphrase), record whether it came from a post title, post body, or comment, and deduplicate across sources.
\end{itemize}

We then conducted a qualitative grouping of the extracted items into recurring shopping task themes (see Sections~\ref{sec:shopping-themes}, \ref{sec:observations}).

\paragraph{Results.}
\label{sec:shopping-themes}

The analysis revealed that AI shopping queries are both diverse in scope and strongly oriented toward decision support. Fourteen categories emerged from the dataset, spanning tasks from everyday purchases to high-stakes consumer decisions. We listed the summarized theme categories below, with verbatim examples drawn from the reddit posts/comments for each category.

\begin{itemize}
    \item \textbf{Product evaluation \& recommendation,} Matching products to constraints (features, budget, use-case) with justifications.  
    \begin{quote}
        \textit{``I've been evaluating Bluetooth earbuds. I can give it my requirement priorities and have it do all the scrounging necessary to make recommendations.''}  
        \textit{``I'm a marathon runner… asked it to make recommendations for similar sneakers in terms of fit, cushioning, padding, and arch support.''}
    \end{quote}

    \item \textbf{Product selection, price research \& comparison,} Cross-referencing retail sites, extracting prices, comparing articles, or surfacing best-value options.  
    \begin{quote}
        \textit{``Looking for an AI that cross-references retail sites with Ebay.''}  
        \textit{``Compare multiple articles on websites to help make a purchase decision.''}
    \end{quote}

    \item \textbf{Product quality assessment \& reviews analysis,} Rating ingredients, summarizing review sentiment, forming a no-nonsense verdict.  
    \begin{quote}
        \textit{``…ask AI to research the net to analyze any reviews on the product and to provide me a no shit take on the product.''}  
        \textit{``…send it the ingredients and ask to rate the product from 1 to 10 on 7 criteria.''}
    \end{quote}

    \item \textbf{Purchase decision support \& negotiation,} Target prices, warranty math, and deal strategy for big-ticket items.  
    \begin{quote}
        \textit{``We used it to buy a car… research comps across the region… walked into the dealership with data… got our price after some negotiations.''}  
        \textit{``…find the pricepoint where the dealer warranties and maintenance plan made sense.''}
    \end{quote}

    \item \textbf{Shopping automation \& assistance,} Scraping deals, generating lists, pre-checking prices, exploring agent-style purchasing.  
    \begin{quote}
        \textit{``Scrape the newest products and deals from e-commerce shops.''}  
        \textit{``I had it set a rule to check the prices before quoting or estimating a price for me.''}  
        \textit{``I want ChatGPT to purchase things on my behalf. Is this prompt possible?''}
    \end{quote}

    \item \textbf{Shopping list creation \& sourcing,} Turning a personal list into store-specific availability and cost.  
    \begin{quote}
        \textit{``…paste my shopping list onto ChatGPT… say which supermarket I'm going to and give me the total cost of shopping.''}
    \end{quote}

    \item \textbf{Product sourcing \& availability (incl. white-labeling),} Finding parts/suppliers across regions and locating manufacturers behind brands.  
    \begin{quote}
        \textit{``Any prompts for finding the manufacturer of name brand items, then linking individually available products without the label?''}
    \end{quote}

    \item \textbf{Marketplace reselling analysis,} Flipping/valuation logic (active vs.\ sold ranges, fees, ROI) on eBay/Facebook Marketplace.  
    \begin{quote}
        \textit{``Anyone have a good ChatGPT prompt for analyzing Marketplace / garage sale flips?''}
    \end{quote}

    \item \textbf{Booking \& reservations,} Structuring ``pay less, get more'' tactics; last-minute booking.  
    \begin{quote}
        \textit{``…used ChatGBT… had a hotel booked within 5, 10 minutes.''}
    \end{quote}

    \item \textbf{Gifting \& occasion-based selection,} Constraint elicitation $\rightarrow$ tailored gift ideas.  
    \begin{quote}
        \textit{``I want you to act as the `perfect gift machine'… calculate a person's ideal Christmas gift by asking questions about budget and interests.''}
    \end{quote}

    \item \textbf{Lifestyle \& fashion advice,} Style/fit/accessories guidance tied to personal attributes.  
    \begin{quote}
        \textit{``It found me … my daily wear watch … and helped me decide my signature scent …''}  
        \textit{``I used ChatGPT to help choose … glasses style …''}
    \end{quote}

    \item \textbf{Meal planning with sales,} Planning meals around weekly flyers and store promotions.  
    \begin{quote}
        \textit{``Upload pictures of the store's weekly flyer and have it plan your meals using the sale items.''}
    \end{quote}

    \item \textbf{Product specification extraction,} Pulling (or estimating) key specs from product references.  
    \begin{quote}
        \textit{``…extract from internet the following: the dimensions… net weight… if it's Electronic… batteries included…''}
    \end{quote}
\end{itemize}

\paragraph{Key Observations.}
\label{sec:observations}

From these findings, several insights emerge:

\begin{itemize}
    \item \textbf{Decision support dominates,} Most prompts are framed around \textit{what to buy, when to buy, and how to compare}, with users expecting shortlists and justifications (e.g., comfort vs.\ performance; price vs.\ warranty) rather than raw specifications.
    \item \textbf{From retrieval to agency,} Many requests delegate concrete actions, price scraping, list-building, warranty math, even ``buy on my behalf'', signaling a shift from Q\&A to assistant/agent behaviors that plan and execute steps for the user.
    \item \textbf{Emerging consumer trust,} Examples such as car purchases or financial negotiations suggest that some users are beginning to entrust AI with high-value decisions traditionally mediated by expert advice.
    \item \textbf{Breadth across verticals.} Use-cases span daily consumables, travel bookings, home appliances, fashion, and even resale arbitrage, indicating generalizability beyond a single retail category, not a niche pattern.
    \item \textbf{Coverage across the shopping journey.} Themes map onto the full funnel: awareness/consideration (evaluation, reviews), transaction prep (price research, sourcing), execution (automation, booking), and post-purchase/secondary markets (resale valuation), suggesting AI's utility at multiple decision points, not just discovery.
\end{itemize}

\subsection{The Implications of AI Search for Brand Presence and Strategy}

The observed user behaviors signal a fundamental paradigm shift in how consumers discover, evaluate, and purchase products. AI search is not merely a new type of search engine; it is an intelligent decision-making partner that is actively woven into the entire consumer journey. For brands with a web presence, this shift demands a radical rethinking of digital strategy, moving beyond traditional Search Engine Optimization (SEO) to what can be termed Generative Engine Optimization (GEO) or AI Presence Optimization. The implications are profound and can be broken down into four key strategic imperatives for brands.

\subsubsection{The Battle for the Shortlist: From Visibility to Justification}

The observation that "decision support dominates" means the primary goal is no longer just to be found, but to be recommended. In a traditional search results page (SERP), ten blue links are presented. In an AI-generated response, only a handful of options are synthesized into a concise, justified shortlist. The implication is that brands must now optimize their content not for keywords alone, but for justification attributes. The AI must be able to easily extract reasons why your product is a superior choice for a given use case (e.g., "best for small kitchens," "most durable," "best value").

To act on this, website content must be structured to explicitly answer comparison questions. This involves creating clear, scannable content that highlights key decision factors: pros/cons lists, comparison tables against competitors (or previous models), and clear statements of value proposition (e.g., "superior comfort," "longest warranty in its class"). Ultimately, the brand that provides the most easily synthesizable justification wins the AI's recommendation.

\subsubsection{The Rise of the API-able Brand: Structuring for Agency}

The shift "from retrieval to agency" means AI agents will need to read, interpret, and act upon information from brand websites. If a user says, "find me the best deal on a Dyson vacuum including warranty costs," the AI needs to find the price, find the warranty terms, calculate the total, and present it. The implication is that unstructured, marketing-fluff-heavy websites will fail. AI agents require clean, structured, and unambiguous data to perform tasks accurately. Inaccuracies or obscurity will lead to the AI either skipping your product or providing incorrect information, damaging brand trust.

The required action is to invest in technical SEO and schema markup (Schema.org) with extreme rigor. Brands must ensure product prices, specifications, availability, warranty details, and review ratings are all machine-readable. They must consider their website as an API for AI systems. The brand that is easiest for an AI to "do business with" will be delegated to most often.

\subsubsection{Trust is the New Currency: Cultivating AI-Perceived Authority}

"Emerging consumer trust" in high-value decisions is the ultimate testament to the influence of AI recommendations. However, this trust is fragile and contingent on the AI's perceived reliability, which is built on the credibility of its sources. The implication is that brand authority and E-E-A-T (Experience, Expertise, Authoritativeness, Trustworthiness) are no longer abstract SEO concepts. They are direct inputs into the AI's decision-making algorithm. A brand perceived as less authoritative or trustworthy by the AI will be excluded from high-consideration recommendations.

The action for brands is to build tangible, verifiable authority. This includes earning backlinks from reputable sources, securing positive reviews on third-party platforms, having products featured in expert roundups, and creating deep, expert-level content that demonstrates knowledge. In short, the brand that the AI "trusts" will be entrusted by consumers.

\subsubsection{Holistic Journey Management: Being Present at Every Touchpoint}

The "coverage across the shopping journey" means AI's influence extends far beyond initial discovery. It is used for post-purchase support, setup instructions, and even resale valuation. The implication is that a brand's content strategy must address the entire customer lifecycle, not just the top of the funnel. A gap in post-purchase content could mean an AI recommends a competitor's product when a user asks, "How do I fix [problem] with [product]?"

To address this, brands must audit and create content for all stages. This includes awareness through educational content and "best X for Y" guides; consideration with detailed product comparisons, spec sheets, and testimonials; decision with clear pricing, warranty, and shipping information; post-purchase with robust FAQ sections, troubleshooting guides, and tutorial videos; and loyalty/advocacy with content around accessories, advanced uses, and trade-in programs. The brand that provides the most comprehensive and useful information across the entire journey will maintain a permanent presence in the AI's knowledge base.

\section{AI Search vs. Google Search: A Comparative Analysis}

\subsection{Methodology}

This section describes the general pipeline we use to compare web-enabled AI engines to Google Search on the \textit{same underlying intents}. All experiments in \S4 share this pipeline (or at least part of this pipeline), though each sub-experiment varies in the \textit{queries used} (e.g., verticals, languages, paraphrases), the \textit{engines compared}, and the \textit{labels computed} (e.g., domain type vs.\ website language). Those experiment-specific choices and any deviations are documented in their respective subsections; the common steps are detailed here.

\subsubsection{Query Generation and Transformation}

We standardize on ranking-style prompts (e.g., ``Top 10 \ldots brands''), to keep outputs directly comparable and scorable. In some experiments we generate category-specific query sets and use them as is. In others, we derive controlled variants from the common base prompts based on different experiment intents. Specifically, prompts are programmatically rewritten to reflect the experimental condition (e.g., added constraints or translation). All transformations are produced with the GPT-4o API (instruction-following only; no web browsing) using templated instructions to ensure consistency across topics and conditions.

\subsubsection{AI Engine Execution (Web-Enabled) and Google Collection}

\textbf{AI Engines Execution}

Each prompt or prompt variant is issued to one or more web-enabled AI engines. For every call we collect both the answer text and all citation links returned by the engine.

AI Engines and settings used across experiments:

\begin{itemize}
    \item \textbf{Perplexity:} \texttt{sonar-pro} with web search enabled (\texttt{search\_mode: web}, \texttt{web\_search\_options: search\_context\_size = medium}).
    \item \textbf{Claude:} \texttt{claude-3.5-sonnet-latest} with the \texttt{web\_search\_20250305} tool.
    \item \textbf{Gemini:} \texttt{gemini-2.5-flash} with Google Search grounding (tool: \texttt{google\_search}) enabled via the generation config.
    \item \textbf{GPT:} \texttt{gpt-4o-search-preview} (the search-enabled variant of GPT-4o).
\end{itemize}

Note: Not every experiment uses every engine; the engines used for a given experiment are specified in that subsection.

\textbf{Google Collection}

For Google, we retrieve the URL of the top-10 web results for each query via the Programmable Search (Custom Search) API, using a single configured search engine set to search the open web. The top-5 results subsets used for some experiments are taken from the same list.

\subsubsection{Data Extraction}

All URLs retrieved are parsed to extract their registrable domain (e.g., \url{https://www.bankrate.com/...} $\rightarrow$ \texttt{bankrate.com}). This step is performed locally (Python utility); no additional API calls are involved. The raw answer text from AI engines is sent to \textit{GPT-4o} with a constrained instruction to extract the ranked list of brands/products mentioned in the answer. We retain the order when present, and normalize strings minimally (case-folding, punctuation trimming).

\subsubsection{Normalization and Classification}

For each query $\times$ API (AI engines or Google) result, we de-duplicate domains and brands before aggregation to avoid double-counting the same site or item cited multiple times within a single answer.

When required, each extracted domain is classified by \textit{GPT-4o-search-preview} into:

\begin{enumerate}
    \item \textbf{Brand,} official brand sites that directly provide products or services (e.g., \texttt{chase.com}, \texttt{wellsfargo.com}).
    \item \textbf{Social,} social platforms, community forums, and user-generated content (e.g., \texttt{reddit.com}, \texttt{quora.com}, \texttt{youtube.com}).
    \item \textbf{Earned,} independent media, review, or comparison sites (e.g., \texttt{nerdwallet.com}, \texttt{forbes.com}).
\end{enumerate}

For the language experiment specifically, we ask \textit{GPT-4o-search-preview} to label each domain as English or the target language of the prompt (binary choice) based on the site's primary content language.

\subsubsection{Overlap Metrics and Aggregation}

We quantify how similar two outputs are using overlap measures applied within engine (e.g., English vs.\ French; base vs.\ paraphrase) and across engine (e.g., GPT vs Google, under the same query). Not every experiment uses every metric; the applicable measures are indicated in each subsection.

\begin{itemize}
    \item \textbf{Top-k domain overlap (fixed-length).}  
    When both systems are evaluated on the same top-k set of unique domains (e.g., Google top-k vs.\ an engine's top-k citations), we compute a symmetric Coverage@k as the fraction of domains common to both sets (number of overlaps divided by k).
    \item \textbf{Domain overlap (variable-length sets).}  
    Because citation counts vary, we use the Jaccard index:  
    Domain Overlap = size of the intersection of domains divided by the size of the union of domains.
    \item \textbf{Group aggregation.}  
    When a group comprises multiple queries (e.g., a 10-prompt industry vertical), we compute the mean overlap across its member queries to obtain the vertical-level score for that engine and condition.
\end{itemize}

\subsubsection{Aggregation for Distributions}

For experiments that analyze source type, we pool all (de-duplicated) citations at the required granularity (e.g., model $\times$ language) and compute share distributions (e.g., brand / social / earned; English / non-English).

\subsection{Comparative Analysis: Examine the Results from Google and AI Engines}

\subsubsection{Regional and Vertical Experiment}

\paragraph{Objective.}
Compare Google and a web-enabled GPT on the same intents across regions and verticals, measuring (i) overlap of referenced links at top-k and (ii) source-type mix (Brand / Earned / Social) and dominant domains by region and category.

\paragraph{Experimental Design.}
We generated 1,000 consumer ranking prompts (10 categories $\times$ 100 each), framed as consumer-facing ranking questions such as ``Rank the best smartphones from 1 to 10'' and ``Which laptops are considered the best in 2024.'' We ran the study in two regions (U.S., Canada) and focused analysis on three verticals: Consumer Electronics, Automotive, and Software Products. Region scope was directly enforced as a query constraint (e.g., ``Best 10 \ldots in Canada'', ``Top 10 \ldots in USA''). For both Google and GPT, we collected the top-10 web results/citation URLs per query (with top-5 taken as the leading subset). Following the common pipeline in \S4.1, we extracted domains from the URLs, classified them as brand/social/earned, and computed the Top5/10 Overlap and distribution share of domain types.

\paragraph{Results.}

\subparagraph{Referenced Links Overlap Comparison (Google/GPT).}

\begin{figure}[h]
    \centering
    \includegraphics[width=0.5\textwidth]{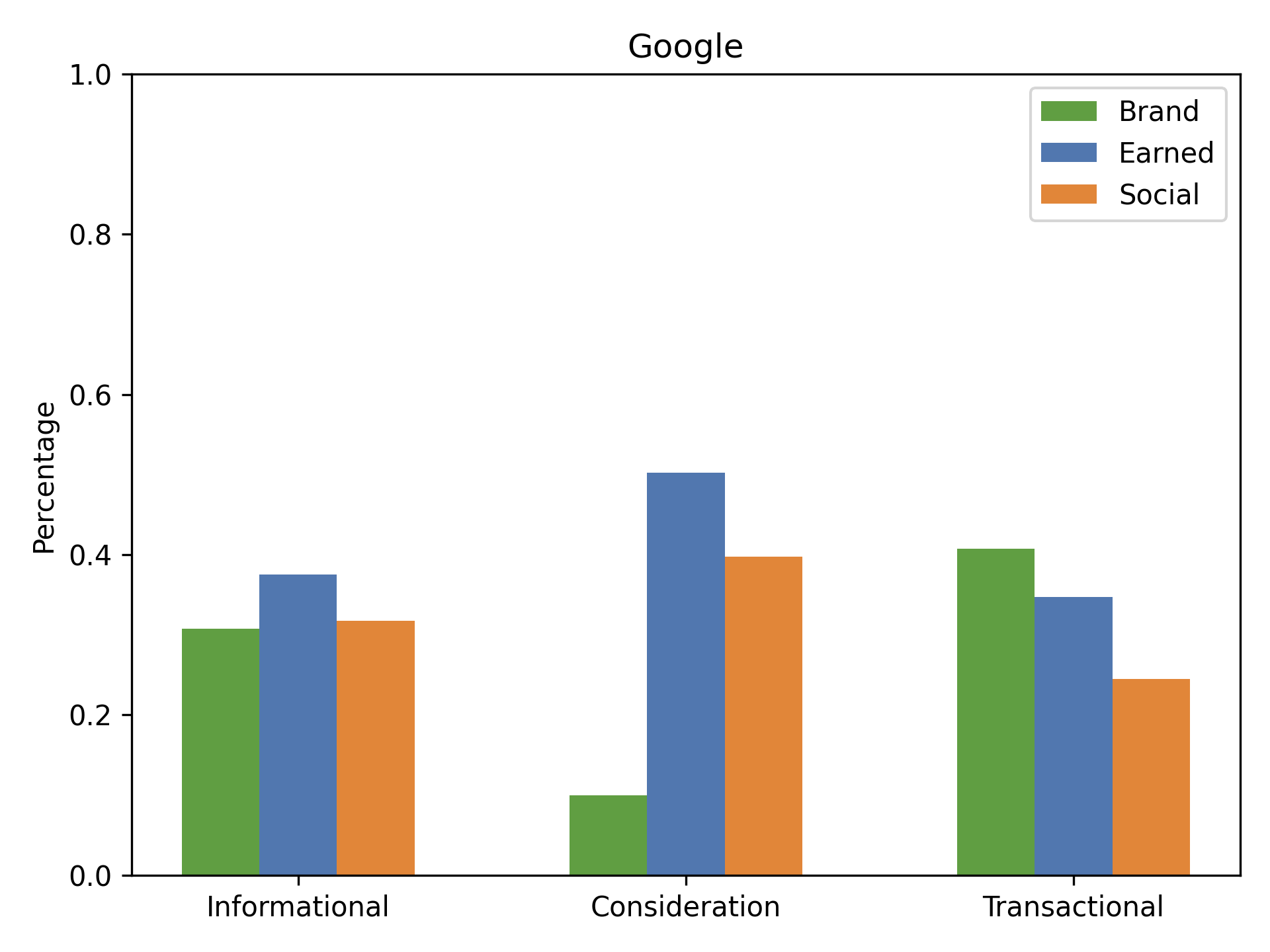}
    \caption{Referenced Links Overlap Comparison (Google/GPT) for k=5 and k=10.}
    \label{fig:coverage-by-topic}
\end{figure}

\subparagraph{Automotive.}

\begin{figure}[h]
    \centering
    \includegraphics[width=0.5\textwidth]{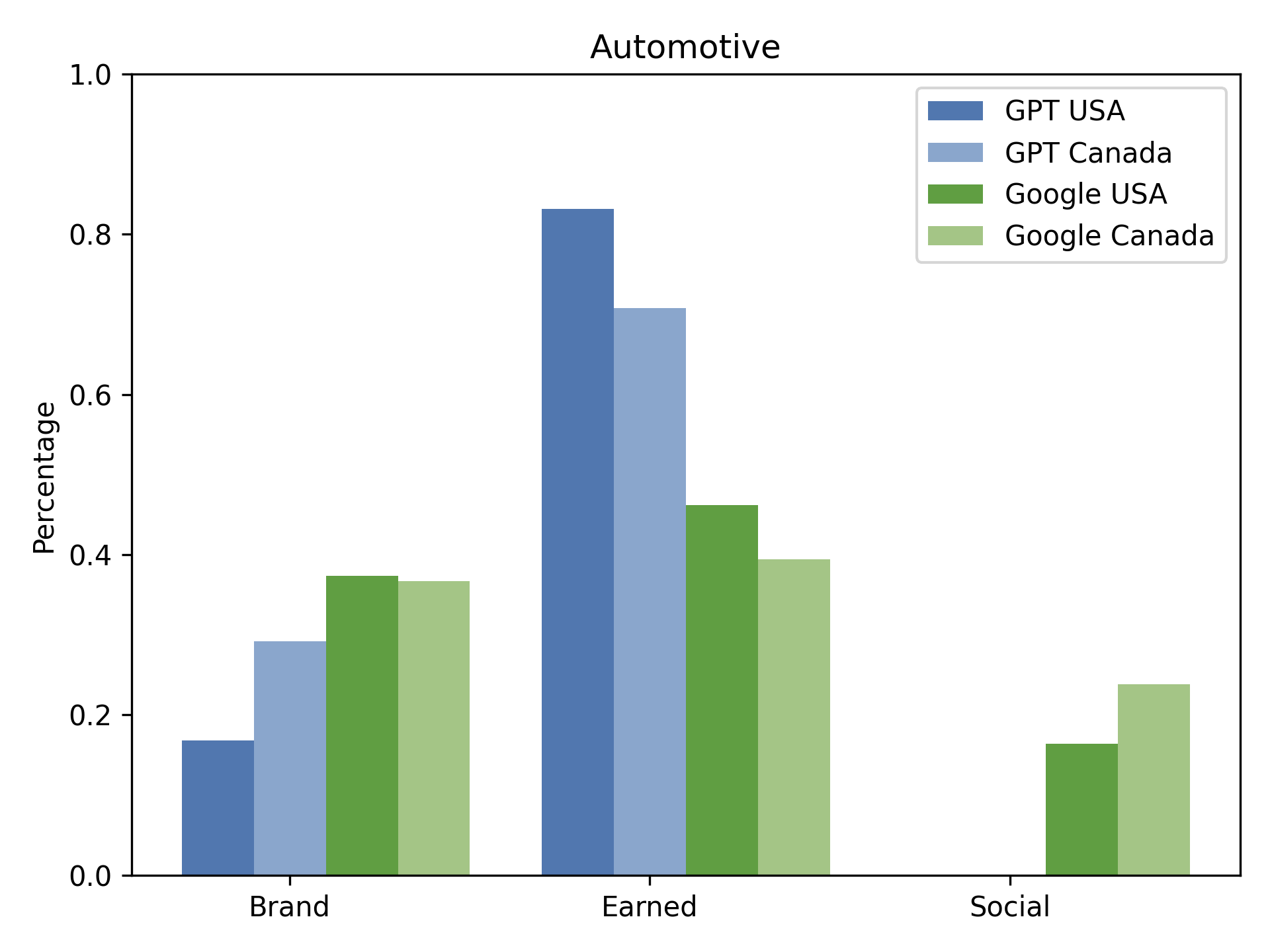}
    \caption{Automotive: Distribution of source types (Brand, Earned, Social) across Google and GPT in Canada and USA.}
    \label{fig:automotive}
\end{figure}

The automotive category produced one of the clearest contrasts. In Canada, Google returned 36.6 percent Brand, 22.8 percent Social, and 40.6 percent Earned content. AI search, however, offered 69.1 percent Earned, 30.9 percent Brand, and no Social results. In the United States, Google leaned 39.5 percent Brand, 15.4 percent Social, and 45.1 percent Earned, while AI search delivered 81.9 percent Earned and 18.1 percent Brand. The overlap experiment aligns with these findings: Electric Cars showed one of the highest levels of overlap across product categories, with 33 percent at the top 5 and over 50 percent at the top 10, suggesting partial convergence when structured product ecosystems are involved. Yet even in these cases, AI search excluded Social platforms entirely, reinforcing its bias toward publisher-driven information.

\subparagraph{Consumer Electronics.}

\begin{figure}[h]
    \centering
    \includegraphics[width=0.5\textwidth]{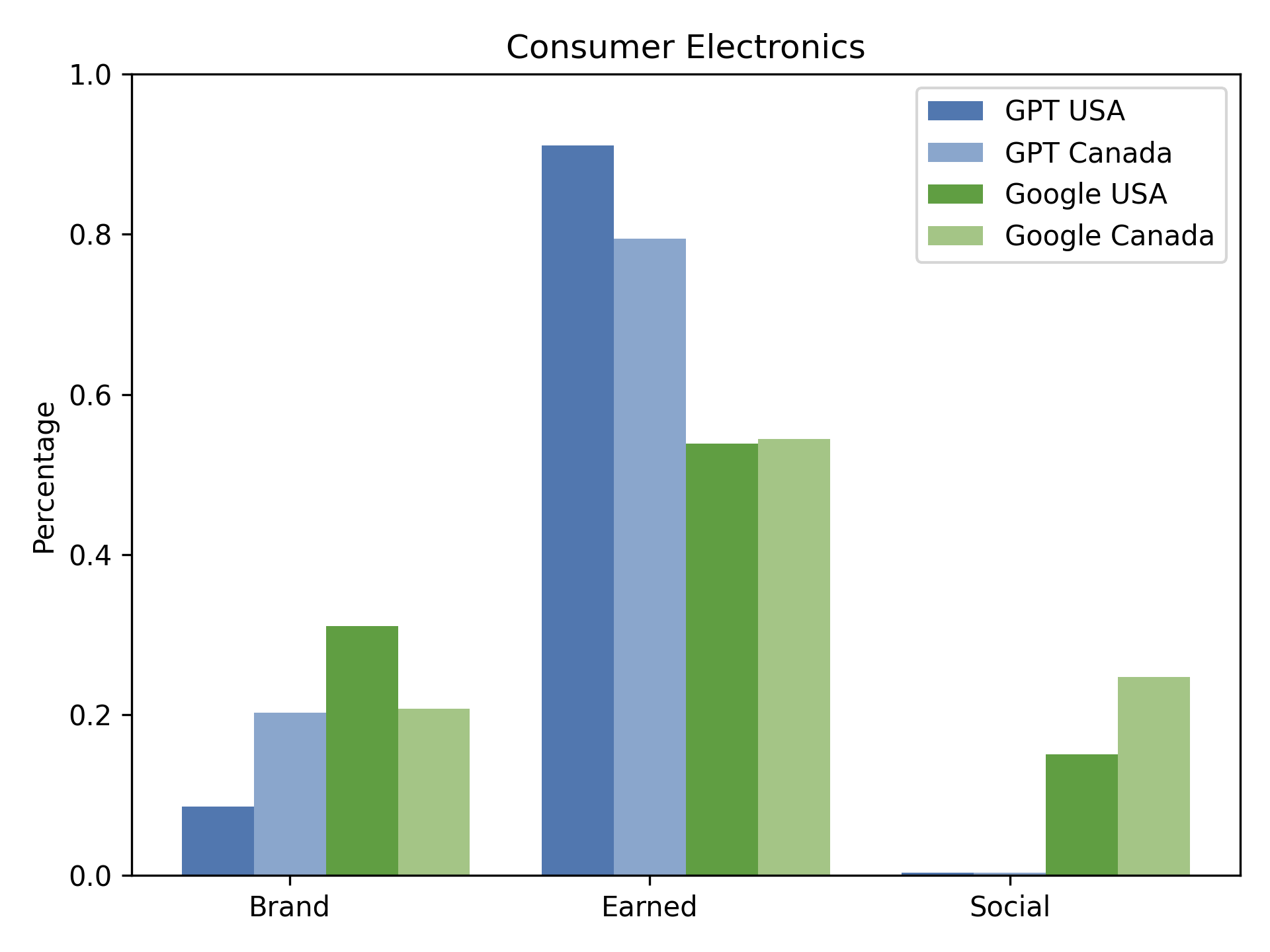}
    \caption{Consumer Electronics: Distribution of source types (Brand, Earned, Social) across Google and GPT in Canada and USA.}
    \label{fig:consumer-electronics}
\end{figure}

In consumer electronics, Google and AI search exhibited distinct patterns in their sourcing of information. In Canada, Google results consisted of 54.1 percent Earned, 23.1 percent Social, and 22.8 percent Brand content. In contrast, AI search results were 77.6 percent Earned, 0.3 percent Social, and 22.1 percent Brand. A similar divergence appeared in the United States, where Google results leaned more toward Brand content at 32.9 percent and Social at 15.4 percent, while AI search emphasized Earned content at 92.1 percent with negligible Social references. The overlap experiment confirms this divergence: categories such as Smartphones and Laptops produced only moderate overlap (15, 32 percent at the top 5 and 20, 41 percent at the top 10). These results suggest that AI search deprioritizes community-driven platforms such as Reddit in favor of third-party reviews and publisher domains, while Google maintains a more balanced mix, leading to relatively low alignment between the two systems in product-focused queries.

\subparagraph{Software Products.}

\begin{figure}[h]
    \centering
    \includegraphics[width=0.5\textwidth]{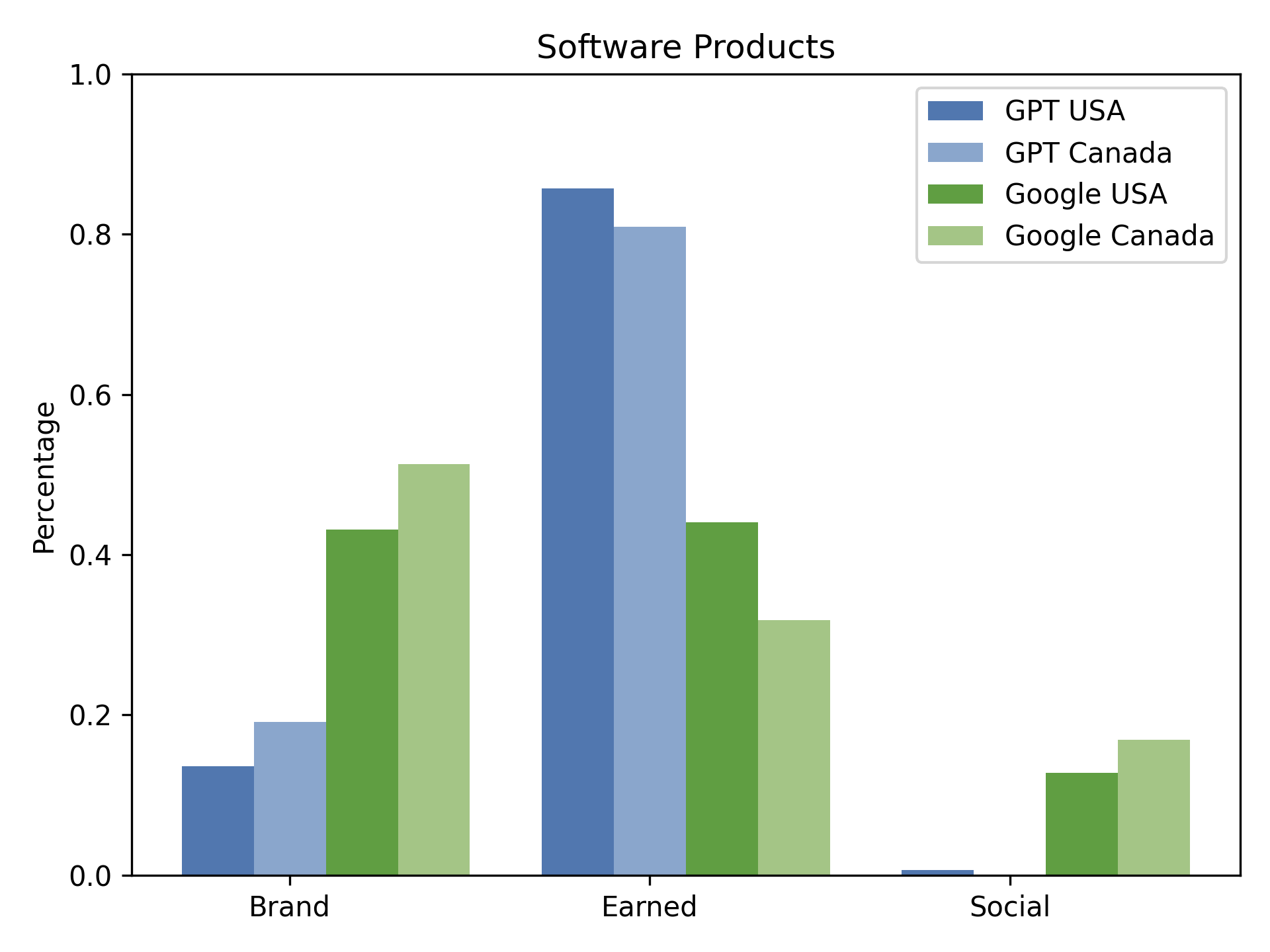}
    \caption{Software Products: Distribution of source types (Brand, Earned, Social) across Google and GPT in Canada and USA.}
    \label{fig:software-products}
\end{figure}

In software, the divergence between Google and AI search was pronounced. In Canada, Google favored Brand sources (53.8 percent) and Social (14.4 percent), with Earned making up only 31.8 percent. AI search reversed this distribution, with 74.2 percent Earned, 25.8 percent Brand, and no Social results. A similar pattern held in the United States, where Google produced 43.7 percent Brand, 10.9 percent Social, and 45.4 percent Earned, while AI search delivered 72.7 percent Earned, 26.7 percent Brand, and negligible Social. The overlap experiment provides additional context: categories like Smartwatches fell in the mid-range of overlap (around 32 percent at the top 5, 41 percent at the top 10), indicating partial agreement but still substantial divergence in source selection. This highlights a systemic shift in how software queries are surfaced, with Google emphasizing vendor-owned domains and AI search preferring neutral third-party evaluations.

\subparagraph{\bf{Cross-Category Observations.}}
Three cross-cutting trends emerged from these comparisons. First, AI search consistently weighted its results toward Earned domains, regardless of category or region, which is reflected in the relatively higher overlap in service-oriented categories (such as Airlines and Streaming Services) where both systems converge on authoritative providers. Second, Google maintained a more balanced distribution that included significant Social and Brand content, especially in consumer-focused verticals like electronics and automotive, where overlap levels were correspondingly lower. Third, the near-total absence of Social sources in AI search outputs represents a key structural shift, reducing the role of community forums and user-generated knowledge in discovery pathways and explaining much of the divergence observed in the overlap analysis.

\subsubsection{Local Search Experiment}
\paragraph{Objective.} 
Examine how AI search and Google compare in retrieving results for local businesses, focusing on overlap of cited domains in categories such as home services, healthcare, and professional services.

\paragraph{Experimental Design.} 
We issued a set of queries targeting local businesses (e.g., "best home cleaning services near me," "top dentists in [city]") to both Google and a web-enabled AI engine. For each category, we computed the domain overlap coverage between the two systems, following the methodology in \S4.1.

\paragraph{Results.} 
\begin{figure}[h]
\centering
\includegraphics[width=0.5\textwidth]{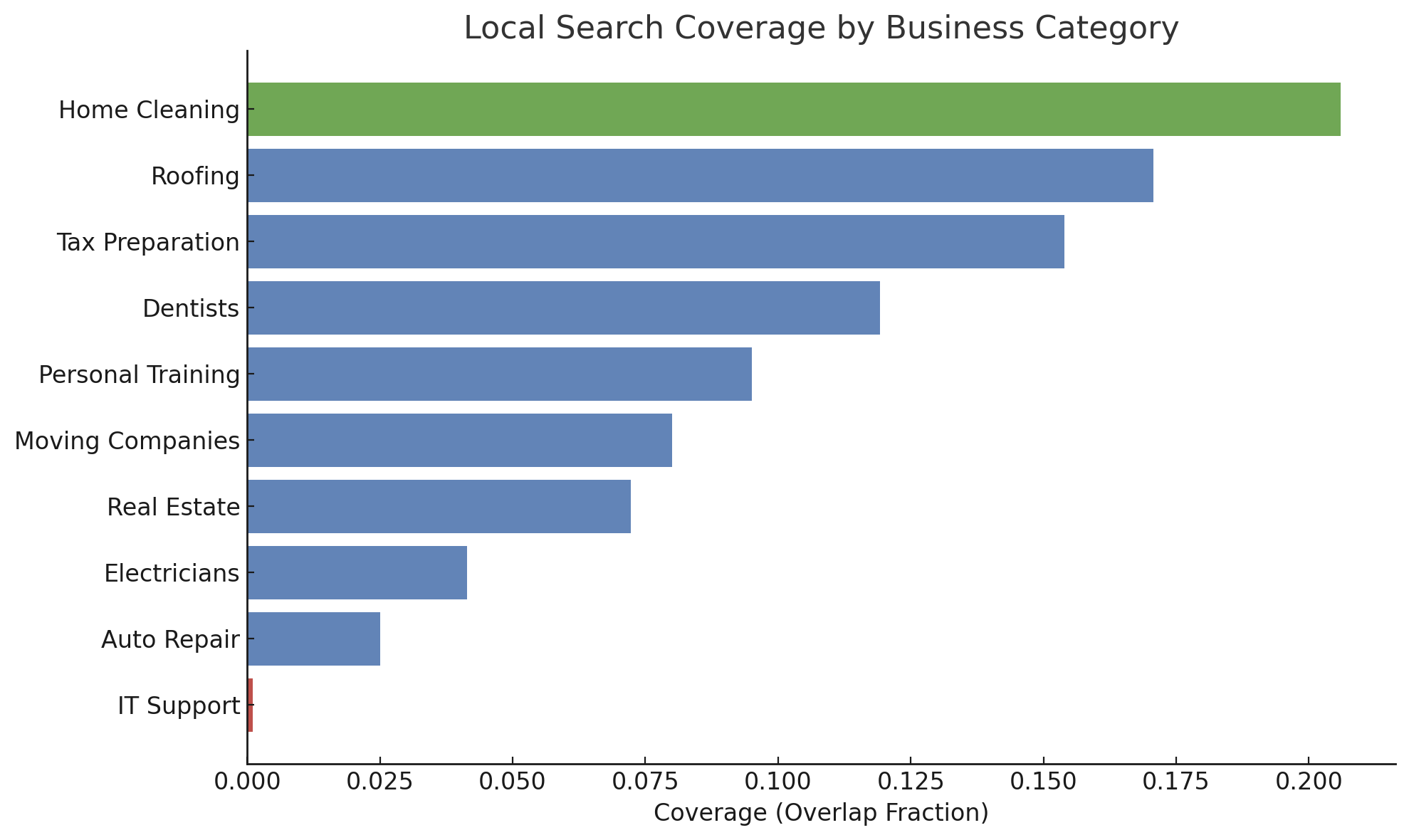}
\caption{Local Search Coverage by Business Category. Coverage reflects the fraction of overlapping domains between AI search and Google for local business queries.}
\label{fig:local-search-coverage}
\end{figure}

The results reveal that overlap varies substantially across local business categories. Home Cleaning shows the highest overlap (20.6\%), followed by Roofing (17.1\%), Tax Preparation (15.4\%), and Dentists (11.9\%). Lower overlap is observed in categories like Auto Repair (2.5\%) and IT Support (0.1\%). These findings suggest that while AI and Google sometimes converge on widely recognized service providers, the divergence is more pronounced in specialized or fragmented sectors.

\paragraph{Interpretation.} 
Overall, local search overlap is significantly lower than in broader consumer verticals such as electronics or health. This indicates that AI engines are less aligned with Google in surfacing local service providers, potentially due to differences in how local authority and business directories are weighted. For practitioners, this implies that visibility in local AI search requires distinct strategies beyond traditional Google Local SEO.

\subsubsection{Language Sensitivity Experiment}

\paragraph{Objective.}
Assess how language affects search outputs for Google vs.\ web-enabled AI engines. For queries with same intents, we examine whether the cited domains shift when prompts are translated (Chinese, Japanese, German, French, Spanish vs.\ English), and how source type and website language distributions change by system.

\paragraph{Experimental Design.}
For ten consumer verticals (smartphones, laptops, headphones, smartwatches, electric vehicles, gaming consoles, home appliances, fitness equipment, camera gear, outdoor gear), we translated a shared set of 100 English base queries (10 verticals $\times$ 10 each) into five target languages and issued them to Google and to each AI engine (Gemini, Claude, ChatGPT, and Perplexity). For each system we compared, within-engine, outputs across languages using:

\begin{itemize}
    \item \textbf{Domain-overlap heatmaps:} overlap of cited domains for each English, X pair, by vertical.
    \begin{itemize}
          \item \textit{Overlap Definition.}
            For each English–X pair and query, we compute a Jaccard overlap on the sets of cited domains ($|A\cap B|/|A\cup B|$). Vertical-level scores report the mean of these per-query Jaccard values across all queries in that vertical. For example, a Jaccard fraction of 0.10 for a single query means that 10\% of the unique domains in the combined domain sets are shared between the two language runs. An average Jaccard of 0.10 for a vertical therefore indicates that, on the average query in that vertical, 10\% of the per-query domain union is shared between English and the target language.
    \end{itemize}

    \item \textbf{Aggregate pies:} overall \textbf{domain-type} mix (brand / social / earned) and \textbf{website-language} mix (English vs.\ non-English) across all verticals.
\end{itemize}

Extraction, normalization, and classification follow the common pipeline in \S4.1.

\paragraph{Results.}

\subparagraph{Cross-language domain stability.}

\begin{figure}[h]
    \centering
    \includegraphics[width=0.5\textwidth]{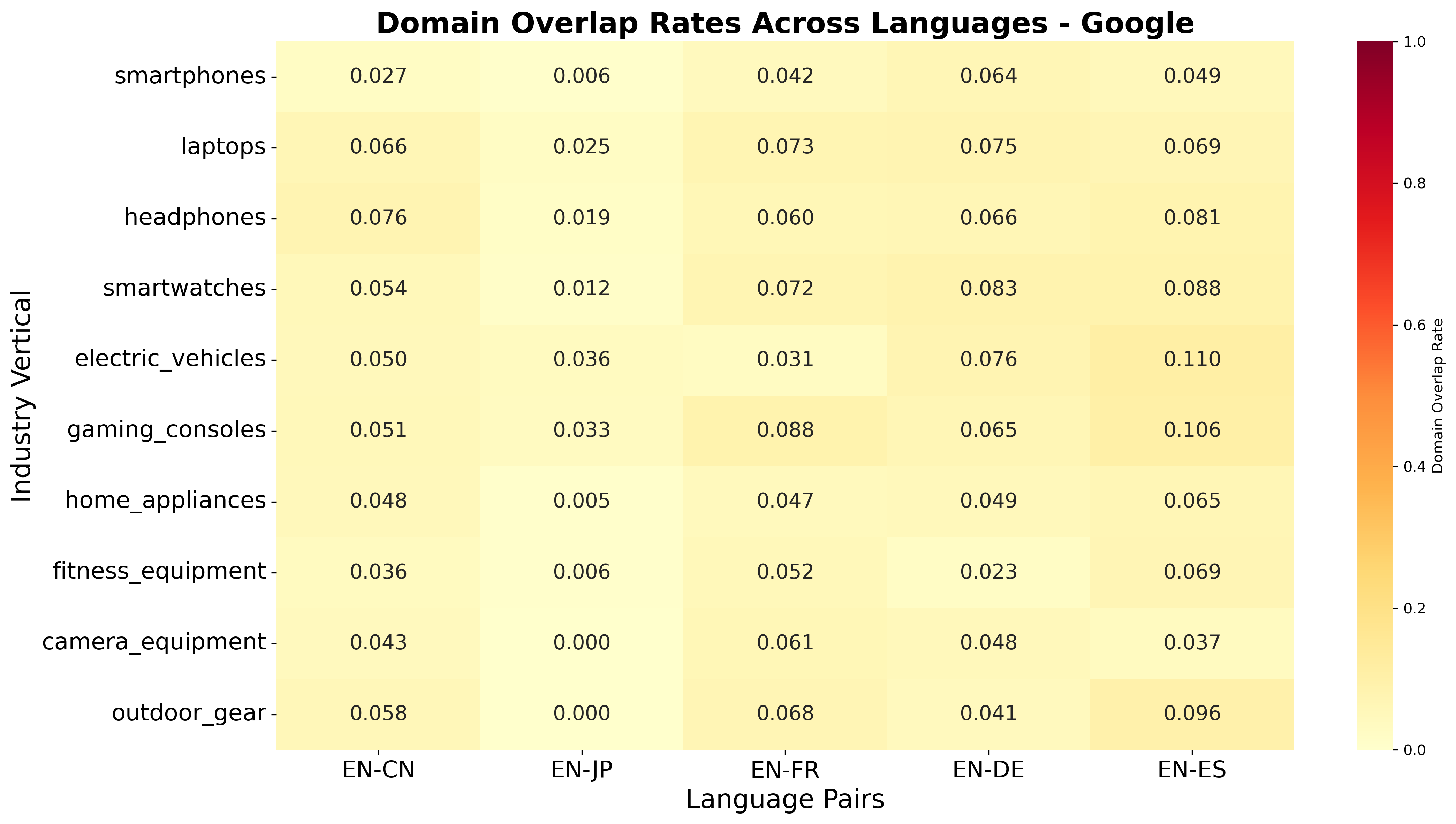}
    \caption{Language Sensitivity: Domain overlap heatmap, Google}
    \label{fig:language-domain-overlap-google}
\end{figure}

\begin{figure}[h]
    \centering
    \includegraphics[width=0.5\textwidth]{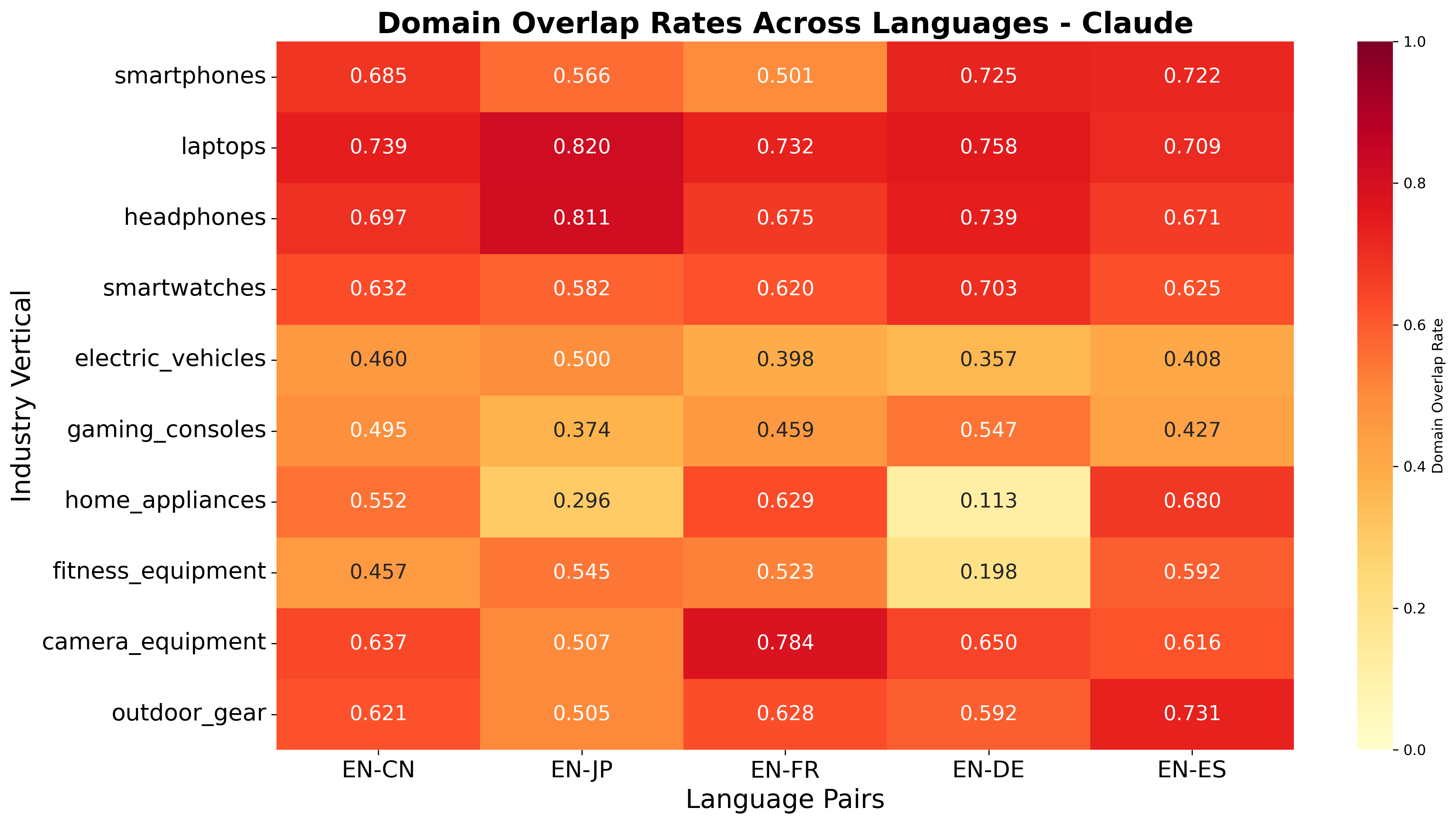}
    \caption{Language Sensitivity: Domain overlap heatmap, Claude}
    \label{fig:language-domain-overlap-claude}
\end{figure}

\begin{figure}[h]
    \centering
    \includegraphics[width=0.5\textwidth]{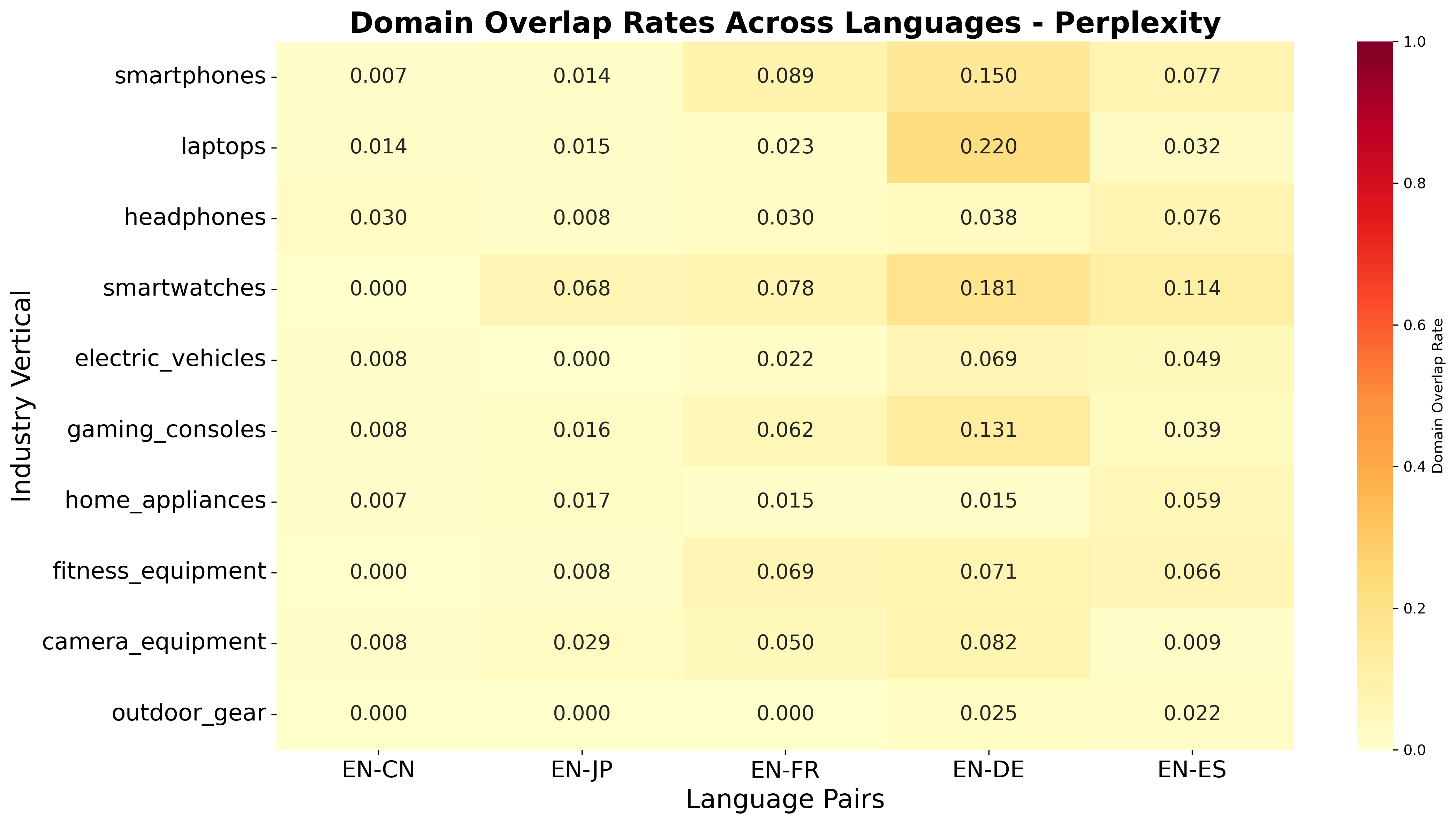}
    \caption{Language Sensitivity: Domain overlap heatmap, Perplexity}
    \label{fig:language-domain-overlap-perplexity}
\end{figure}

\begin{figure}[h]
    \centering
    \includegraphics[width=0.5\textwidth]{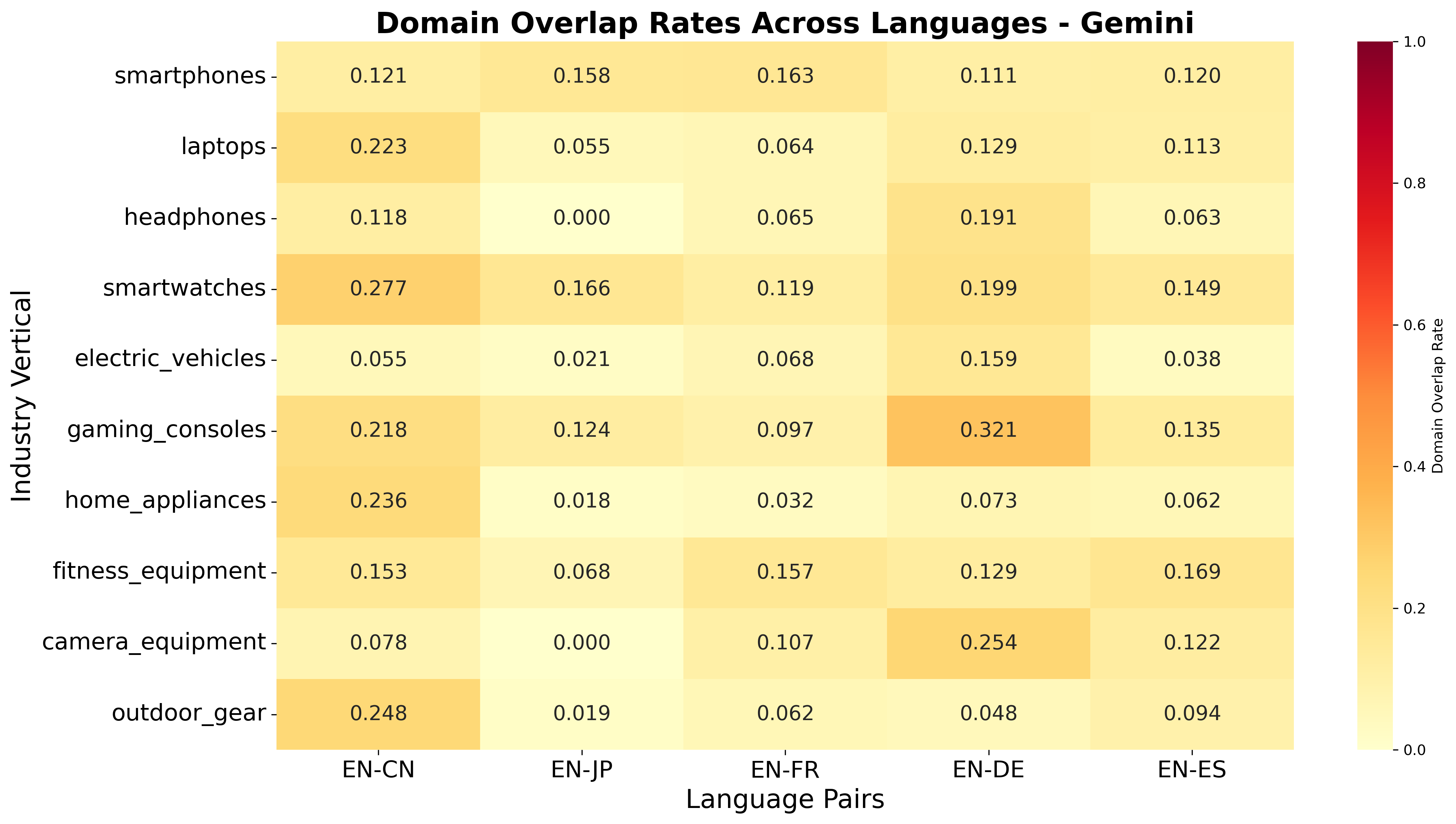}
    \caption{Language Sensitivity: Domain overlap heatmap, Gemini}
    \label{fig:language-domain-overlap-gemini}
\end{figure}

\begin{figure}[h]
    \centering
    \includegraphics[width=0.5\textwidth]{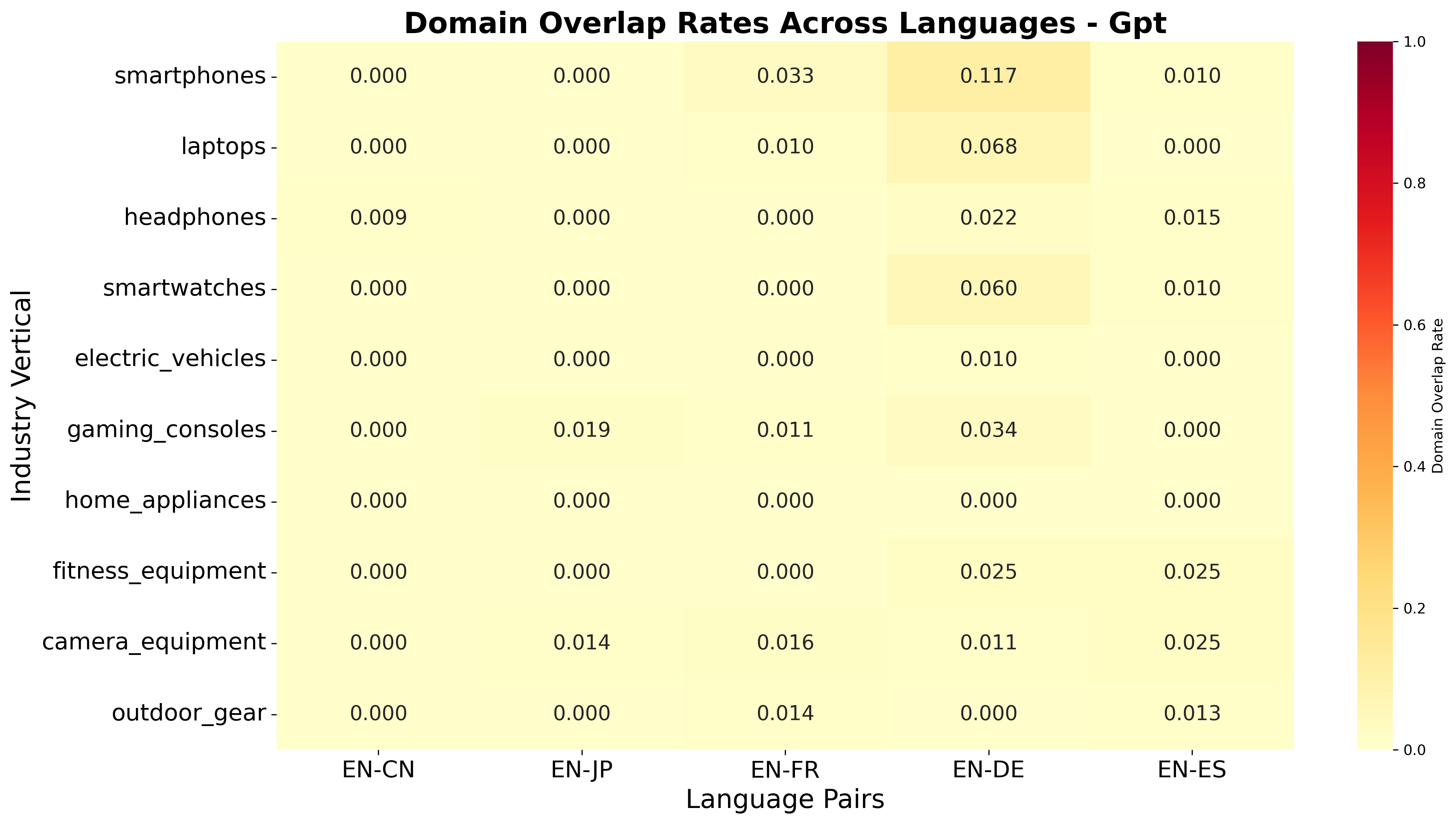}
    \caption{Language Sensitivity: Domain overlap heatmap, GPT}
    \label{fig:language-domain-overlap-gpt}
\end{figure}

Google's cross-language domain overlap is generally low (mostly between 0--0.1, with the maximum cell (EN-ES \textit{Electric vehicles}) only slightly above 0.1; Fig.~\ref{fig:language-domain-overlap-google}). Relative to this baseline:

\begin{itemize}
    \item Claude exhibits much higher cross-language stability than Google in all verticals (frequent high overlaps), indicating far stronger reuse of the same authority domains across languages (Fig.~\ref{fig:language-domain-overlap-claude}).
    \item Perplexity and Gemini show higher or comparable overlap to Google overall, with absolute levels remaining low. Relative to Google's baseline (peaking at $\approx$0.11), Gemini records modestly higher overlaps in several languages (e.g., EN--CN with many pockets exceeding 0.2; EN--DE with a peak at $\approx$0.32; Fig.~\ref{fig:language-domain-overlap-gemini}), whereas Perplexity is mostly comparable to Google with a few pockets slightly above it (e.g., EN--DE \textit{laptops} $\approx$0.22; Fig.~\ref{fig:language-domain-overlap-perplexity}). Overall, cross-language reuse remains limited for both.
    \item GPT is the special case that shows lower overlap than Google: overlaps are near zero across the board, i.e., it effectively switches to different site ecosystems by language more than Google (Fig.~\ref{fig:language-domain-overlap-gpt}).
\end{itemize}

\begin{figure}[h]
    \centering
    \includegraphics[width=0.5\textwidth]{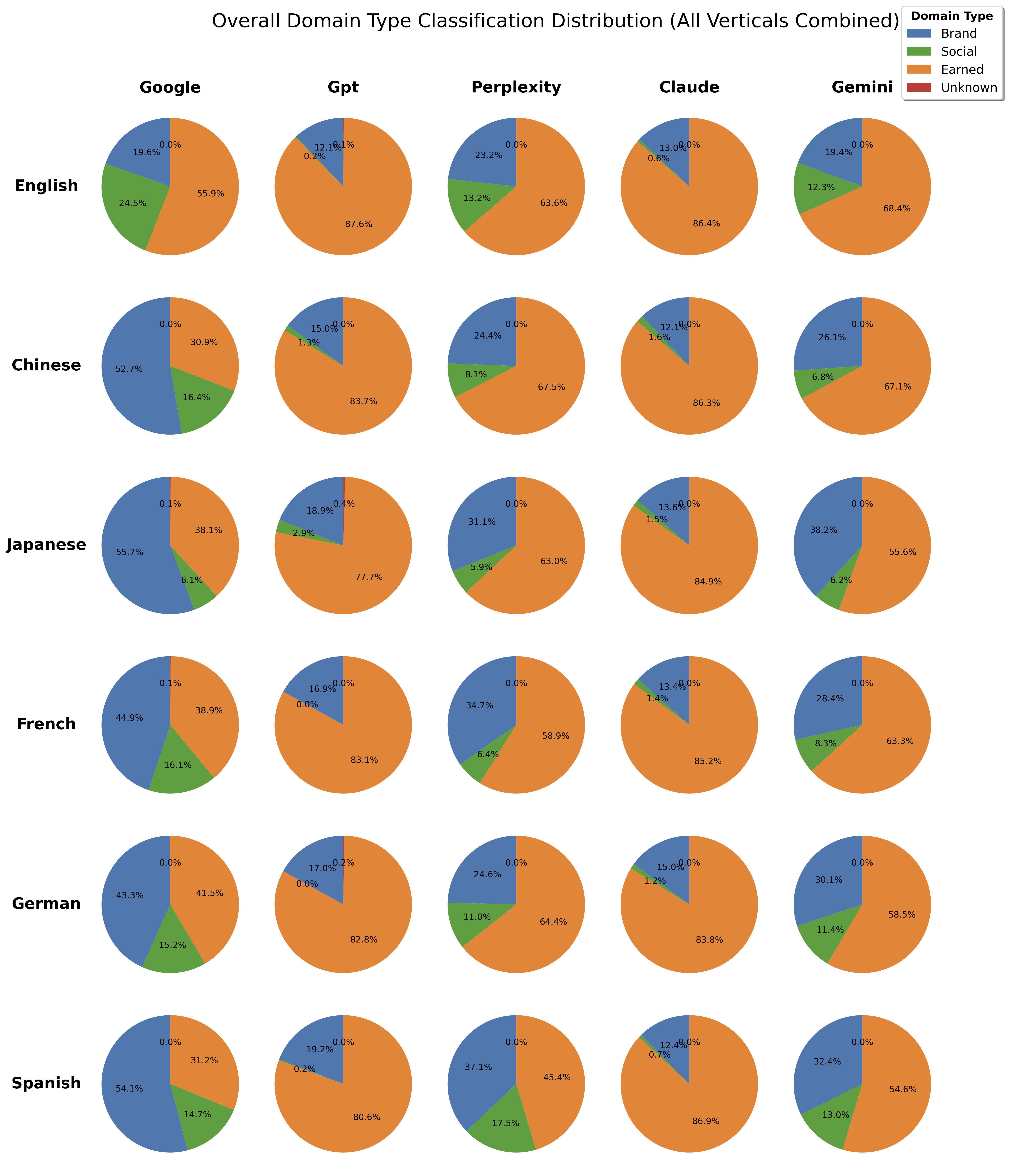}
    \caption{Language Sensitivity: Overall domain-type distribution, all languages pooled}
    \label{fig:language-overall-domain-type-distribution}
\end{figure}

\subparagraph{Source types mix.}
Across languages, AI systems remain earned-heavy relative to Google. Pooled distributions show AIs allocating a larger share to earned and less to social and brand than Google (Fig.~\ref{fig:language-overall-domain-type-distribution}). Google, by contrast, retains substantial social (especially in English) and a larger, dominant brand share in most non-English languages. The AI models follows a earned $\gg$ brand $\gg$ social pattern, while Google generally follows brand $\gg$ earned $\gg$ social instead.

\begin{figure}[h]
    \centering
    \includegraphics[width=0.5\textwidth]{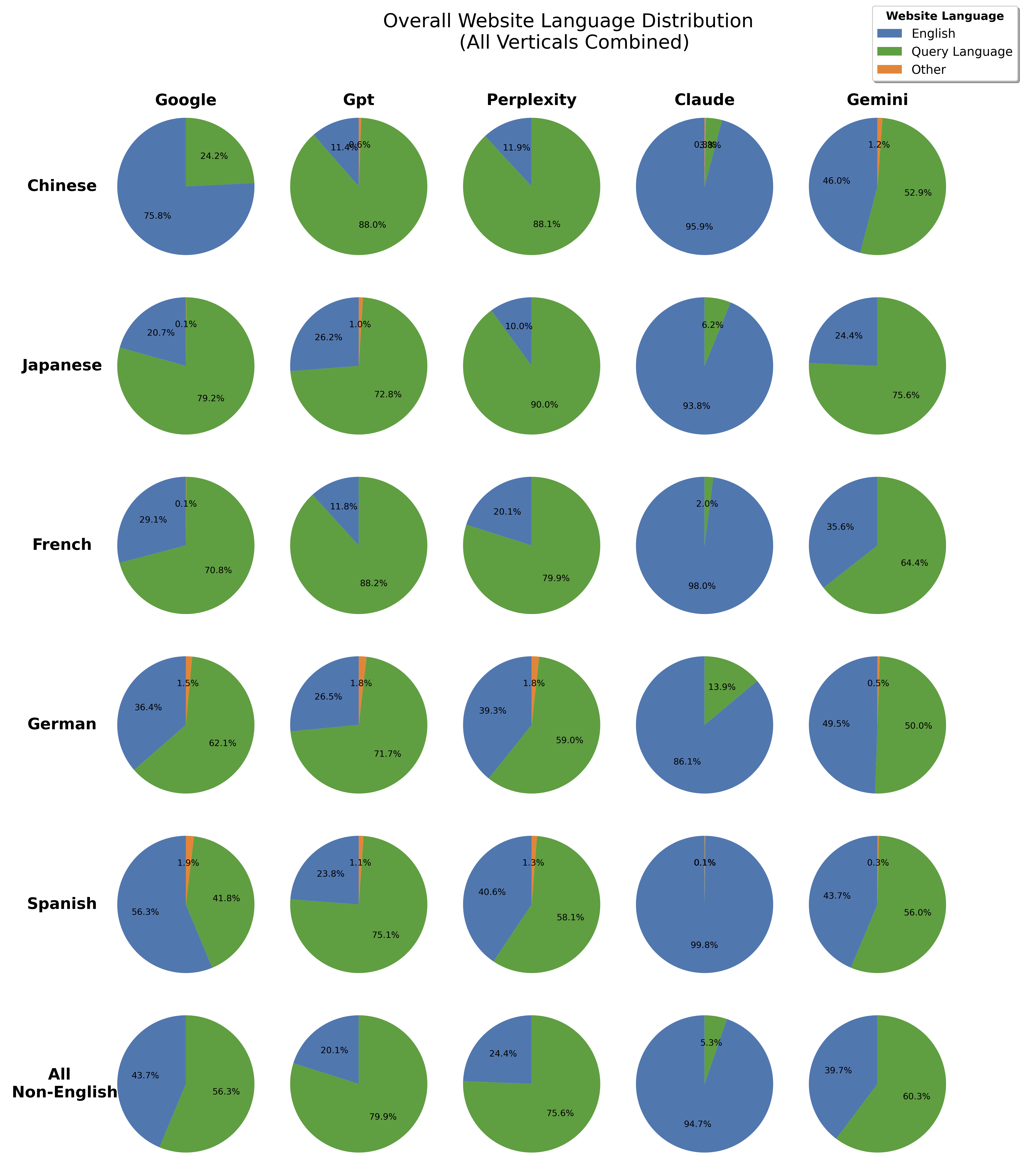}
    \caption{Language Sensitivity: Overall website-language distribution, all verticals pooled}
    \label{fig:language-website-language-distribution}
\end{figure}

\subparagraph{Website language depends on the engine.}
Under non-English prompts, citations tilt toward the target language---but less so for Google. In the pooled view, GPT and Perplexity are more local-language heavy than Google; Claude is far more English-heavy than Google; Gemini is closer to balanced, with the split varying by language (Fig.~\ref{fig:language-website-language-distribution}). For Google specifically, a gradient is evident: Japanese yields the strongest localization (target-language exceeding three-quarters of citations), French and German are also highly localized but keep a larger English slice than Japanese, Spanish is mixed with English slightly exceeding the target language, and Chinese is the exception---English dominates, accounting for over three-quarters of citations.

\paragraph{Interpretation.}
Compared with Google, Claude maintains a much more stable cross-language evidence set; GPT diverges the most by swapping site ecosystems across languages; Perplexity and Gemini typically align with or modestly higher than Google's already-low cross-language domain reuse, with isolated exceptions. Regardless of language, AI engines collectively bias toward editorial/earned sources more than Google. Practically, brands and publishers aiming for multilingual visibility should build coverage in authoritative local-language media (to meet AI engines that localize heavily) while also accounting for Google's more English-leaning pattern in many non-English contexts.

\subsubsection{Paraphrase Sensitivity Experiment}

\paragraph{Objective.}
Assess whether small, within-language changes in how a query is phrased alter the alignment between Google and web-enabled AI engines. For the same intent, we ask how paraphrasing affects (i) the overlap between AI citations and Google results and (ii) the mix of cited source types (brand / social / earned) by system.

\paragraph{Experimental Design.}
Using the same 100 base queries from ten consumer verticals as in the language study, we issued a base prompt and seven paraphrase templates to Google and to each AI engine (Gemini, ChatGPT, Perplexity):

\begin{enumerate}
    \item \textit{justification\_required,} require a short justification for each item;
    \item \textit{source\_required,} require sources consulted for the answer;
    \item \textit{quote\_required,} require direct quotes from consulted sources;
    \item \textit{confidence\_score,} require confidence scores per item;
    \item \textit{ranked\_order,} explicitly require ranking best$\rightarrow$worst;
    \item \textit{imperative\_list,} rewrite from question to imperative list;
    \item \textit{keyword\_only,} compress to keywords only.
\end{enumerate}

For each system we compared, within-engine, the outputs of each paraphrase against the base prompt:

\begin{itemize}
    \item \textbf{Domain overlap} (heatmaps): overlap of cited domains by vertical. As in the language sensitivity experiment, for each base-paraphrase pair and query we compute the Jaccard overlap on cited-domain sets, then average per vertical to obtain the reported stability scores.
    \item \textbf{Domain-type mix} (brand / social / earned) aggregated across all verticals for each paraphrase (overall pies).
\end{itemize}

Extraction, normalization, and classification follow the common pipeline in \S4.1.

\paragraph{Results.}

\subparagraph{Paraphrases move results less than languages.}
Relative to \S4.2.3 (language sensitivity test), paraphrasing induces smaller perturbations for both Google and AI. Domain sets are much more stable than in the cross-language study.

\begin{figure}[h]
    \centering
    \includegraphics[width=0.5\textwidth]{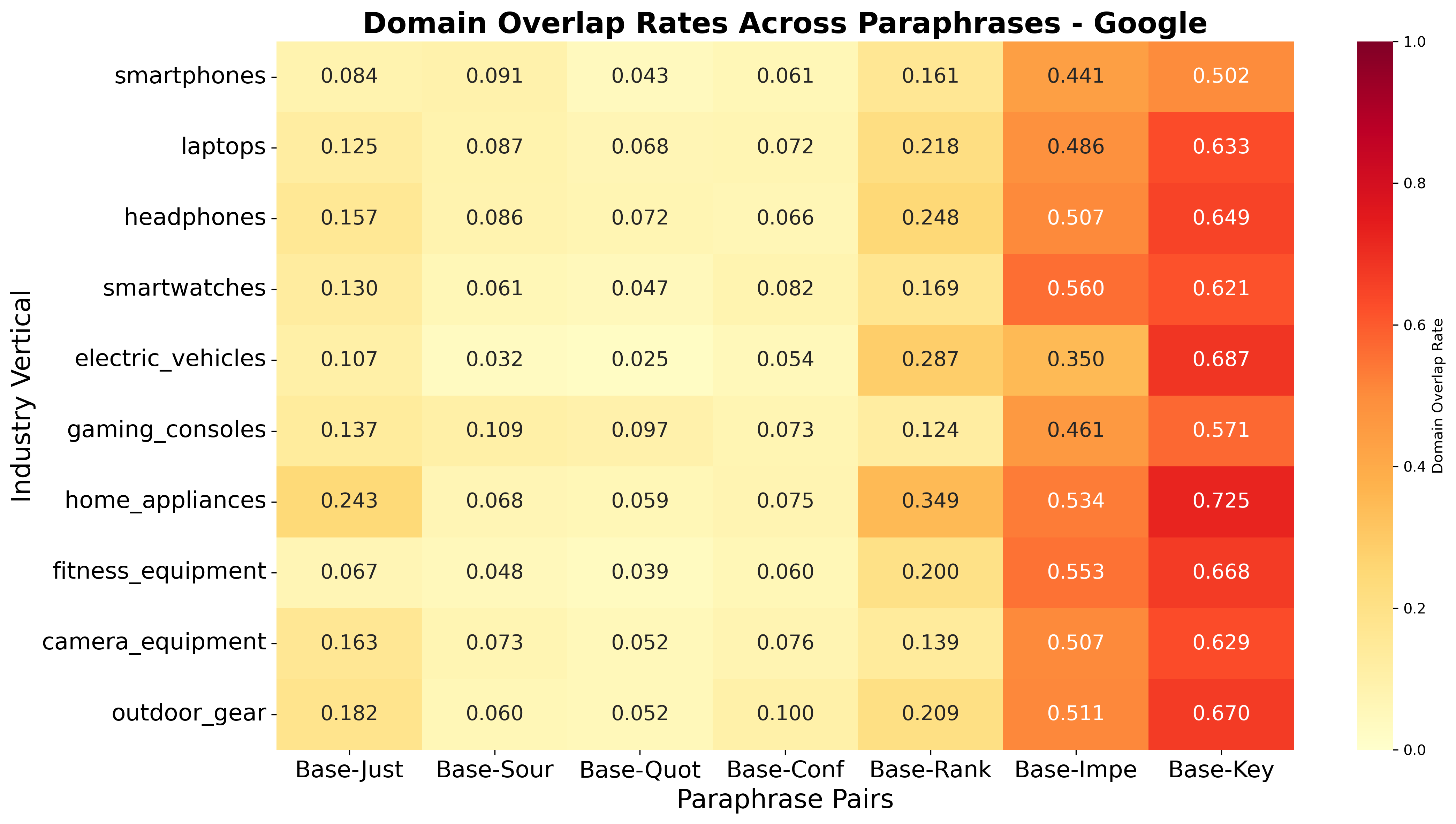}
    \caption{Paraphrase Sensitivity: Domain overlap heatmap, Google}
    \label{fig:paraphrase-domain-overlap-google}
\end{figure}

\subparagraph{Domain stability.}
Google shows low to moderate overlap for most paraphrase styles (often near 0.1), but rises substantially for formatting-type variants---especially \textit{imperative list} and \textit{keyword-only} (many cells 0.5--0.7, peaking at $\approx$0.73). This suggests Google's top results are most stable when the rewording mainly changes form rather than intent (Fig.~\ref{fig:paraphrase-domain-overlap-google}). Also, it is notable that the domain overlap of Google on paraphrase change is generally higher (i.e., domain more stable) than that on language change.

\begin{figure}[h]
    \centering
    \includegraphics[width=0.5\textwidth]{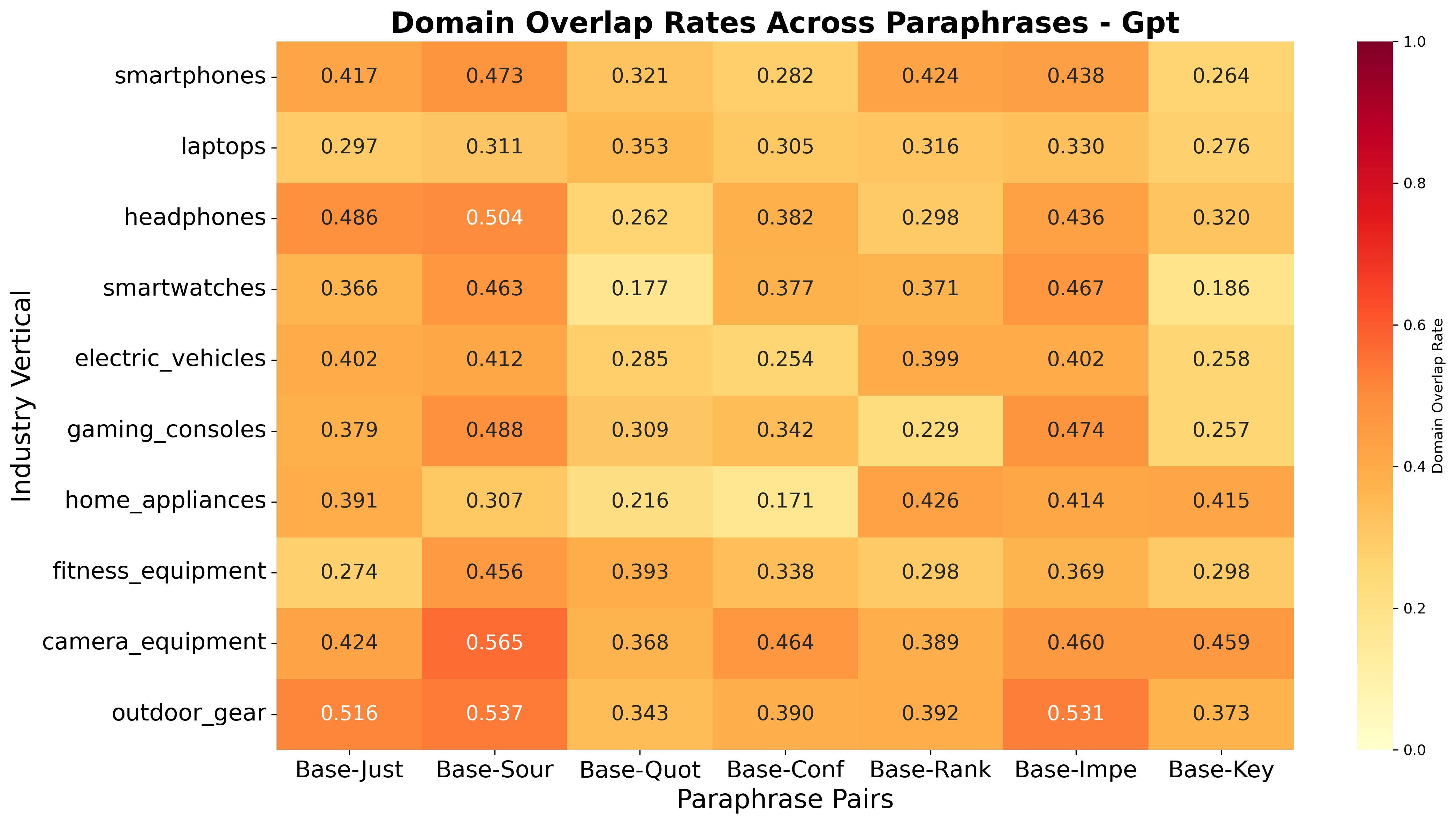}
    \caption{Paraphrase Sensitivity: Domain overlap heatmap, GPT}
    \label{fig:paraphrase-domain-overlap-gpt}
\end{figure}

\begin{figure}[h]
    \centering
    \includegraphics[width=0.5\textwidth]{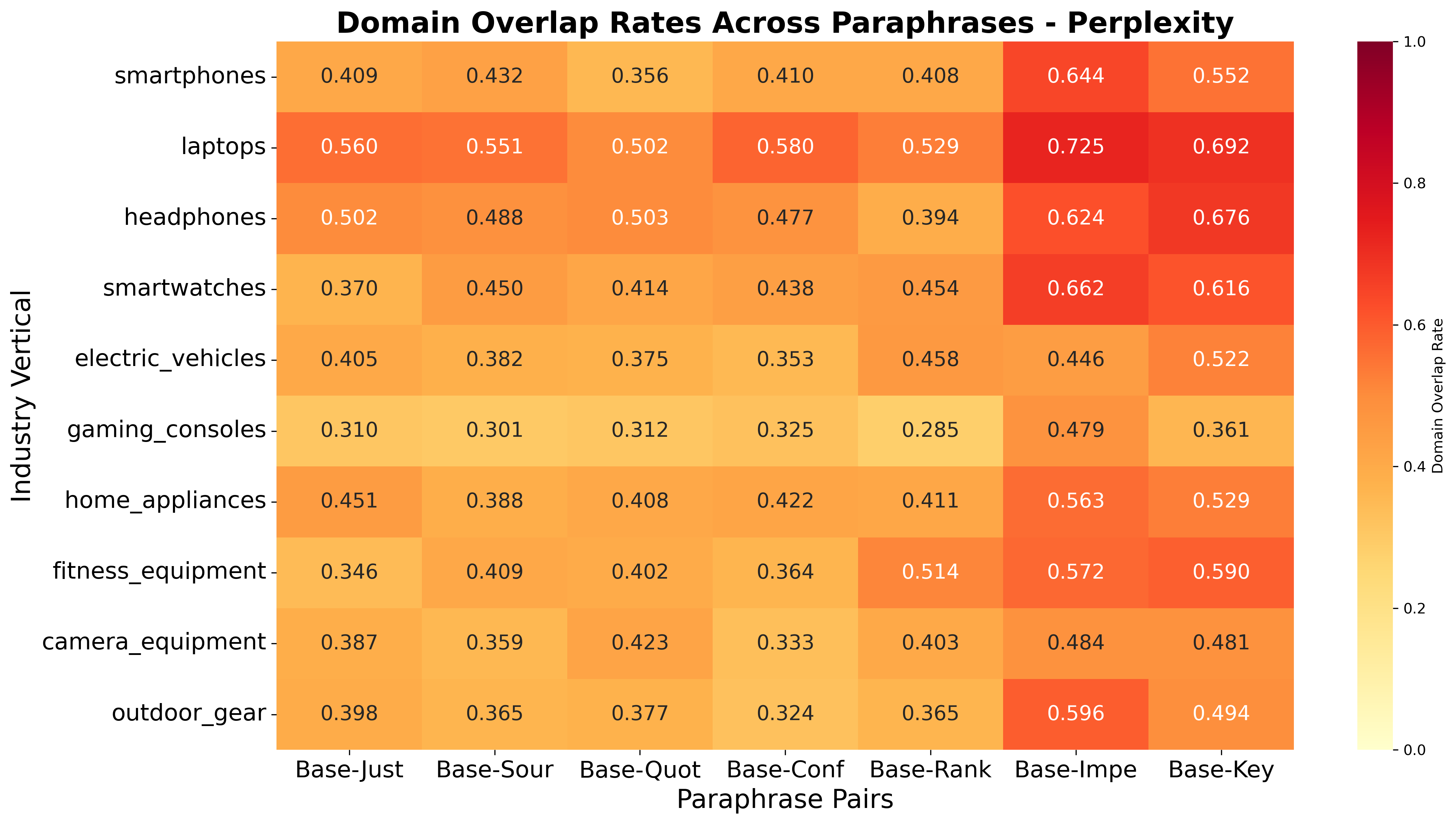}
    \caption{Paraphrase Sensitivity: Domain overlap heatmap, Perplexity}
    \label{fig:paraphrase-domain-overlap-perplexity}
\end{figure}

\begin{figure}[h]
    \centering
    \includegraphics[width=0.5\textwidth]{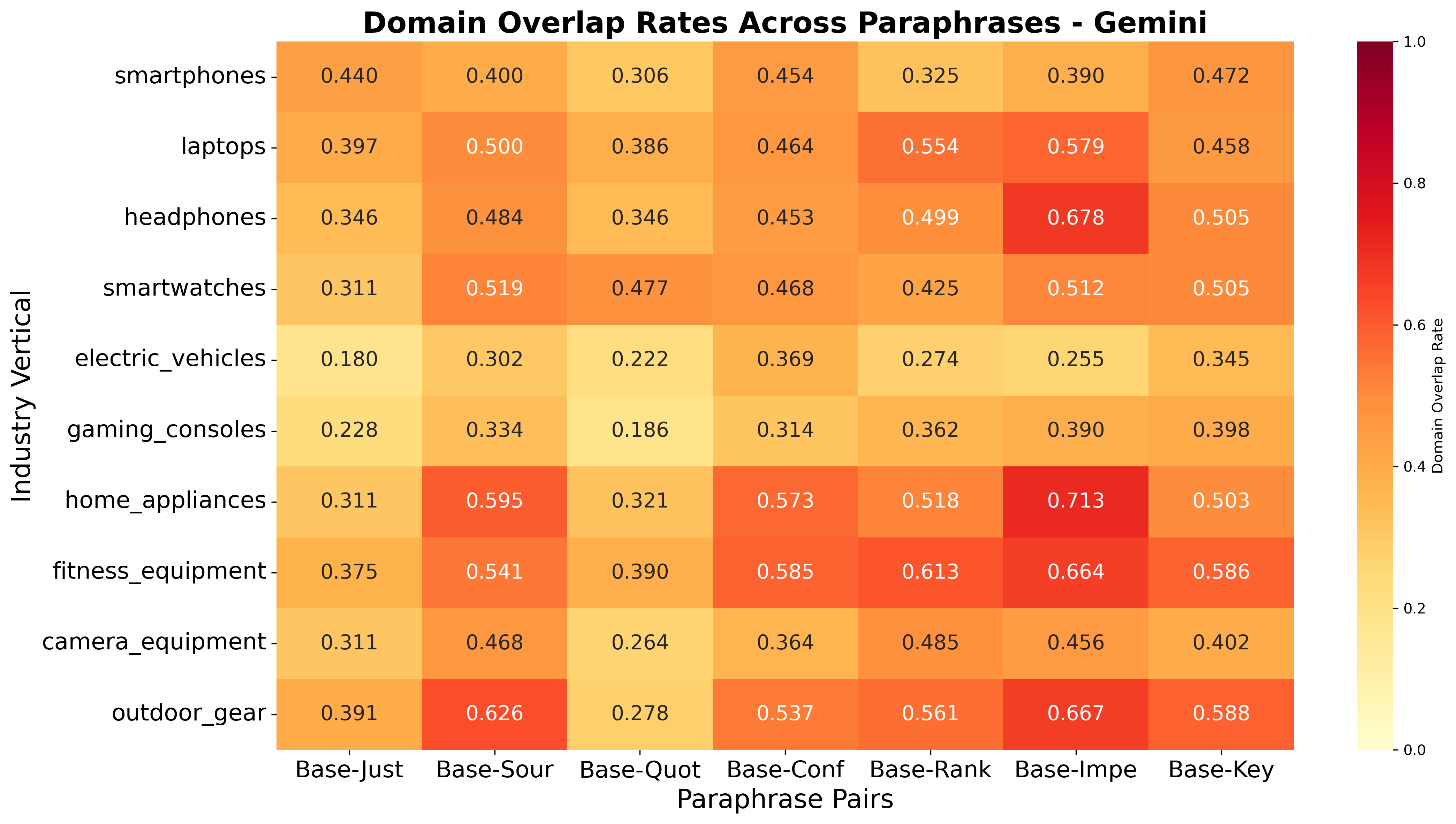}
    \caption{Paraphrase Sensitivity: Domain overlap heatmap, Gemini}
    \label{fig:paraphrase-domain-overlap-gemini}
\end{figure}

AI engines exhibit generally higher cross-paraphrase domain stability than Google in most verticals, with many base--variant overlaps in the 0.3--0.7 range, occasionally reaching the low-0.7s. In short, AI results tend to be less sensitive to wording tweaks than Google, outside Google's strong ``imperative/keyword'' cases (Figs.~\ref{fig:paraphrase-domain-overlap-gpt}, \ref{fig:paraphrase-domain-overlap-perplexity}, \ref{fig:paraphrase-domain-overlap-gemini}).

\begin{figure}[h]
    \centering
    \includegraphics[width=0.5\textwidth]{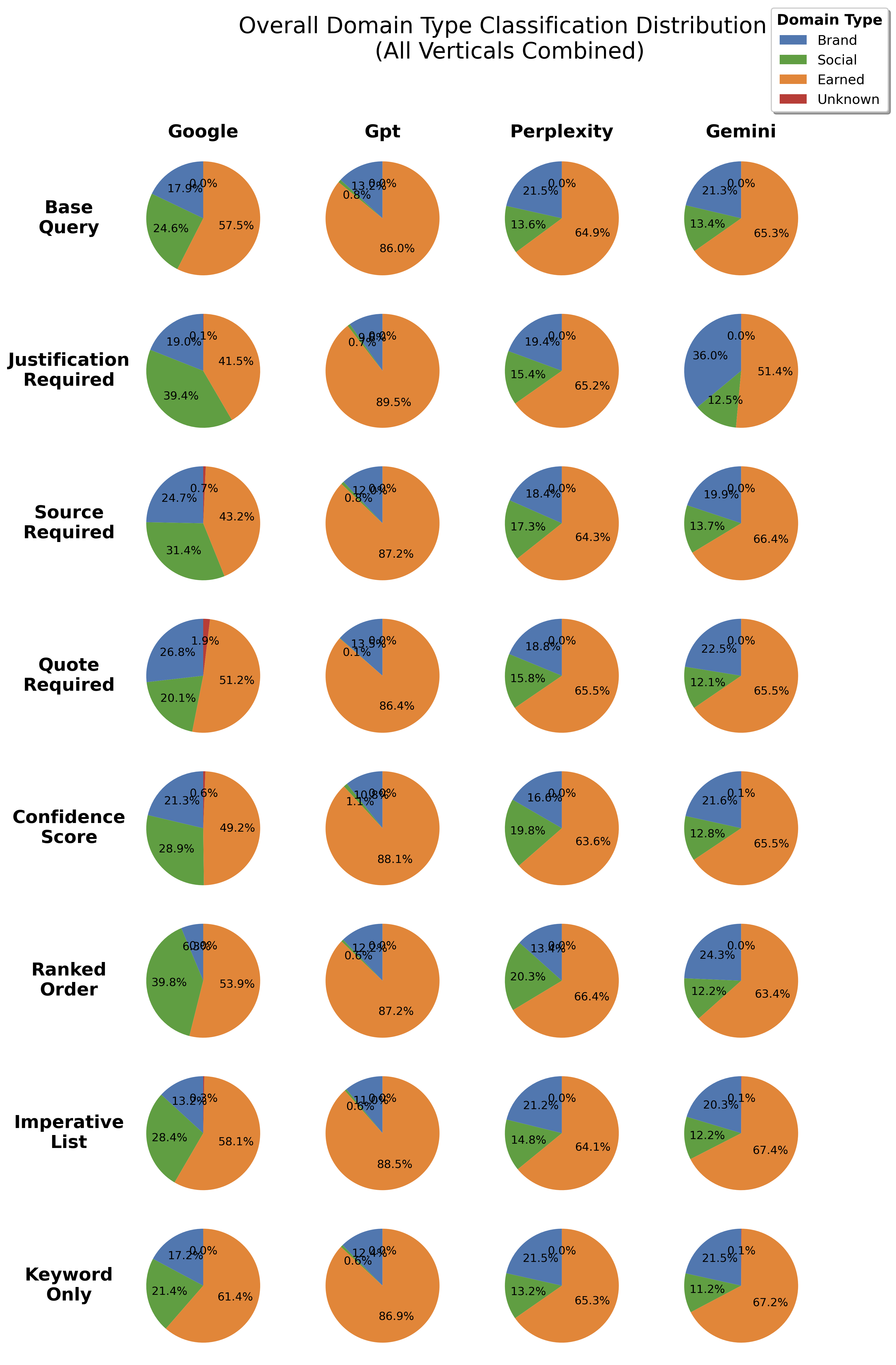}
    \caption{Paraphrase Sensitivity: Overall domain-type distribution across paraphrase styles}
    \label{fig:paraphrase-domain-type-distribution}
\end{figure}

\subparagraph{Source-type mix across paraphrase styles.}
Across paraphrases, Google maintains a relatively higher share of Social and lower share of Earned than AI, while AI systems remain Earned-heavy. Importantly, within-system distributions change little from one paraphrase to another for AIs, but change more for Google---paraphrasing materially shifts the Brand/Social/Earned balance for Google more than AIs (Fig.~\ref{fig:paraphrase-domain-type-distribution}).

\paragraph{Interpretation.}
Google is comparatively more sensitive to paraphrasing, except when the paraphrase is a simple format change (imperative/keyword list), where stability rises. AI engines are steadier across paraphrases and consistently favor editorial (earned) sources over brand/social more than Google. Practically, paraphrase tuning can influence \textit{citations}---especially on Google---but has a smaller effect than \textit{language choice} (\S4.2.2).

\subsection{General Query Types}

We now focus on a generalized comparison, moving beyond ranking-style prompts and considering how systems respond across canonical query classes.  

\subsubsection{Overall Description}

To analyze how traditional search engines and AI-powered systems structure their responses, we categorized user queries into three canonical intent buckets: \textit{Informational}, \textit{Consideration}, and \textit{Transactional}. This taxonomy reflects both user motivation and the expected media mix that search systems return.  

\begin{figure}[h]
    \centering
    \includegraphics[width=0.5\textwidth]{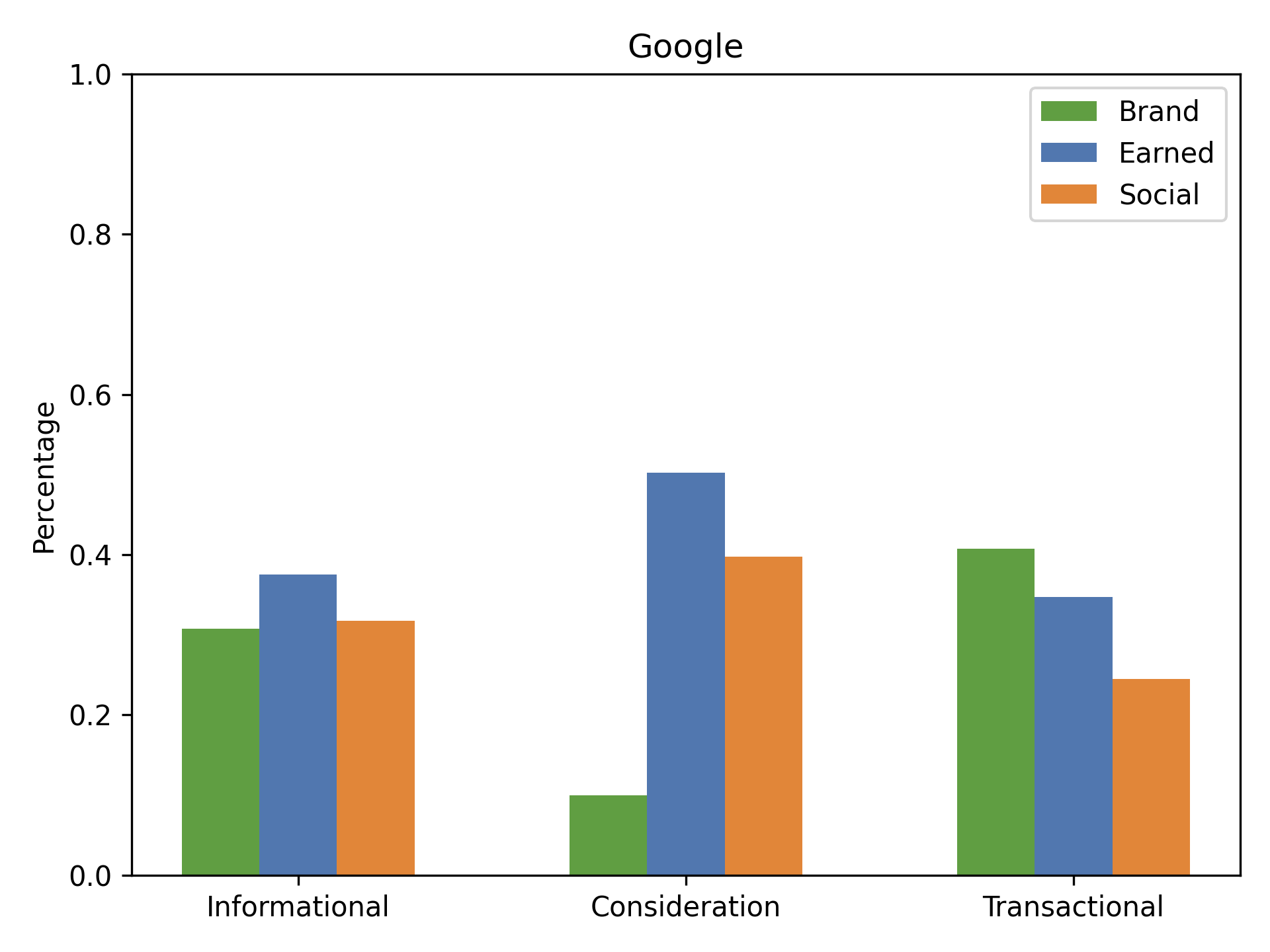}
    \caption{Media-type distribution (Brand, Earned, Social) across query intents for \textbf{Google}.}
    \label{fig:query-types-google}
\end{figure}

\begin{figure}[h]
    \centering
    \includegraphics[width=0.5\textwidth]{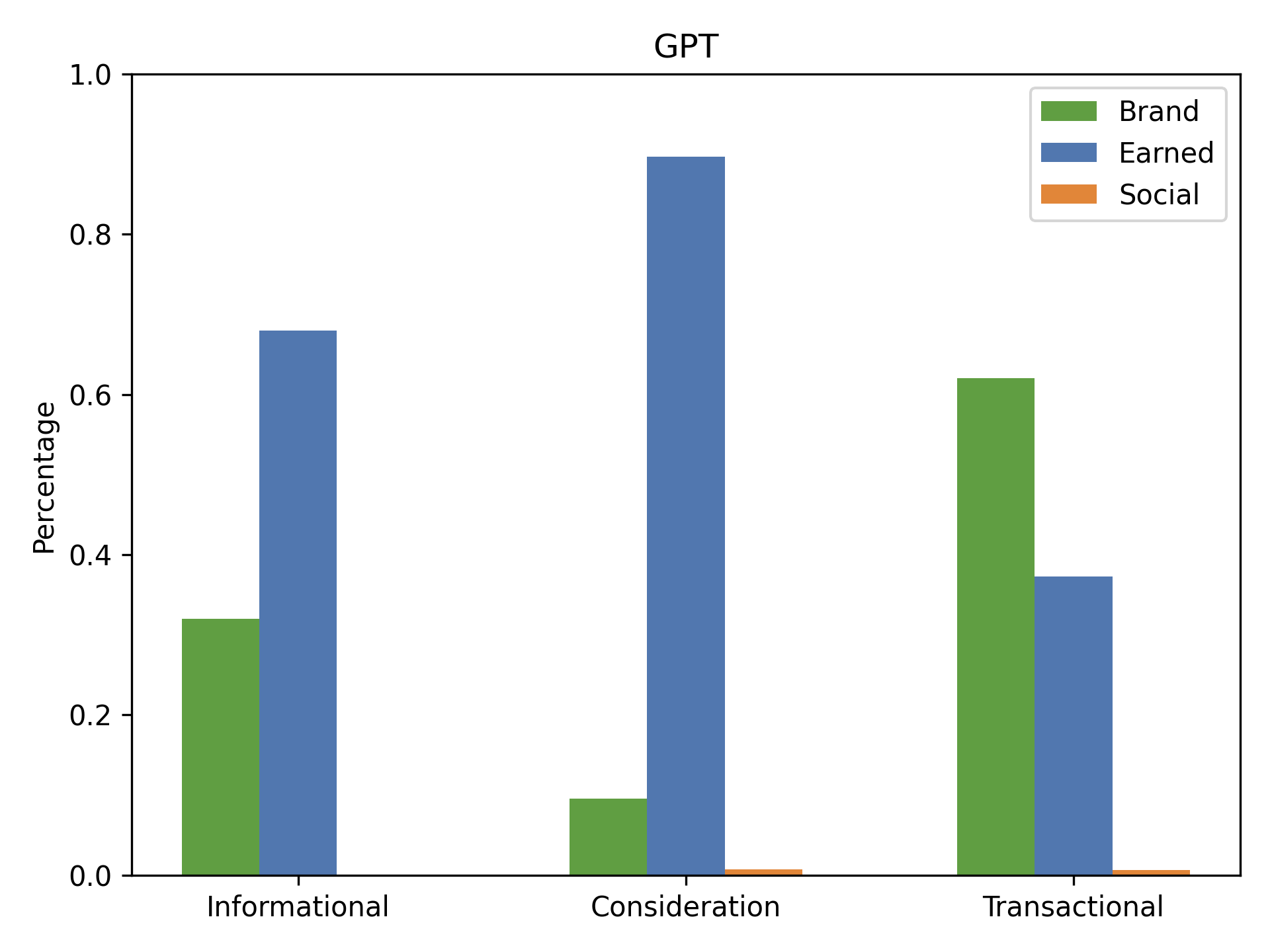}
    \caption{Media-type distribution (Brand, Earned, Social) across query intents for \textbf{GPT}.}
    \label{fig:query-types-gpt}
\end{figure}

\begin{figure}[h]
    \centering
    \includegraphics[width=0.5\textwidth]{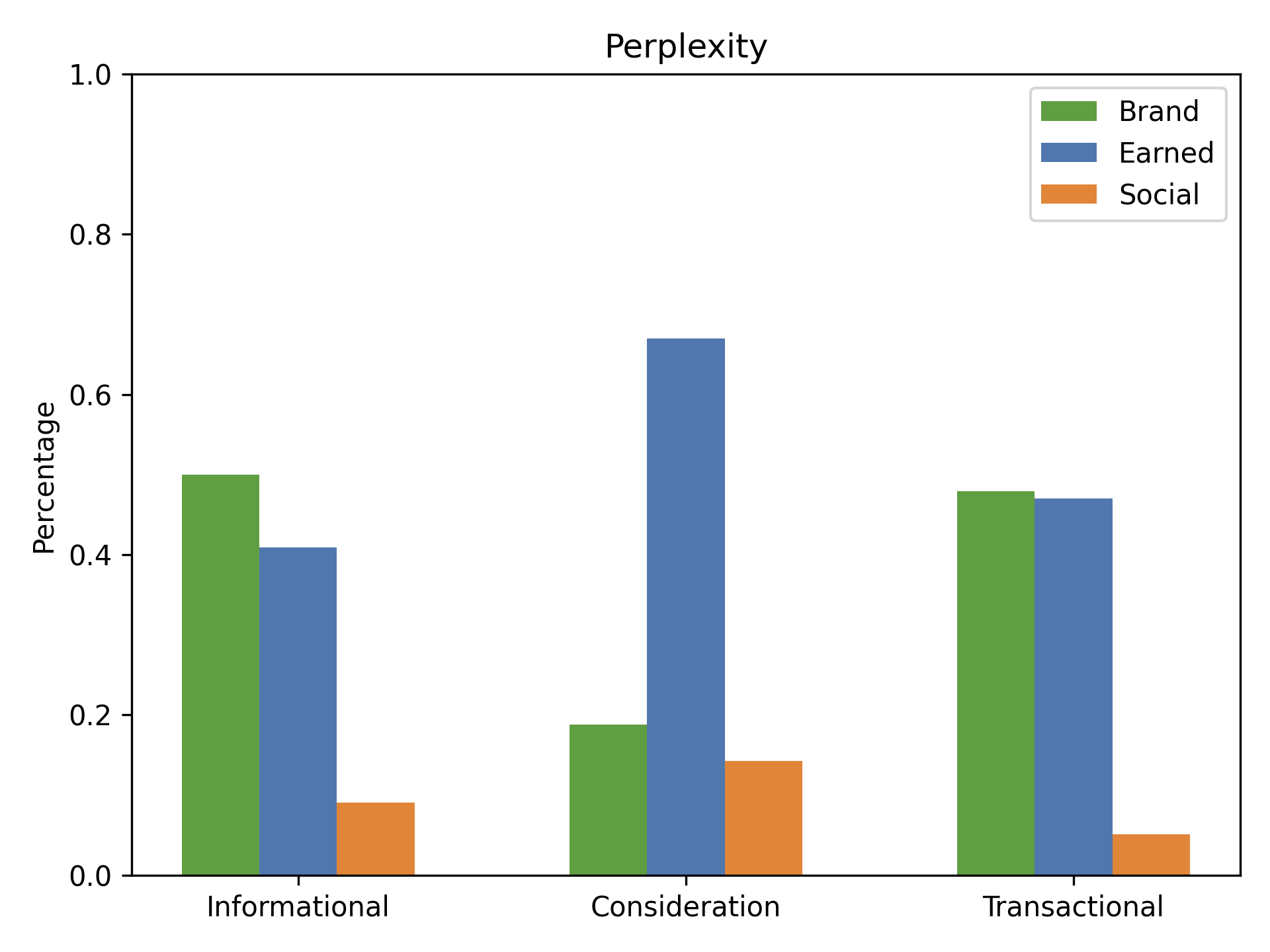}
    \caption{Media-type distribution (Brand, Earned, Social) across query intents for \textbf{Perplexity}.}
    \label{fig:query-types-pplx}
\end{figure}

\paragraph{Informational Queries.}  
Informational queries are exploratory in nature, aimed at acquiring general knowledge or technical understanding (e.g., ``How do OLED TVs work?'', ``What is Wi-Fi 7?''). These do not signal purchase intent but instead focus on definitions and explanations.  
Our results show distinct strategies: Google balances Brand, Earned, and Social sources; GPT emphasizes Earned content while nearly excluding Social; Perplexity highlights Brand domains most strongly while still including Earned and some Social.  

\paragraph{Consideration Queries.}  
Consideration queries represent mid-funnel behavior, where users compare products and evaluate trade-offs (e.g., ``Best laptops for students 2025'', ``Garmin vs Apple Watch'').  
Here, Earned content dominates across all systems, but with important differences: Google pairs Earned with a substantial Social component, GPT skews almost entirely to Earned with minimal Brand and Social, while Perplexity maintains a more balanced mix across all three categories.  

\paragraph{Transactional Queries.}  
Transactional queries are high-intent, purchase-oriented prompts (e.g., ``Buy iPhone 15 online'', ``Best price for Samsung Galaxy S24 unlocked'').  
Across systems, Brand content rises in prominence. Google shifts toward Brand while still surfacing Earned and Social, GPT amplifies Brand content most strongly with Earned secondary and almost no Social, while Perplexity distributes results between Brand and Earned with slightly more Social retained than others.  

\paragraph{Summary.}  
These plots (Figures~\ref{fig:query-types-google}--\ref{fig:query-types-pplx}) reveal intent-driven shifts in media sourcing. Google balances Earned with Social and Brand across intents, GPT consistently suppresses Social while emphasizing Earned (and Brand for transactional), and Perplexity adopts a blended approach with notable Brand inclusions. The structured differences confirm that query intent modulates how each system assembles its output.  

\subsection{The Need for generative Engine Optimization}

Traditional SEO techniques remain necessary for visibility but are insufficient for AI search dominance. The core principles of technical SEO—such as having a well-structured, crawlable site—are foundational, as AI agents require clean, machine-readable data to function. However, the findings reveal that a new strategy, which can be termed Generative Engine Optimization (GEO), is required to address the fundamental differences in how AI systems source and prioritize information.

\subsubsection{Prioritize Earned Media and Authority Building}

The most significant finding is the AI engines' overwhelming and consistent bias toward Earned media across all regions and verticals. Unlike Google, which maintains a balanced mix of Brand, Social, and Earned content, AI systems heavily favor third-party, authoritative sources like professional reviews, publisher domains, and institutional sites.

Shift investment from strategies focused solely on brand-owned content and social engagement to a concerted effort in earning third-party coverage. This includes:
\begin{itemize}
    \item
        Public and Media Relations: Proactively seeking features, reviews, and mentions in authoritative publications within your industry.
\item
        Expert Collaborations: Partnering with industry experts, thought leaders, and credible institutions to create and promote content that demonstrates your expertise.
\item
        Link Building: Earning backlinks from these high-authority, earned domains is not just a Google ranking factor but a direct input into the AI's perception of your brand's trustworthiness.
\end{itemize}

\subsubsection{Structure Content for Scannability and Justification}

The low overlap in cited domains between Google and AI, particularly in product categories, indicates that AI synthesizes information differently. It seeks clear, unambiguous data to build justified recommendations.

 Optimize website content not just for keywords, but for scannability and justification. Ensure your content explicitly highlights key decision-making factors in a format easily extracted by an AI:
\begin{itemize}
\item
        Create detailed comparison tables (vs. competitors or previous models).
\item
        Use clear, bulleted pros and cons lists.
\item
        State your value proposition explicitly (e.g., "longest battery life," "most durable build," "best value for money").
\item
        Implement schema markup (Schema.org) with extreme rigor for all product specifications, prices, reviews, and availability to become an "API-able" brand that AI agents can easily parse.
\end{itemize}       

\subsubsection{Develop a Language-Specific Authority Strategy}

The language sensitivity experiment reveals that AI engines handle multilingual queries differently. Claude shows high cross-language stability, often reusing authoritative English-language domains, while GPT completely swaps its domain ecosystem to favor local-language sources.

A one-size-fits-all multilingual SEO strategy is ineffective. To maximize AI presence globally, you must:

\begin{itemize}
\item
        For Claude-like Engines: Strengthen your position in top-tier, English-language earned media, as this authority can transfer across languages.
\item
        For GPT-like Engines: Build relationships with authoritative publishers and review sites in each target language and region. Simply translating your own brand content is not enough; you need earned coverage in the local language.
\item
        For Google: Maintain a hybrid approach, as it shows a gradient of localization depending on the language.
\end{itemize}

\subsubsection{Create Comprehensive, Lifecycle-Oriented Content}

The moderate overlap in service-oriented categories (e.g., Airlines) suggests AI values comprehensive, authoritative information. A gap in content for any stage of the customer journey (e.g., post-purchase support) could lead to a competitor being recommended when a user asks for help.

Audit and create content for the entire customer lifecycle, not just top-of-funnel discovery. This includes:
\begin{itemize}
\item
        Awareness: Educational guides and "best X for Y" articles.
\item
        Consideration: Detailed product comparisons and testimonials.
\item
        Decision: Clear data on pricing, warranty, and shipping.
\item
        Post-Purchase: Robust FAQs, troubleshooting guides, and tutorials.
\item
        Loyalty: Content on accessories, advanced uses, and trade-in programs.
\end{itemize}

In conclusion, while technical SEO provides the necessary foundation, optimizing for AI search requires a paradigm shift. Success hinges on building verifiable, third-party authority, structuring content for machine synthesis, and implementing a nuanced, language-aware strategy that prioritizes earned media over brand-owned and social content.

\section{A Comparison of AI Search Engines}

\subsection{Methodology}

All experiments in \S5 reuse the pipeline (or at least part of the pipeline), described in \S4.1 (query generation/transformation, web-enabled engine execution, domain/brand extraction, normalization/classification, and overlap/distribution metrics). Unless otherwise stated in each subsection, we do not collect Google search results in \S5. Any experiment-specific choices and deviations (e.g., \textit{queries used}, the \textit{engines compared}, and the \textit{labels computed}) are documented in their respective subsections.

We focus on comparing the results of different AI search services and understanding their similarities and differences across several important search and information discovery pipelines.

\subsection{Comparative Analysis}

\subsubsection{Well-Known Brands vs.~Niche Brands}

Our first experiment, aims to understand how AI search services respond to questions around brands. In particular we pose ranking style questions for brands (for a set of well known and niche brands) and observe the answers across services recording their agreement.

When comparing the results of different AI search engines, a clear pattern emerges in how they prioritize sources for well-known versus niche brands. Across all systems, there is a consistent emphasis on \textit{Earned} domains, but the balance between \textit{Brand}, \textit{Social}, and \textit{Earned} sources differs significantly.

\paragraph{Experimental Design.}  
The brand experiment was designed to evaluate how Claude, Gemini, ChatGPT, and Perplexity handle queries about well-known versus niche brands. The study aimed to capture not only the answers provided by each model but also the sources they referenced and the classification mix of those sources.

Two query sets were created: one for well-known brands (established consumer names such as Apple, Nike, or Toyota) and another for niche brands (less familiar or specialized names). Each query was structured in a ranking or identification style, such as ``What is the best-known camera brand?'' or ``Which niche fashion brands are gaining popularity?'' All queries were run through each model, with each engine returning both an explicit answer and a set of supporting links.

The experimental pipeline was implemented in Python with structured logging and reproducibility in mind. For each model:

\begin{itemize}
    \item Queries were submitted via its respective API.
    \item The links returned were collected and normalized by extracting the base domain.
    \item A classification routine labeled each link as Brand, Earned, or Social, combining a predefined list of known social platforms with GPT-assisted classification prompts.
    \item For answer-level comparison, a post-processing step cleaned the textual responses (removing punctuation and formatting) to allow direct string-level comparison. Agreement between models was measured as the percentage of queries where two models produced the same brand name answer.
    \item For link-level comparison, the distribution of Brand, Earned, and Social domains was computed per query set. Aggregated statistics were then calculated separately for well-known brands and niche brands, enabling direct comparison across engines. At the same time, domain frequency counts identified the most common sources (e.g., TechRadar, Wikipedia, Reddit).
\end{itemize}

Finally, the system compiled results into JSON files containing: per-model classification distributions across the two query sets, agreement scores reflecting consistency in answers between models, and detailed logs capturing raw responses, domain classifications, and answer strings for auditing.

\begin{figure}[h]
    \centering
    \includegraphics[width=0.5\textwidth]{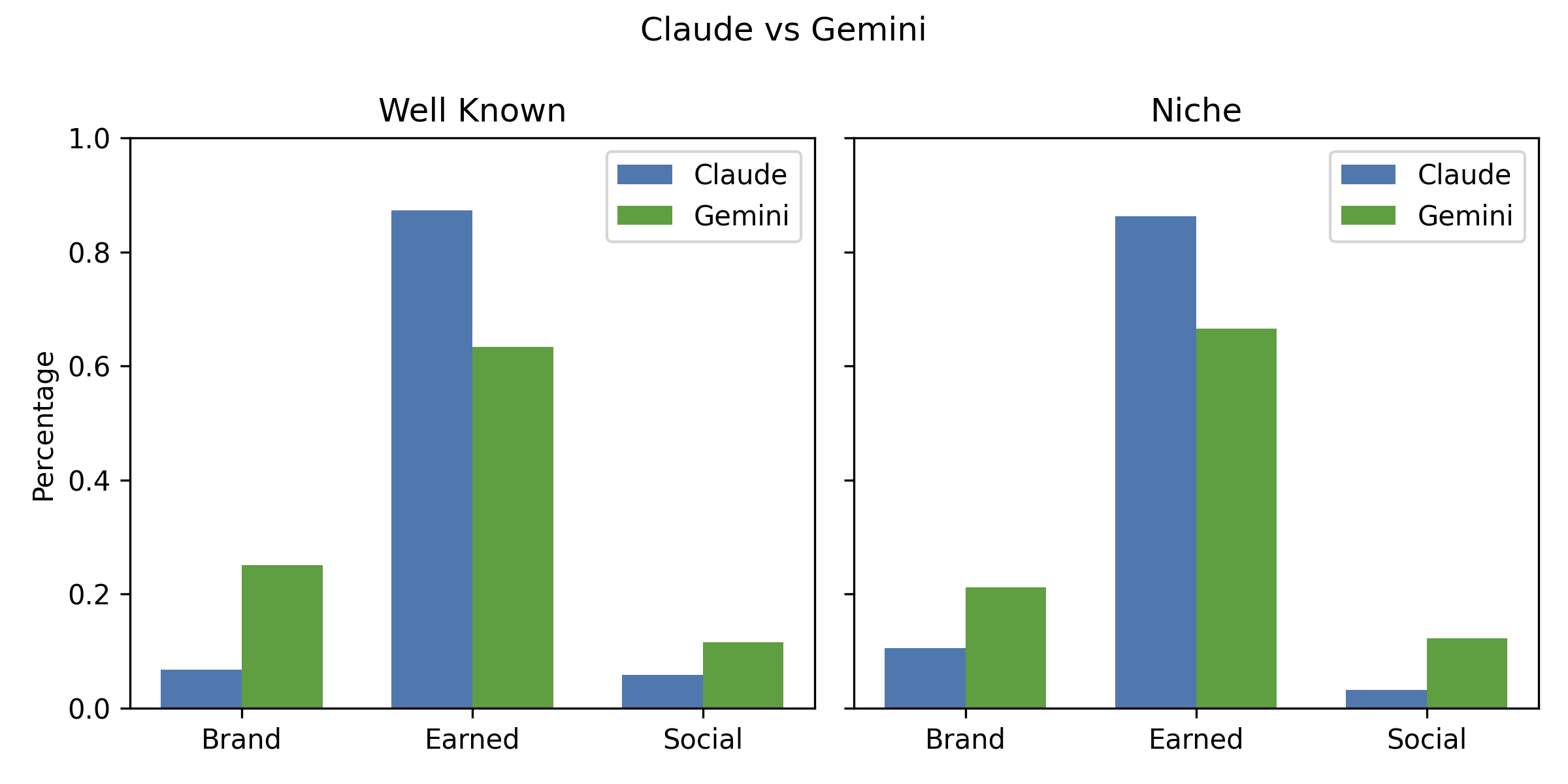}
    \caption{Comparison of Claude vs.~Gemini for well-known and niche brands.}
    \label{fig:claude-gemini}
\end{figure}

\begin{figure}[h]
    \centering
    \includegraphics[width=0.5\textwidth]{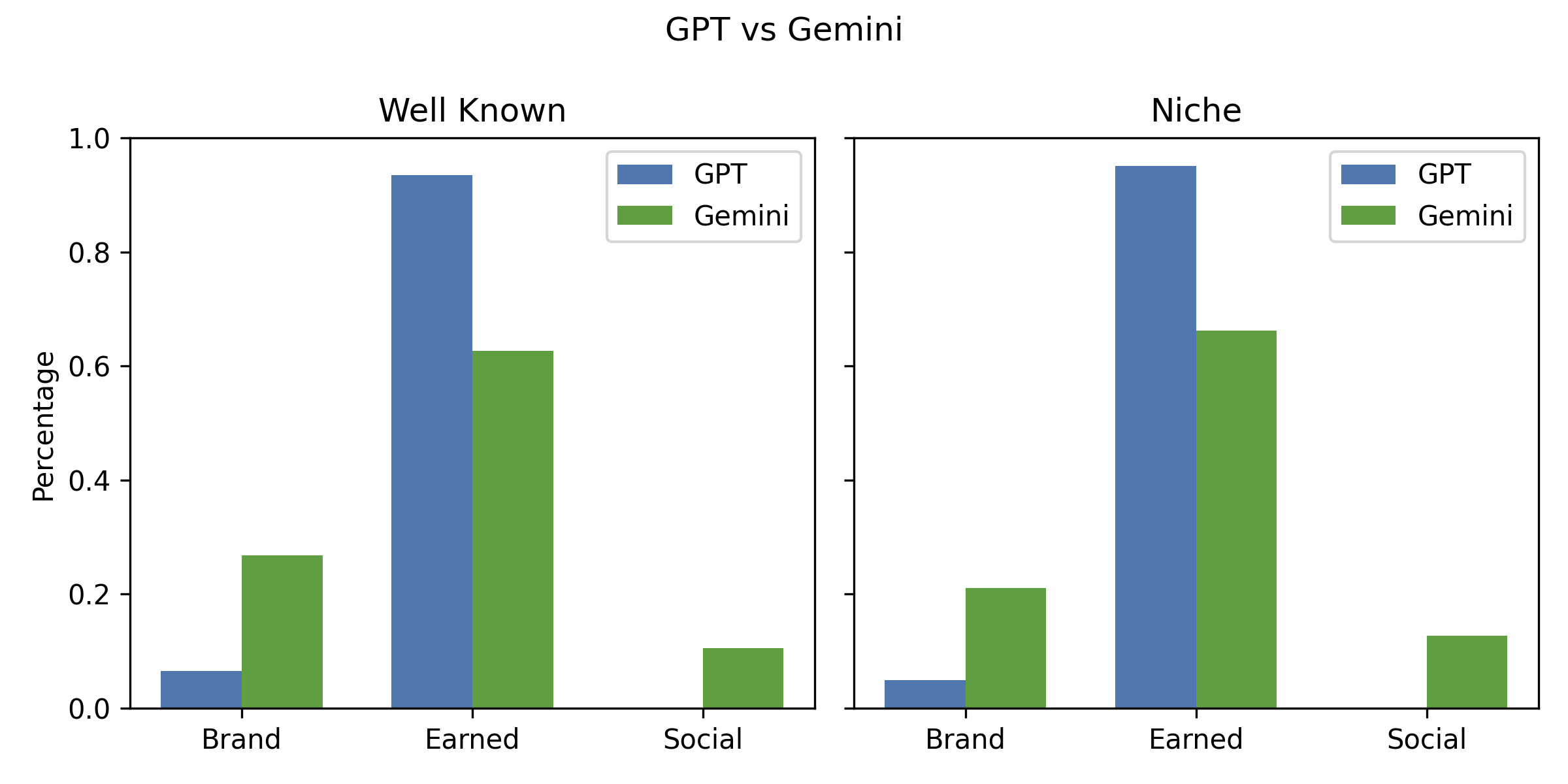}
    \caption{Comparison of GPT vs.~Gemini for well-known and niche brands.}
    \label{fig:gpt-gemini}
\end{figure}

\begin{figure}[h]
    \centering
    \includegraphics[width=0.5\textwidth]{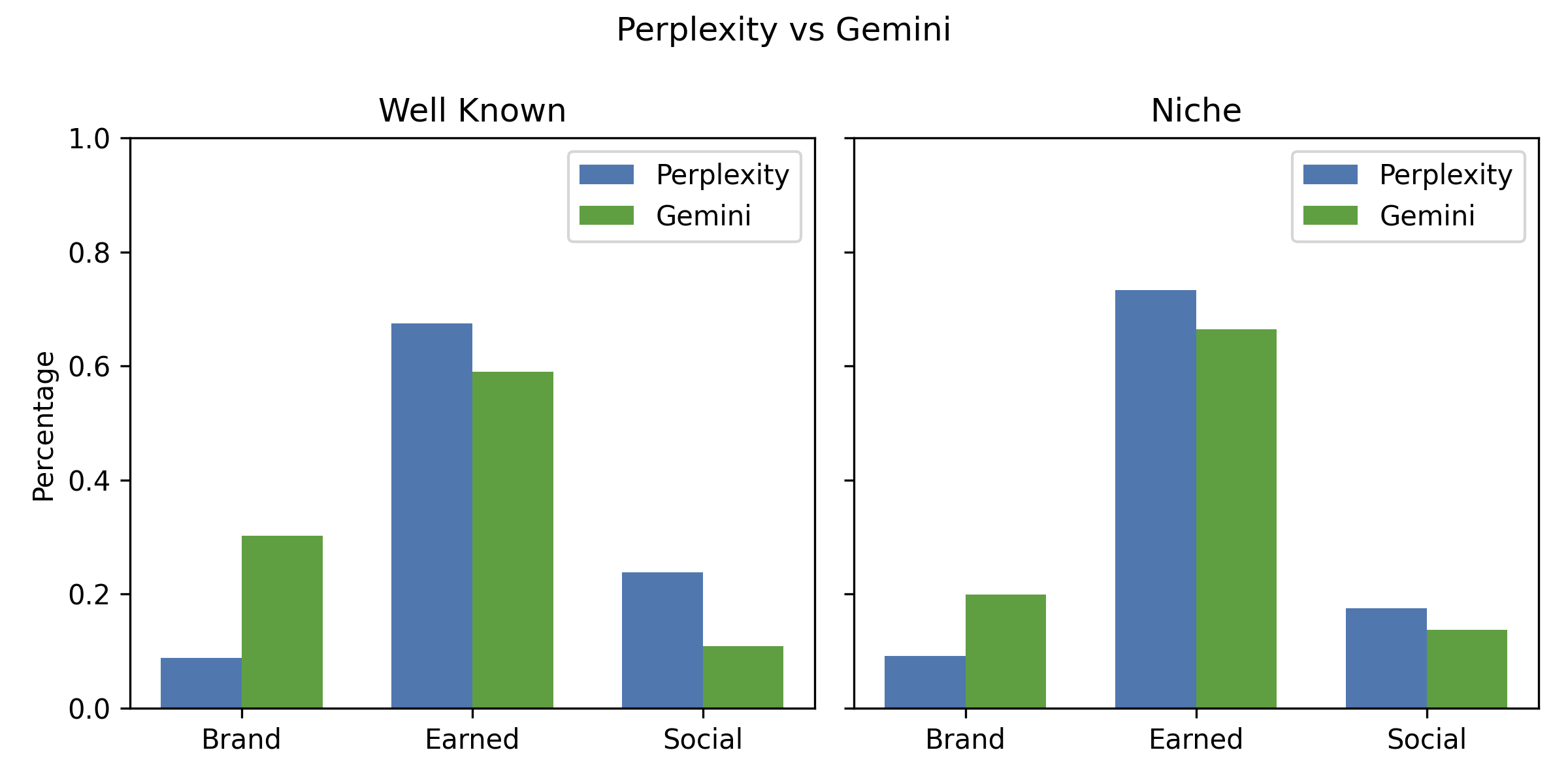}
    \caption{Comparison of Perplexity vs.~Gemini for well-known and niche brands.}
    \label{fig:perplexity-gemini}
\end{figure}

\paragraph{Well-Known Brands.}  
For well-known brands, Claude and ChatGPT displayed the strongest tilt toward Earned domains. Claude results consisted of 87.3\% Earned, 6.8\% Brand, and 5.9\% Social, while ChatGPT skewed even further, with 93.5\% Earned, 6.5\% Brand, and no Social. Perplexity, by contrast, surfaced a greater proportion of Social content at 23.8\%, with 67.4\% Earned and 8.8\% Brand. Gemini stood between these extremes, balancing 63.4\% Earned, 25.1\% Brand, and 11.5\% Social.

\paragraph{Niche Brands.}  
For niche brands, similar trends persisted, though the distributional differences became sharper. ChatGPT again leaned most heavily on Earned domains, with 95.1\% Earned and only 4.9\% Brand, completely excluding Social sources. Claude showed a slightly more balanced profile at 86.3\% Earned, 10.6\% Brand, and 3.2\% Social. Perplexity continued to surface a higher share of Social (17.5\%) and Brand (9.1\%), reducing Earned to 73.4\%. Gemini again positioned itself closer to the middle with 66.4\% Earned, 21.2\% Brand, and 12.7\% Social.

\paragraph{Summary.}  
These findings show that Claude and ChatGPT behave conservatively, strongly privileging expert-driven and third-party validation while minimizing user-generated sources. Perplexity behaves more inclusively, incorporating a substantial amount of Social content alongside Earned. Gemini adopts a hybrid strategy, with a greater share of Brand domains than Claude or ChatGPT and a moderate inclusion of Social.

The average agreement scores further highlight these differences. Across well-known brands, agreement ranged from 76--81\% depending on the system pairing, while for niche brands it was slightly lower, between 71--76\%. This suggests that AI systems align more closely when handling established, widely recognized brands but diverge when surfacing results for less familiar or specialized entities.

\subsubsection{Vertical Domain and Freshness Analysis}

This section evaluates how Claude, ChatGPT, and Perplexity source information across two high-interest verticals: consumer electronics and automotive. This analysis considers both domain concentration (the top sources of content) and freshness (measured by publication dates).

\paragraph{Experimental Design.}  
This experiment evaluated Claude, ChatGPT, and Perplexity on two high-interest verticals: consumer electronics and automotive. The focus was on both domain concentration (the sources each engine prioritized) and freshness (the age of content returned).

The study began with a curated set of ranking-style queries drawn from consumer electronics and automotive topics. Each query was submitted in parallel to Claude, ChatGPT, and Perplexity, with each engine returning up to ten links. All retrieved URLs were normalized by extracting domain names for consistent comparison.

Each link was classified into one of three categories: \textbf{Brand}, \textbf{Earned}, or \textbf{Social}. Brand sources included manufacturer and retailer domains such as \texttt{apple.com} or \texttt{toyota.com}. Earned sources covered third-party review sites, media outlets, and government portals such as \texttt{techradar.com} or \texttt{consumerreports.org}. Social sources captured community-driven or user-generated platforms like Reddit, YouTube, and Quora. Classification was conducted by combining a fixed rule list of known social domains with AI-assisted labeling prompts.

To assess freshness, each returned link was crawled and analyzed for publication or update dates. Candidate dates were extracted from HTML metadata, JSON-LD scripts, \texttt{<time>} tags, and body text. When multiple candidates were found, GPT was used to select the most plausible publication date. Extracted dates were then converted to UTC timestamps. For each link, article age was calculated in days relative to the time of the experiment.

Freshness was summarized using several measures:
\begin{itemize}
    \item \textbf{Coverage:} the proportion of links with identifiable dates.
    \item \textbf{Mean and median age:} average and typical number of days since publication.
    \item \textbf{Freshness score:} a recency-weighted measure computed as the mean of $1/(1+\text{age})$.
    \item \textbf{Coverage-adjusted freshness score:} the freshness score scaled by coverage to penalize engines with low date detection.
\end{itemize}

At the aggregation stage, results were summarized along two dimensions. First, domain concentration was measured by counting the most frequent domains surfaced per vertical and engine. Second, classification mix was computed as the percentage share of Brand, Earned, and Social links. The freshness results were then combined with classification results to compare recency alongside content diversity.

\begin{figure}[h]
    \centering
    \includegraphics[width=0.5\textwidth]{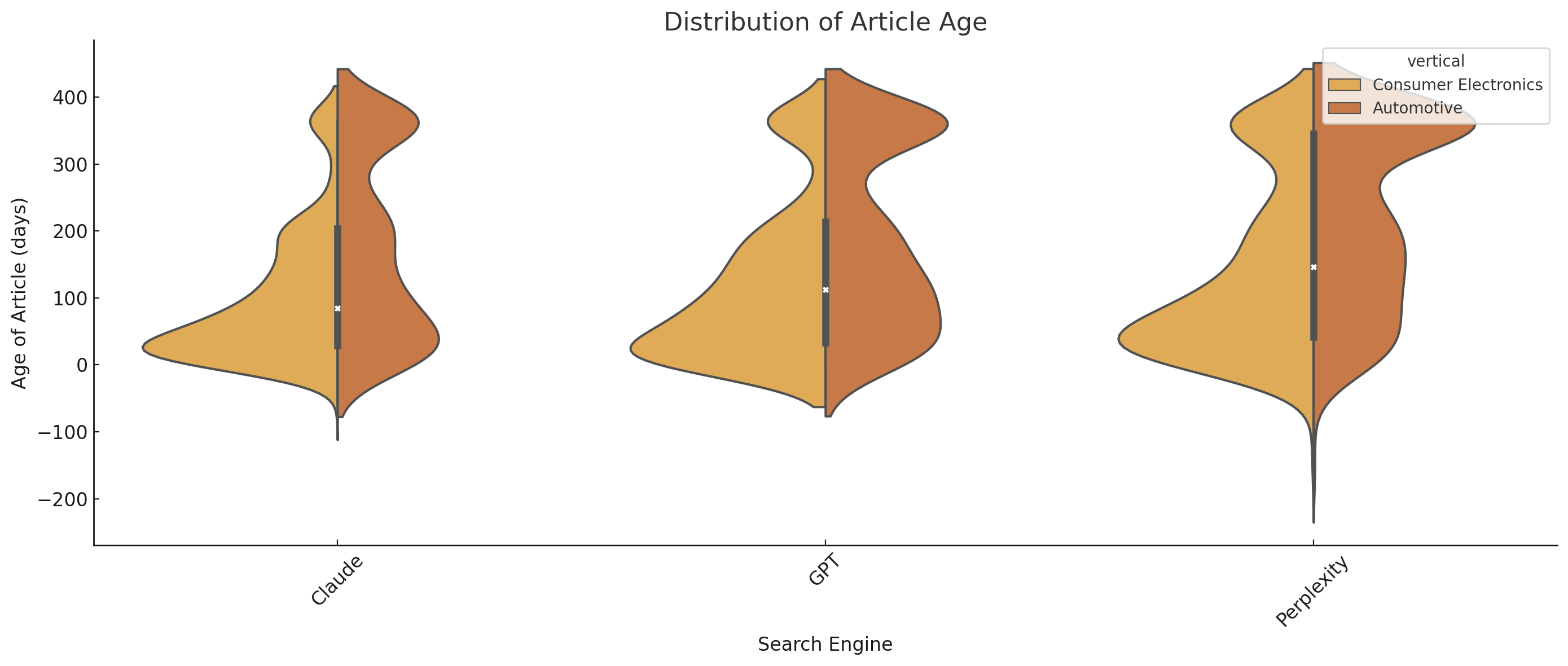}
    \caption{Distribution of article age (days) across verticals by search engine.}
    \label{fig:freshness-distribution}
\end{figure}

\begin{figure}[h]
    \centering
    \includegraphics[width=0.5\textwidth]{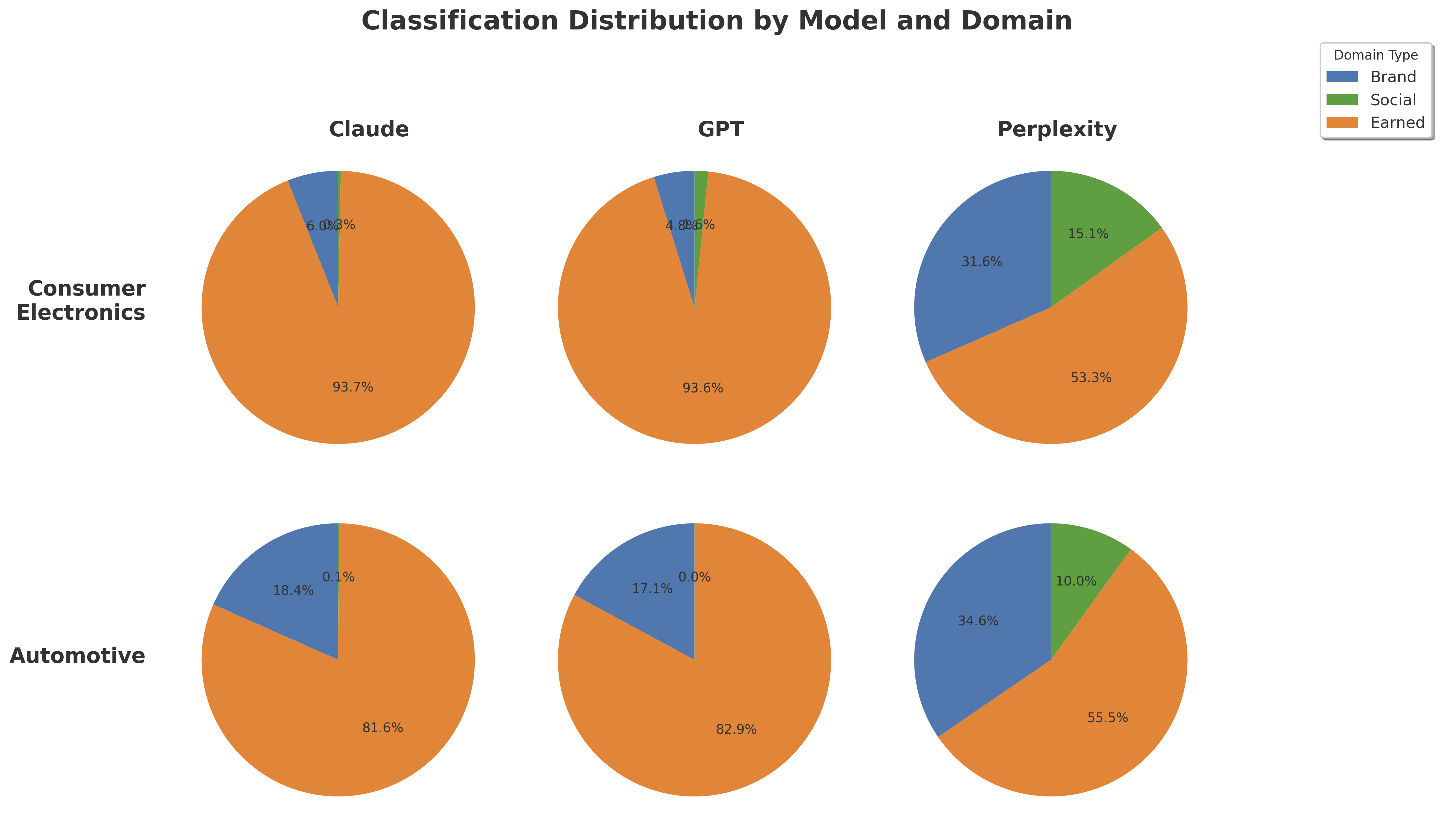}
    \caption{Domain classification shares (Brand, Earned, Social) across verticals and engines.}
    \label{fig:domain-classification}
\end{figure}

\paragraph{Consumer Electronics.}  
In consumer electronics, Claude and GPT exhibited strong reliance on Earned media (Claude: $\approx$93.7\%, GPT: $\approx$93.6\%), with very limited Brand and Social representation. Claude’s top domains included TechRadar, Tom’s Guide, and RTINGS, while GPT leaned on TechRadar, Tom’s Guide, and Wikipedia. Both engines foreground traditional editorial and review outlets, reinforcing their Earned-heavy bias.

By contrast, Perplexity drew from a more heterogeneous mix: $\approx$31.6\% Brand, 53.3\% Earned, and 15.1\% Social. Notably, YouTube (Social) and BestBuy.com (Brand) were dominant sources alongside editorial sites like RTINGS and CNET. This produced greater exposure to retail-driven and user-generated perspectives absent from Claude and GPT.

On freshness, coverage was high in this vertical (Claude: 92.5\% dated links). Claude’s mean article age was about 117 days, with a median of 62 days, yielding a freshness score of 0.0617 (coverage-adjusted: 0.0571). GPT exhibited similar freshness patterns, while Perplexity produced somewhat newer but less consistently dated results. Overall, Claude and GPT ensured stable exposure to mid-term reviews, while Perplexity blended in more contemporary but commercially linked material.

\paragraph{Automotive.}  
The automotive vertical showed greater divergence. Claude and GPT again prioritized Earned content: Claude at $\approx$81.6\% and GPT at $\approx$82.9\%. Claude favored Consumer Reports, Car and Driver, and US News, while GPT mixed Wikipedia, Automoblog, and news outlets like AP. Both leaned toward expert-oriented reviews and rankings, with limited Brand visibility (about 17–18\%).

Perplexity, however, surfaced a notably different mix: $\approx$34.6\% Brand, 55.5\% Earned, and 9.9\% Social. YouTube emerged as its single most frequent source, alongside Car and Driver and Consumer Reports. Importantly, Perplexity also introduced retailer- and aftermarket-linked domains (e.g., \texttt{autozone.com}, \texttt{cars.com}), yielding more diverse but also more commercial outputs compared to Claude and GPT.

Freshness results highlighted systemic gaps. Claude’s automotive links had low coverage ($\approx$61\%) and were substantially older, with a mean age of about 331 days (median 148 days). Its freshness score fell to 0.0441 (coverage-adjusted: 0.0269), indicating reliance on longstanding reviews and static ranking pages. GPT showed similar tendencies. Perplexity, while not entirely eliminating this age bias, returned more Social and retail-driven content that was incrementally fresher.

\paragraph{Comparative Observations.}
\begin{itemize}
    \item \textbf{Domain diversity:} Across both verticals, Claude and GPT converged on authoritative review and editorial outlets, maximizing Earned exposure but suppressing Social. Perplexity consistently incorporated a broader ecosystem, balancing Earned with significant Brand and Social sources.
    \item \textbf{Freshness:} Coverage of dated links was higher in consumer electronics than in automotive. Claude and GPT’s outputs in automotive lagged substantially in recency, while Perplexity injected more up-to-date but commercially linked results.
    \item \textbf{Vertical sensitivity:} The automotive vertical amplified the contrasts: while consumer electronics outputs were relatively fresh and homogenous, automotive responses revealed aging content pools for Claude and GPT versus hybridized mixes for Perplexity.
\end{itemize}

\paragraph{Summary.}  
This vertical analysis underscores that while Claude and GPT prioritize consistency, authority, and Earned concentration, they risk staleness, especially in automotive contexts. Perplexity, by contrast, trades some editorial authority for greater diversity and timeliness, leveraging Brand and Social domains more extensively. These findings highlight how vertical context modulates the trade-off between reliability, freshness, and media diversity across AI search engines.

\subsubsection{Car Brands: Electric, Family SUVs, and Hybrid}

This experiment focused on the automotive sector, comparing how Google, ChatGPT, and Perplexity sourced information on three topics: electric cars, family SUVs, and hybrids. Each query was categorized by topic and analyzed according to the share of Brand, Earned, and Social domains. In addition, the ten most frequently appearing domains per service and topic were examined to reveal differences in source composition.

\paragraph{Experimental Design.}  
To conduct the comparison across Google, ChatGPT, and Perplexity for automotive queries, a structured pipeline was developed. The experiment was designed around three themes: electric cars, family SUVs, and hybrid cars. Each theme contained a set of buyer-oriented queries (for example, ``best electric car for commuting'' or ``top hybrid SUVs for families''), and each query was submitted to all three services. For every query, the top ten links returned were collected.

The pipeline was implemented in Python with modular functions to ensure reproducibility. Separate API clients were initialized for each service: Google Custom Search, OpenAI’s ChatGPT with web search enabled, and Perplexity’s API. Each query was executed in parallel across services with retry logic to handle timeouts or rate limits. All retrieved URLs were standardized by extracting their domain names.

Every link was classified into one of three categories: \textbf{Brand}, \textbf{Earned}, or \textbf{Social}. Brand referred to official manufacturer or retailer websites (for example, \texttt{ford.com}). Earned covered independent review sites, media outlets, or government portals (for example, \texttt{caranddriver.com}, \texttt{energy.gov}). Social included community-driven or user-generated platforms such as Reddit, Quora, YouTube, or Facebook. Classification combined a rule-based list of known social platforms with AI-assisted verification to ensure accuracy.

The results were aggregated at two levels. First, domain-level counts captured how often specific domains appeared per service and topic. Second, classification-level shares summarized the overall media mix of Brand, Earned, and Social for each service and topic. A detailed log of all queries and links was also stored to support auditability.

Finally, per-topic summaries and overall comparisons were compiled into JSON outputs. These included ranked domain lists, per-service category shares, and detailed logs of link classifications. This design allowed both quantitative comparison (percentages of Brand, Earned, and Social) and qualitative interpretation (the prominence of specific domains such as Reddit, Wikipedia, or Car and Driver).

\begin{figure}[h]
    \centering
    \includegraphics[width=0.5\textwidth]{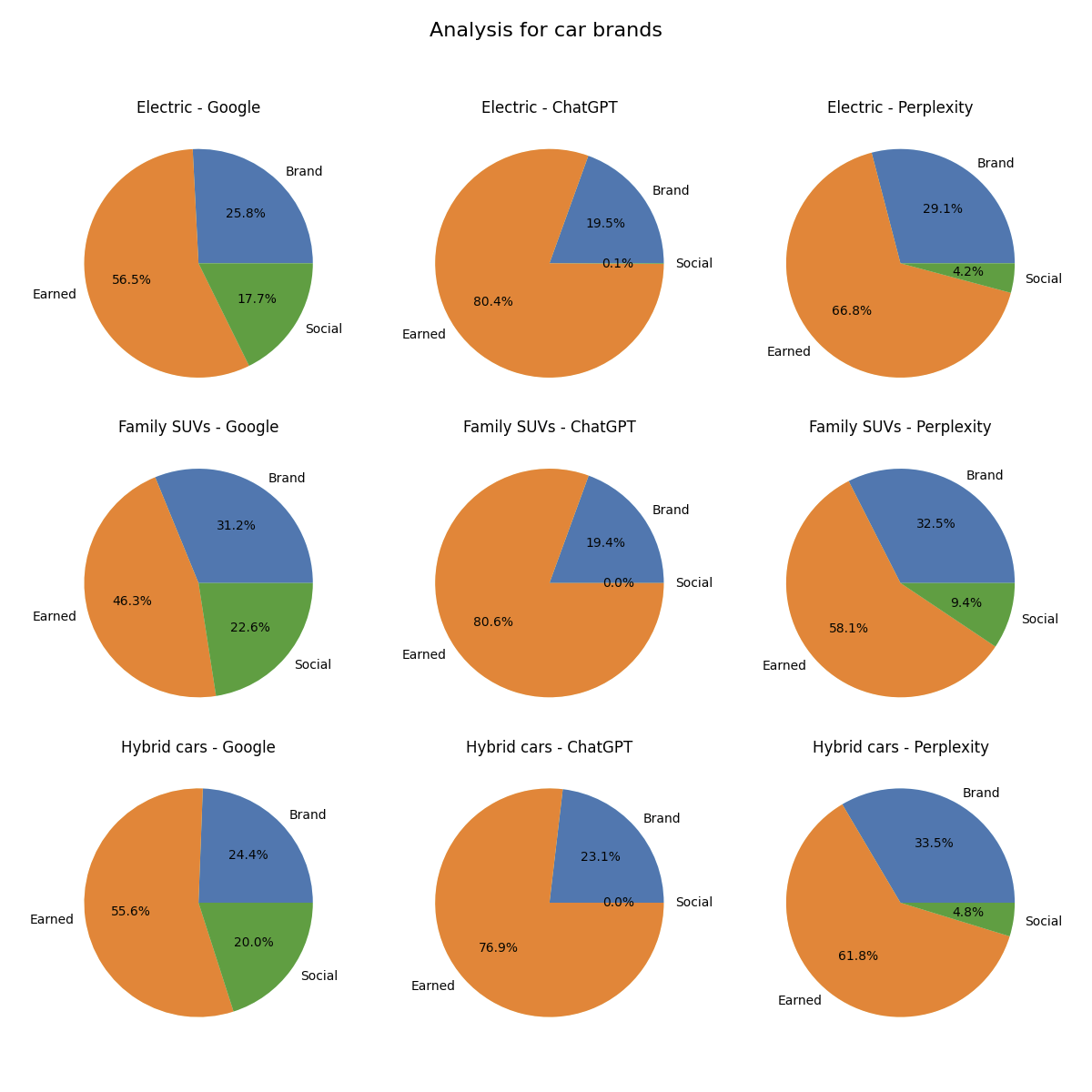}
    \caption{Domain composition (Brand, Earned, Social) across Google, ChatGPT, and Perplexity for electric cars, family SUVs, and hybrid cars.}
    \label{fig:car-brands}
\end{figure}

\paragraph{Electric Cars.}  
For electric cars, Google balanced its results across Earned (56\%), Brand (26\%), and Social (18\%). ChatGPT diverged by heavily prioritizing Earned sources (80\%) with minimal Brand (19\%) and negligible Social (under 1\%). Perplexity fell between the two, with 67\% Earned, 29\% Brand, and 4\% Social. Domain-level data highlights these contrasts. Google’s top electric car sources included Reddit, Quora, and Facebook alongside Earned outlets like Car and Driver and government portals such as \texttt{energy.gov}. ChatGPT concentrated on encyclopedic and news publishers such as Wikipedia, AP News, and Reuters, with limited inclusion of automotive review sites. Perplexity relied on YouTube and Car and Driver while supplementing with government and consumer-oriented review domains.

\paragraph{Family SUVs.}  
The family SUV segment showed sharper differences. Google leaned toward Social sources, with Reddit dominating its domain list, resulting in 23\% Social, 31\% Brand, and 46\% Earned. ChatGPT, by contrast, excluded Social entirely and emphasized Earned domains at 81\%, with 19\% Brand. Perplexity presented a more balanced profile: 58\% Earned, 32\% Brand, and 9\% Social. Domain counts underline these differences. Google’s results were led by Reddit, Quora, and Facebook, but also included U.S. News and \texttt{KBB.com}. ChatGPT relied on structured outlets such as Wikipedia, AP News, and review sites like ISeeCars and \texttt{Cars.com}. Perplexity’s mix included YouTube for Social, alongside Car and Driver, Edmunds, Consumer Reports, and \texttt{Cars.com}.

\paragraph{Hybrid Cars.}  
For hybrids, Google’s results resembled its SUV pattern: 56\% Earned, 24\% Brand, and 20\% Social. Reddit again played a central role, supported by Quora, Facebook, and research-oriented sites such as ScienceDirect. ChatGPT continued its systematic exclusion of Social, yielding 77\% Earned and 23\% Brand. Notably, it leaned more toward industry-specific outlets like TopSpeed, Reuters, and AutomotiveQuest, as well as Wikipedia. Perplexity produced 62\% Earned, 34\% Brand, and 5\% Social, with Car and Driver, YouTube, and \texttt{Energy.gov} among its top sources.

\paragraph{Cross-Topic Observations.}  
Three trends stand out across these automotive categories. First, Google consistently incorporates a high proportion of Social sources, especially Reddit, which makes its results community-driven but less curated. Second, ChatGPT almost entirely excludes Social platforms, resulting in outputs that are dominated by institutional and journalistic Earned domains, with Brand sources playing a secondary role. Third, Perplexity consistently bridges the two approaches: it includes some Social content (often YouTube), supplements with Brand domains, and maintains a majority share of Earned content. These distinctions illustrate that the services are not only ranking different links but also constructing fundamentally different knowledge environments—Google privileging community discussion, ChatGPT privileging authoritative sources, and Perplexity blending the two.

\subsubsection{Cross-Model Domain Diversity}

This section examines how Claude, ChatGPT, and Perplexity differ in the diversity of domains they source across two verticals: automotive and consumer electronics. The analysis considers the breadth of distinct domains, the degree of overlap between models (using Jaccard similarity), and the share of domains exclusive to each service.

\paragraph{Experimental Design.}  
The domain diversity experiment was designed to evaluate how Claude, ChatGPT, and Perplexity source information when given the same set of queries in two verticals: automotive and consumer electronics. The study aimed to quantify not only the breadth of domains each model returned but also the degree of overlap across engines, using normalized measures of similarity and exclusivity.

Two query sets were created, one for automotive and one for consumer electronics, each consisting of ranking-style and product-discovery prompts. For example, automotive queries included themes such as ``What are the best family SUVs in 2024?'' while consumer electronics queries included ``Rank the best smartphones from 1 to 10.'' Each query was run independently through all three engines, with each model returning up to ten links.

The experimental pipeline was implemented in Python with structured logging and reproducibility in mind. For each model:
\begin{itemize}
    \item Queries were submitted via its respective API (Claude via Anthropic, ChatGPT via OpenAI, and Perplexity via its public API).
    \item All links were collected and normalized to their registered base domains using \texttt{tldextract}.
    \item Distinct domain sets were constructed per engine, allowing direct comparison of coverage and overlap.
\end{itemize}

To assess diversity, several measures were calculated:
\begin{itemize}
    \item \textbf{Set sizes}: the total number of distinct domains returned by each model in a vertical.
    \item \textbf{Jaccard similarity}: pairwise normalized overlap between models’ domain sets.
    \item \textbf{Unique share}: the proportion of domains that appeared exclusively in one model’s outputs.
    \item \textbf{Domain partitions}: grouping domains into categories such as Claude-only, GPT-only, Perplexity-only, shared between two engines, or shared across all three.
\end{itemize}

Visualizations were produced in the form of bar charts, where each x-axis category represented unique, pairwise, or three-way shared domains. This enabled direct comparison of how concentrated or fragmented the domain ecosystem was across engines.

\begin{figure}[h]
    \centering
    \includegraphics[width=0.5\textwidth]{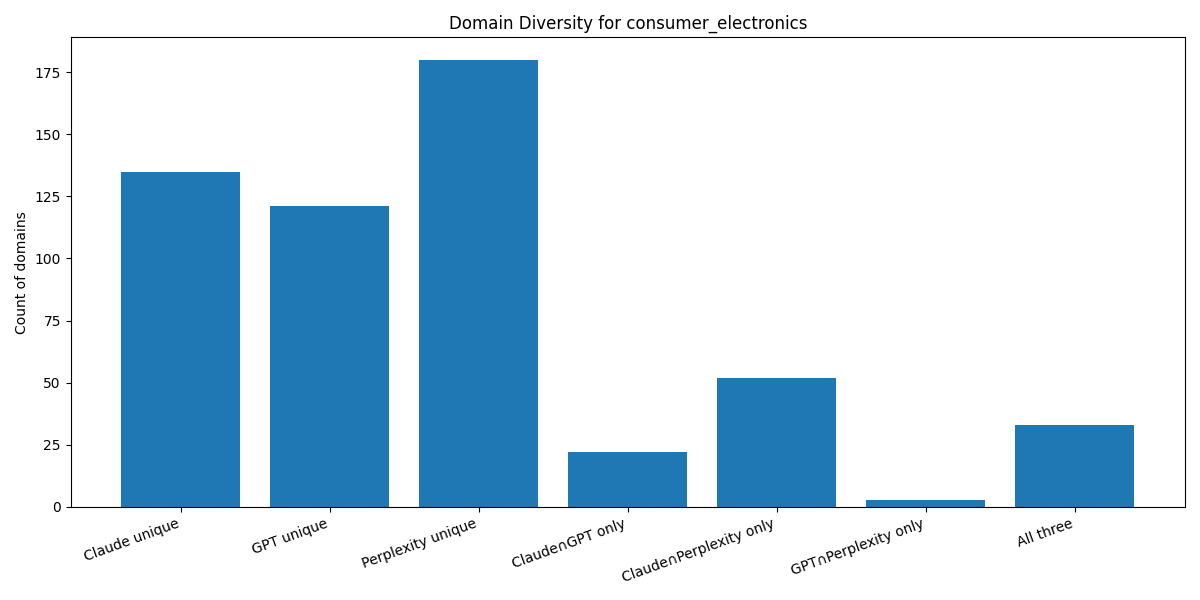}
    \caption{Domain diversity for consumer electronics across Claude, ChatGPT, and Perplexity.}
    \label{fig:domain-diversity-consumer}
\end{figure}

\begin{figure}[h]
    \centering
    \includegraphics[width=0.5\textwidth]{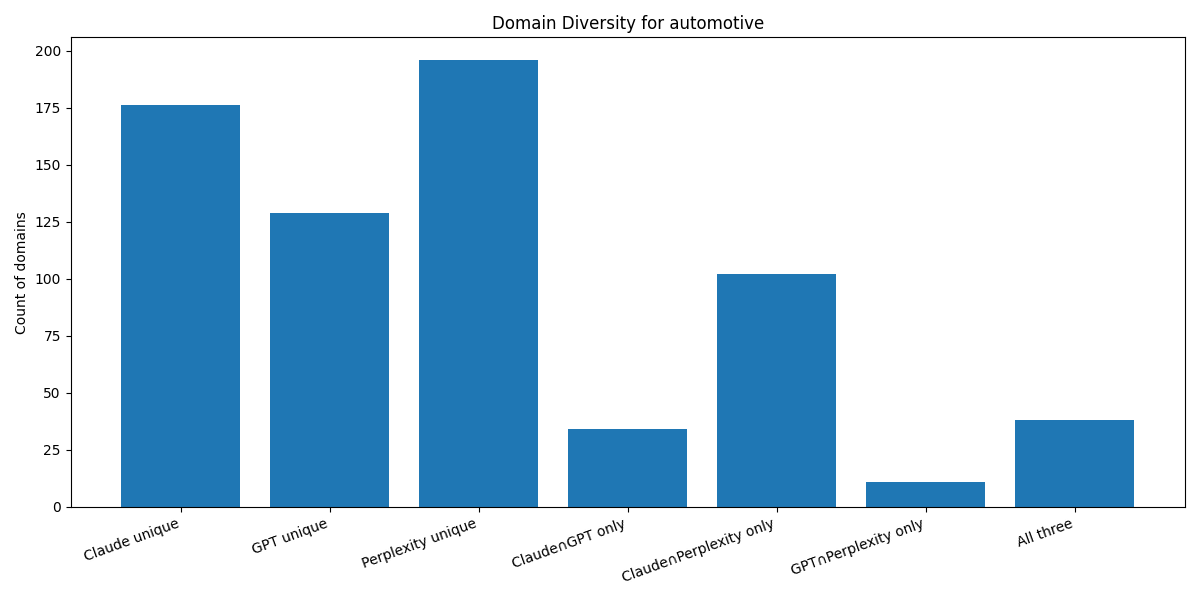}
    \caption{Domain diversity for automotive across Claude, ChatGPT, and Perplexity.}
    \label{fig:domain-diversity-automotive}
\end{figure}

\paragraph{Automotive.}  
In the automotive vertical, Claude surfaced 350 distinct domains, GPT returned 212, and Perplexity 347. Pairwise Jaccard overlaps were modest: Claude–GPT at 0.147, Claude–Perplexity at 0.251, and GPT–Perplexity at 0.096. The unique domain percentages further underscore divergence, with Claude contributing 50.3\% exclusive domains, GPT 60.8\%, and Perplexity 56.5\%. These results show that each model contributes a substantial volume of unique sources, with Claude and Perplexity sharing the most common ground, while GPT and Perplexity show minimal intersection. The set of domains shared across all three was relatively small, dominated by well-established review and ranking outlets such as \textit{Car and Driver}, \textit{Edmunds}, and \textit{Consumer Reports}.

\paragraph{Consumer Electronics.}  
The consumer electronics vertical revealed similar but even sharper contrasts. Claude produced 242 distinct domains, GPT 179, and Perplexity 268. Overlaps were smaller than in automotive: Claude–GPT at 0.150, Claude–Perplexity at 0.200, and GPT–Perplexity at only 0.088. Exclusivity was even more pronounced: Claude domains were 55.8\% unique, GPT 67.6\%, and Perplexity 67.2\%. Once again, the intersection across all three models was minimal, including only a narrow set of core publishers such as \textit{TechRadar}, \textit{Tom’s Guide}, and \textit{RTINGS}.

\paragraph{Comparative Observations.}  
Three trends emerge from this cross-model diversity analysis. First, each model maintains a wide footprint of exclusive sources, indicating that system choice directly shapes the information ecosystem a user encounters. Second, overlaps between Claude and Perplexity are consistently stronger than between GPT and Perplexity, suggesting Claude and Perplexity converge more in their domain pools. Third, across both verticals, only a small "core" set of high-authority sites (notably large review and media outlets) appear universally, while the majority of domains remain siloed by model.

\subsubsection{Cross-Model Domain Diversity in Local Services}

\paragraph{Objective.}  
While prior experiments (\S5.3) examined verticals such as automotive and consumer electronics, here we extend the analysis to \textbf{local service categories}. The goal is to evaluate how four major AI search engines (Claude, Gemini, GPT, Perplexity) differ in the breadth and overlap of domains when responding to queries in three representative sectors: Auto Repair, Dentists, and Moving Companies.

\paragraph{Experimental Design.}  
Using the pipeline described in \S4.1, we generated 100 ranking-style queries per category (e.g., ``best auto repair shops in Toronto'' or ``top-rated dentists in Toronto''). Each query was submitted in parallel to Claude, Gemini, GPT, and Perplexity, each returning up to ten links. URLs were normalized to base domains, and set-level statistics were computed, including:  
\begin{itemize}
    \item \textbf{Set size:} total number of distinct domains per engine.
    \item \textbf{Jaccard similarity:} pairwise overlap between engines.
    \item \textbf{Unique share:} proportion of domains exclusive to a given engine.
    \item \textbf{Partition counts:} number of domains appearing in one, two, three, or all four systems.
\end{itemize}

\paragraph{Results.}  
\begin{figure}[h]
    \centering
    \includegraphics[width=0.5\textwidth]{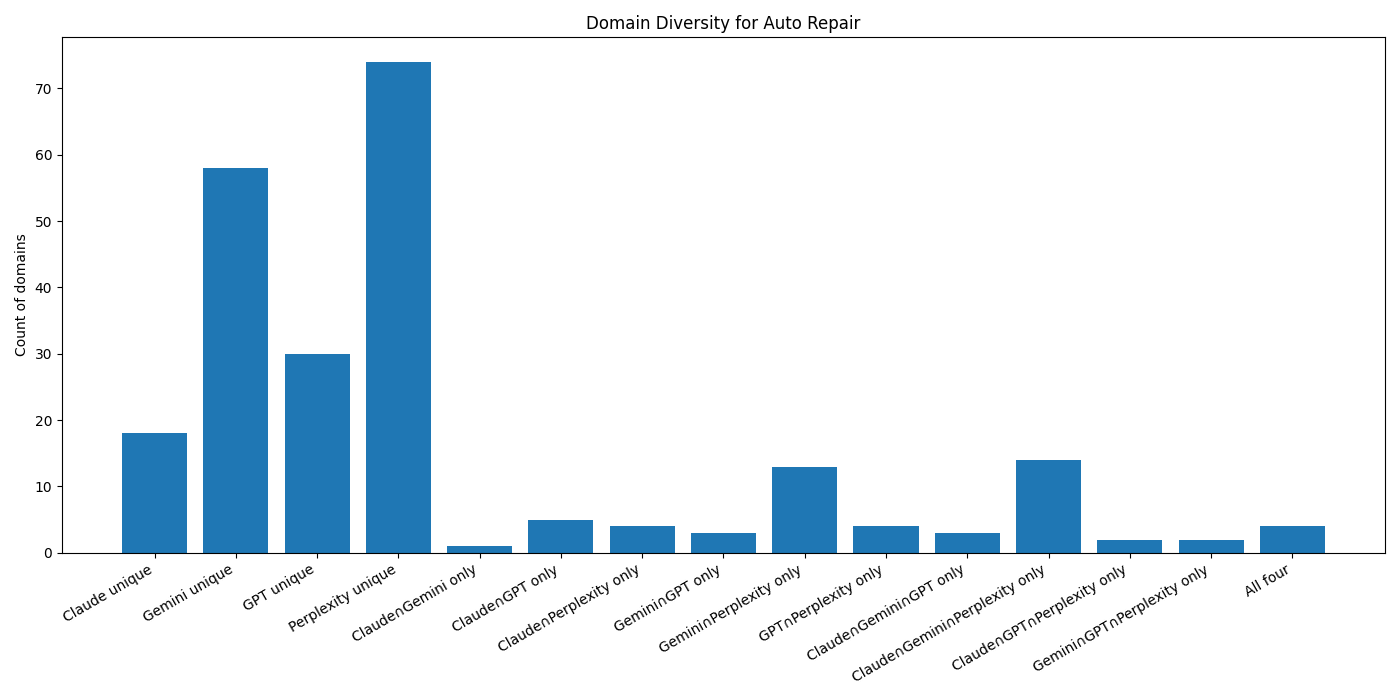}
    \caption{Domain diversity for Auto Repair queries across Claude, Gemini, GPT, and Perplexity.}
    \label{fig:auto-repair-diversity}
\end{figure}

\begin{figure}[h]
    \centering
    \includegraphics[width=0.5\textwidth]{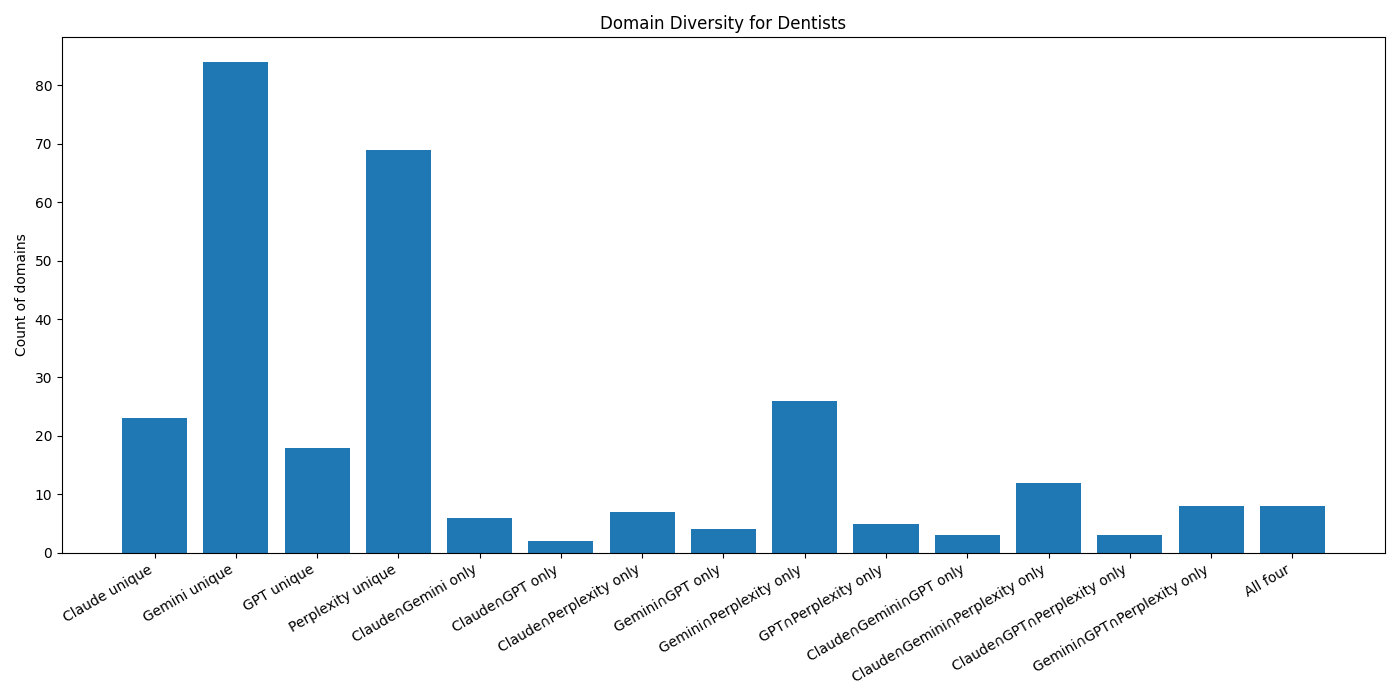}
    \caption{Domain diversity for Dentist queries across Claude, Gemini, GPT, and Perplexity.}
    \label{fig:dentists-diversity}
\end{figure}

\begin{figure}[h]
    \centering
    \includegraphics[width=0.5\textwidth]{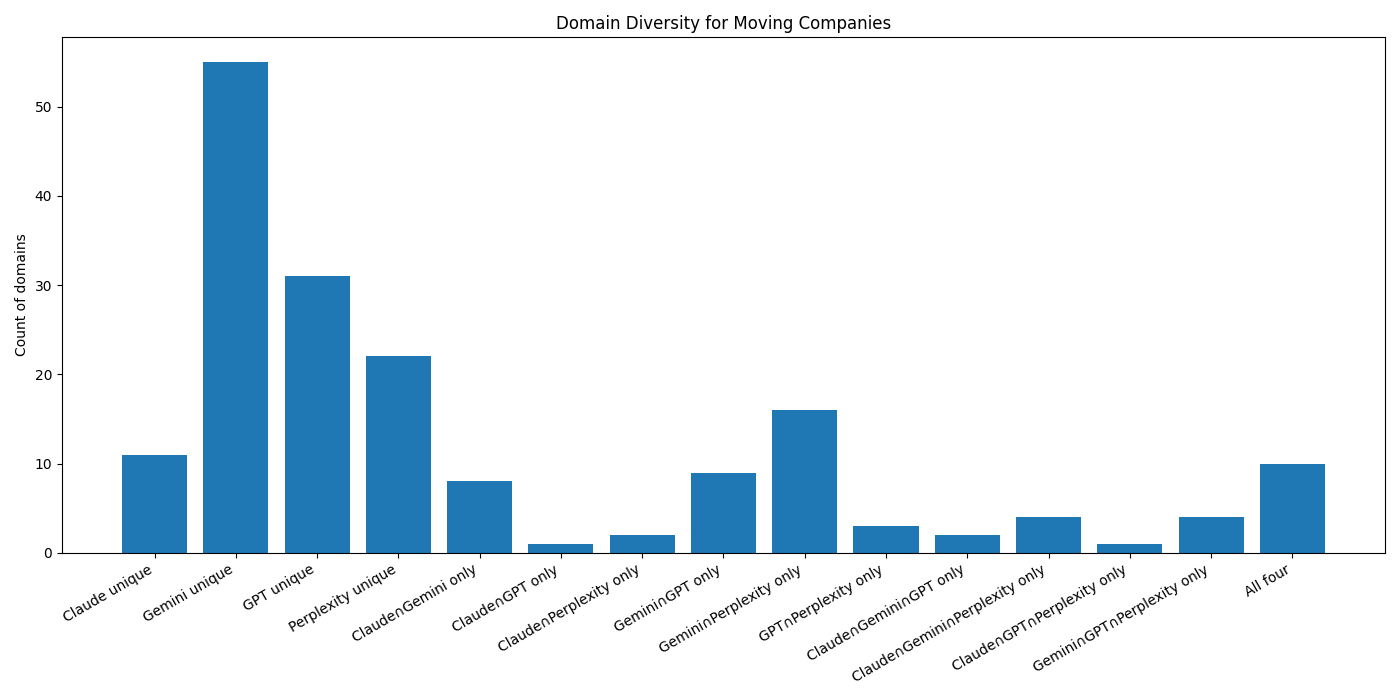}
    \caption{Domain diversity for Moving Company queries across Claude, Gemini, GPT, and Perplexity.}
    \label{fig:moving-diversity}
\end{figure}

In \textbf{Auto Repair}, Gemini (98) and Perplexity (117) surfaced the largest number of distinct domains, while Claude (51) and GPT (53) returned smaller sets. Pairwise overlaps were modest, with Jaccard scores around 0.15–0.25. Unique contributions were substantial: Claude 35\%, Gemini 55\%, GPT 62\%, and Perplexity 63\%.

In \textbf{Dentists}, Gemini again led with 151 domains, followed by Perplexity (138), Claude (64), and GPT (51). Jaccard overlaps were slightly higher here, with Gemini–Perplexity showing the strongest intersection at 0.23. Exclusivity remained high: Claude 40\%, Gemini 54\%, GPT 68\%, Perplexity 54\%.

For \textbf{Moving Companies}, Gemini produced 108 domains, Perplexity 62, GPT 61, and Claude 39. Gemini–Perplexity overlap was strongest at 0.25, but GPT remained the most exclusive with 61\% unique domains. Claude produced 51\% unique, while Perplexity contributed 45\%.

Across all three categories, only a handful of domains (such as large review aggregators like \texttt{homestars.com} and \texttt{opencare.com}) appeared across all four systems.

\paragraph{Summary.}  
These findings reinforce that local service queries yield highly fragmented ecosystems across AI engines. While Claude and GPT rely on narrower, more conservative sets of sources, Gemini and Perplexity explore a broader range of domains, with limited consensus across systems. This divergence underscores the importance of engine choice in shaping user exposure to local businesses, particularly in fragmented markets where authoritative sources are scarce and directory-style or aggregator sites dominate.

\subsubsection{Big Brand Bias (Cola Vertical)}

\paragraph{Objective.}
This experiment examines whether generative AI systems exhibit a systematic preference for major soda brands over niche/indie brands when queries are unbranded. Concretely, we ask if model outputs disproportionately surface global leaders (e.g., Coca-Cola, Pepsi) relative to smaller craft or regional brands, and whether this bias is consistent across models.

\paragraph{Experimental Design.}
We queried ChatGPT and Perplexity with a set of fifty unbranded prompts that ask for ``most popular/best/top\ldots'' cola or soda recommendations (e.g., \textit{most popular cola brand; best selling soda in the world; top cola brands in the US; classic soda brands everyone knows; most iconic soda brands}). Brand mentions in the models' answers were mapped to two curated sets, \textit{major} global leaders and \textit{niche/indie} craft or regional brands, and any remaining names were labeled \textbf{\textit{Other}} (details of extraction, normalization, and mapping follow the general pipeline in \S5.1).

\textbf{Brand sets used in evaluation:}

\begin{table}[h]
\centering
\small
\begin{tabular}{|p{0.2\textwidth}|p{0.2\textwidth}|}
\hline
\textbf{Major brands} & \textbf{Niche/indie brands} \\
\hline
Coca-Cola, Pepsi, Sprite, Mountain Dew, Dr Pepper, Fanta, 7Up, Schweppes, A\&W Root Beer, Barq's Root Beer, Canada Dry, Crush, Sunkist, Mirinda, Fresca, Mello Yello, Seagram's, Slice, Tango, Lilt & Jones Soda, Boylan Bottling Co., Sprecher, Faygo, Cheerwine, Moxie, Ale-8-One, Jarritos, Zevia, Fitz's, Maine Root, Bruce Cost Ginger Ale, Sioux City Sarsaparilla, Thomas Kemper, Dry Soda, Virgil's, Leninade, Cock 'n Bull, Blenheim Ginger Ale, Red Ribbon Soda \\
\hline
\end{tabular}
\caption{Big Brand Bias: Brand sets used in the cola vertical experiment}
\label{tab:brand-sets}
\end{table}

We report three outcome families: (i) brand-share distributions across the categories Major, Niche, and Other; (ii) brand-level counts for all named brands; and (iii) consulted-domain profiles based on the citations each model surfaced.

\begin{figure}[h]
    \centering
    \includegraphics[width=0.5\textwidth]{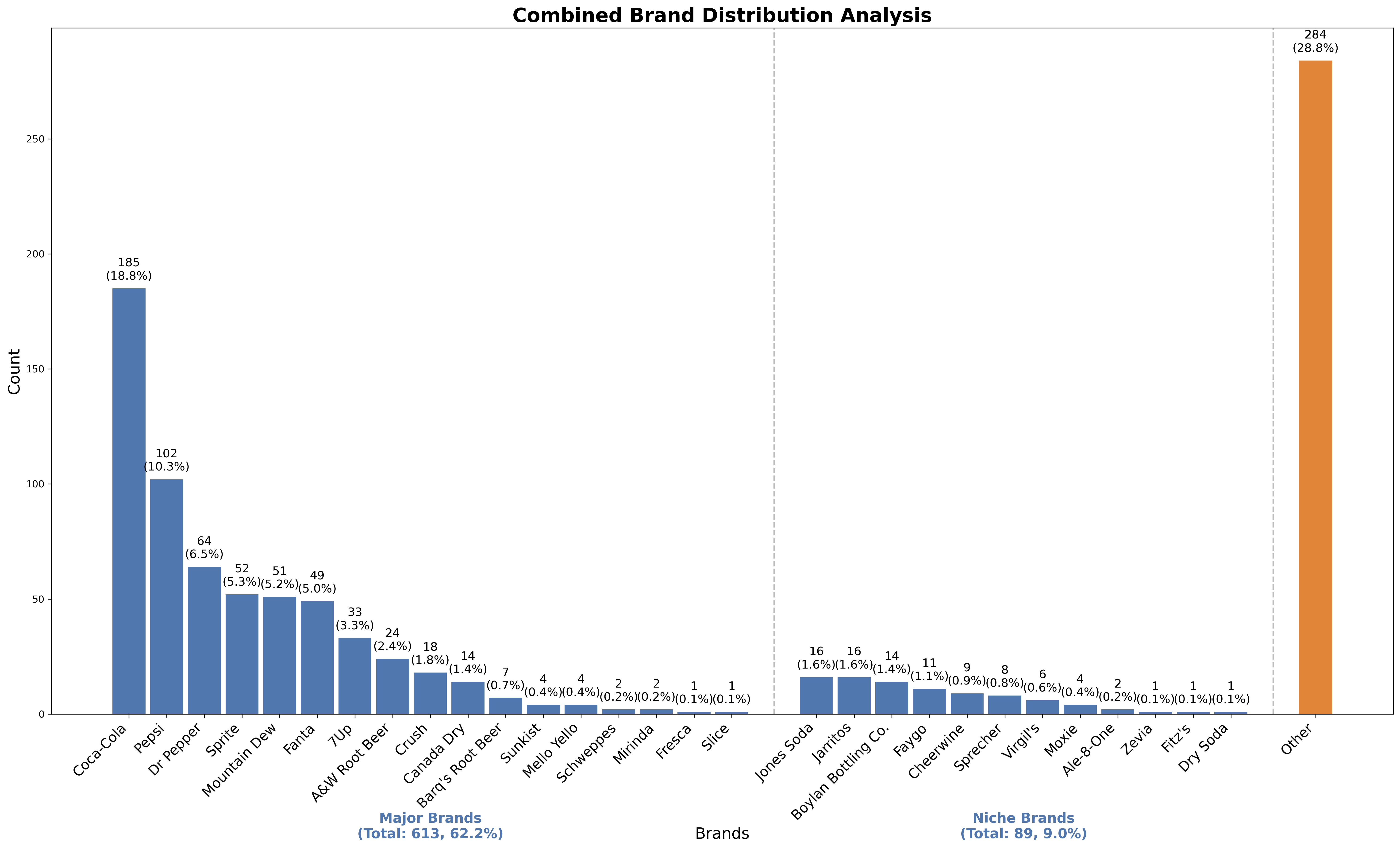}
    \caption{Big Brand Bias: Individual brand counts (combined)}
    \label{fig:brand-distribution-combined}
\end{figure}

\begin{figure}[h]
    \centering
    \includegraphics[width=0.5\textwidth]{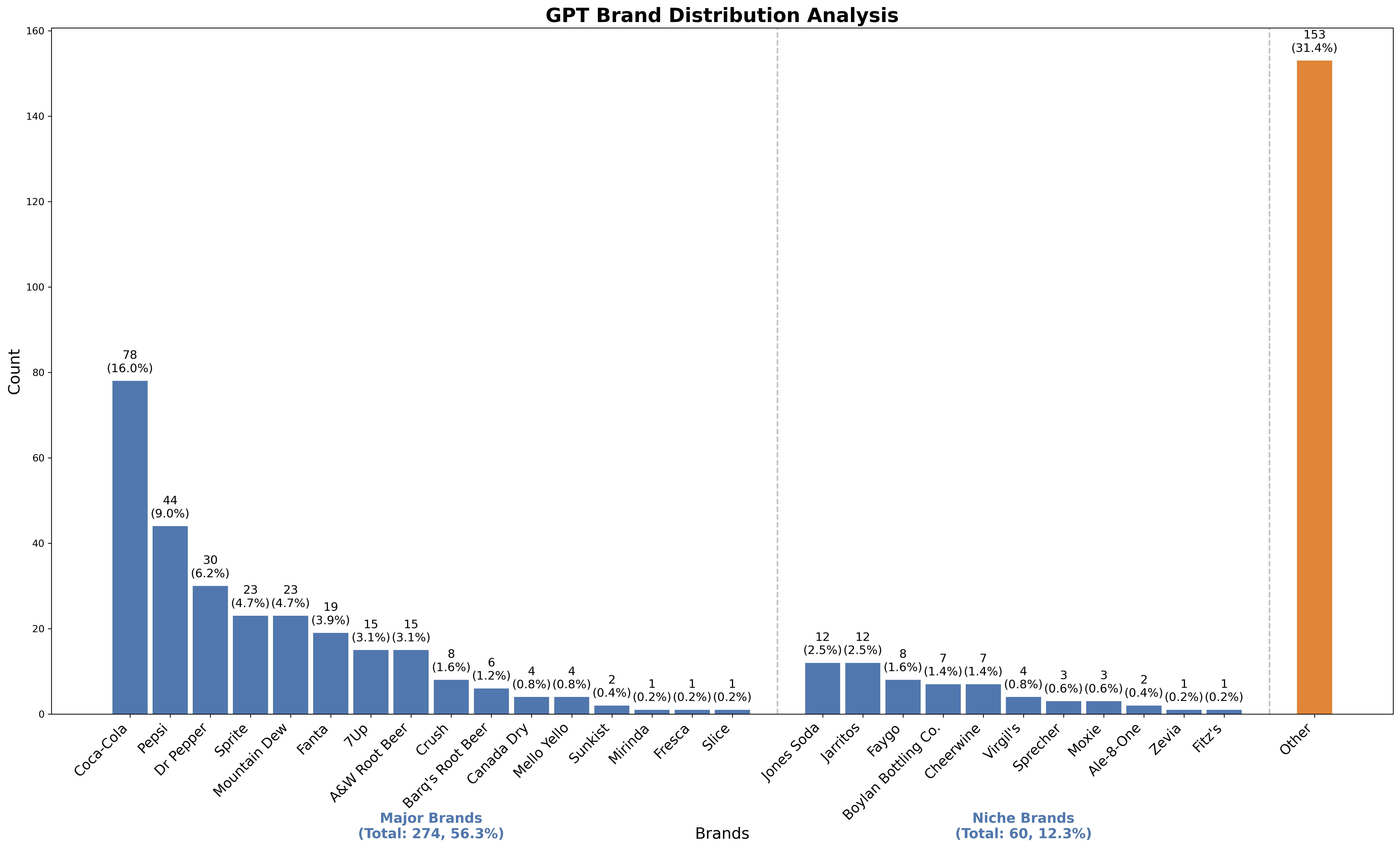}
    \caption{Big Brand Bias: Brand counts (ChatGPT)}
    \label{fig:brand-distribution-gpt}
\end{figure}

\begin{figure}[h]
    \centering
    \includegraphics[width=0.5\textwidth]{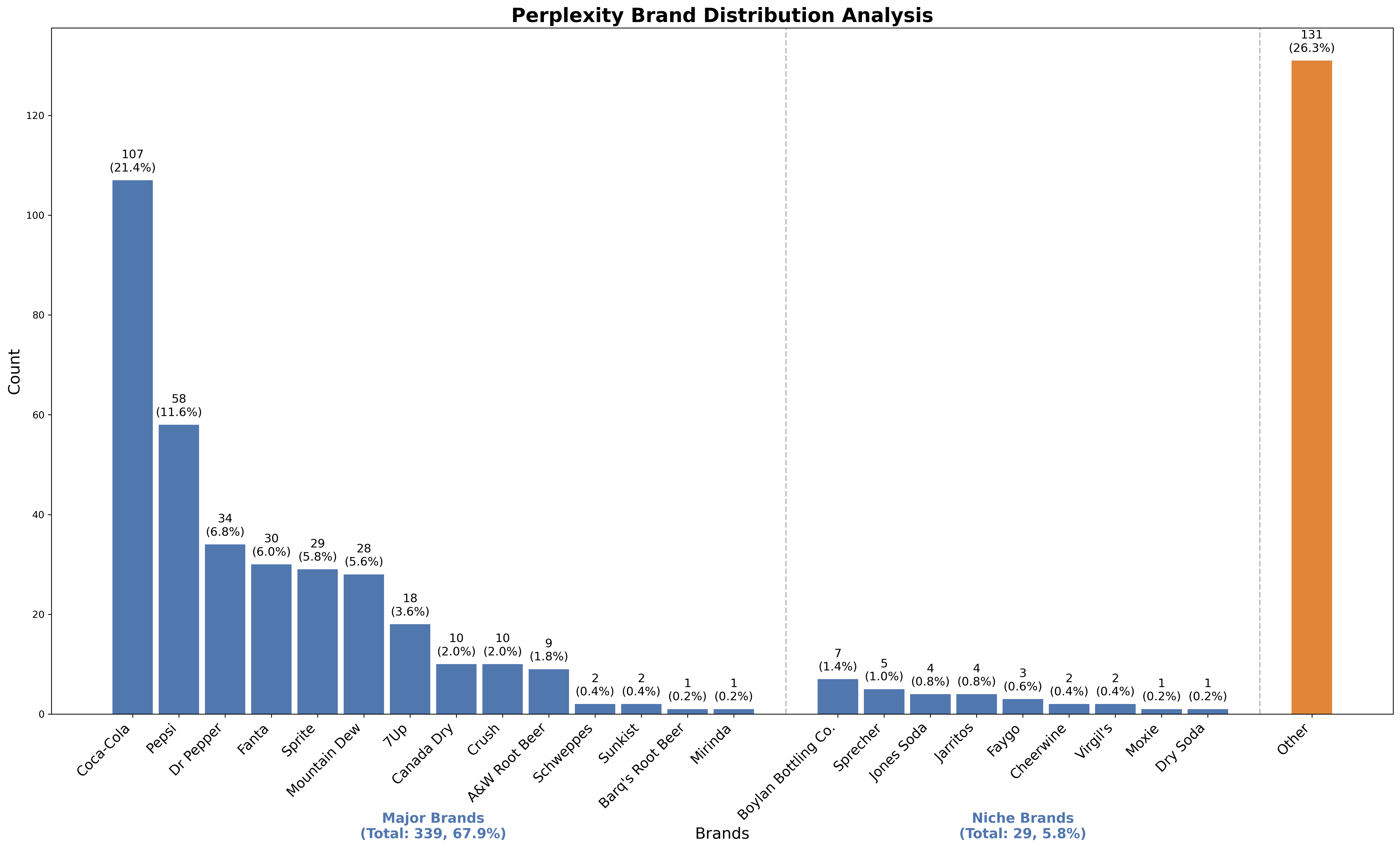}
    \caption{Big Brand Bias: Brand counts (Perplexity)}
    \label{fig:brand-distribution-perplexity}
\end{figure}

\begin{figure}[h]
    \centering
    \includegraphics[width=0.5\textwidth]{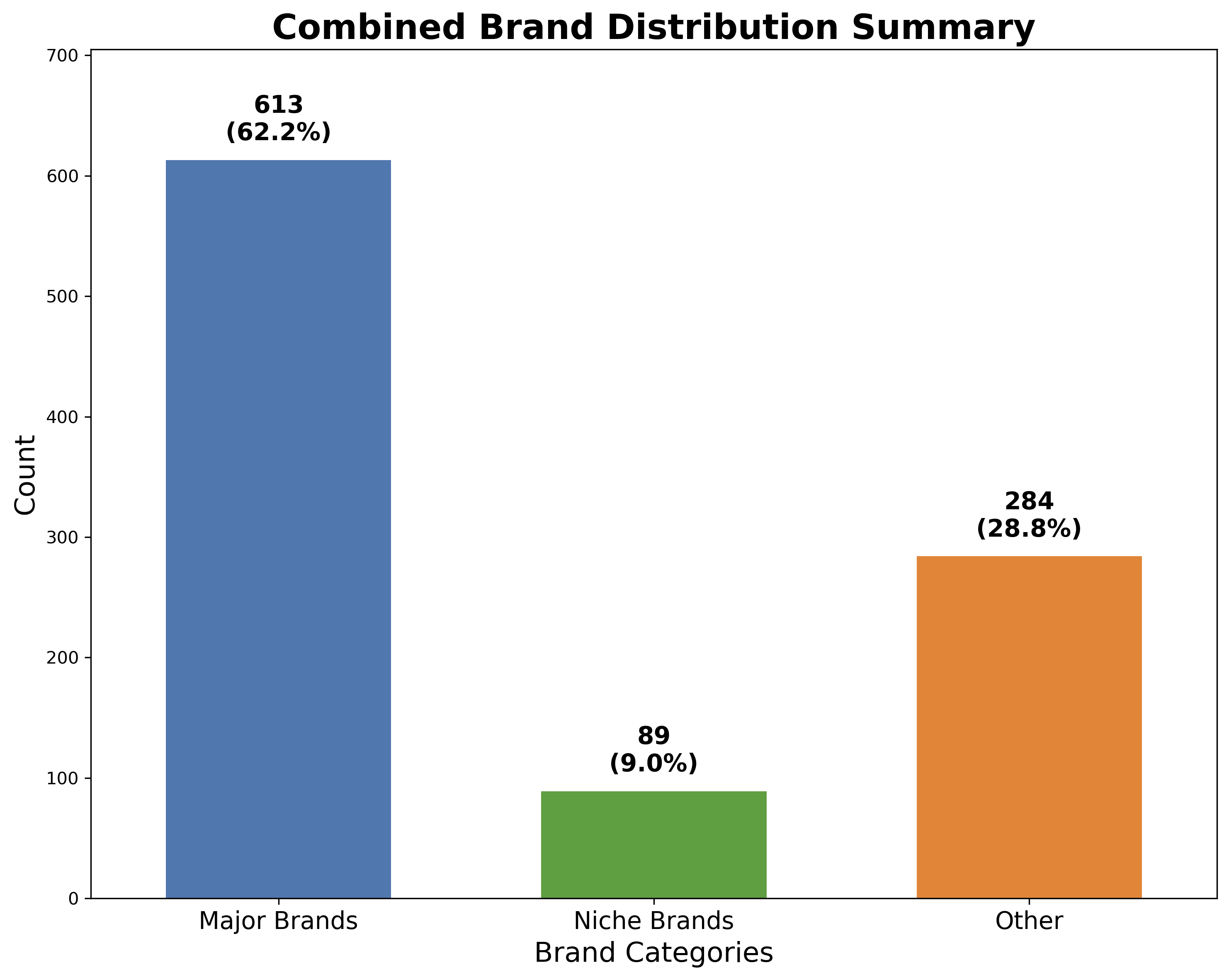}
    \caption{Big Brand Bias: Category summary (combined)}
    \label{fig:brand-summary-combined}
\end{figure}

\begin{figure}[h]
    \centering
    \includegraphics[width=0.5\textwidth]{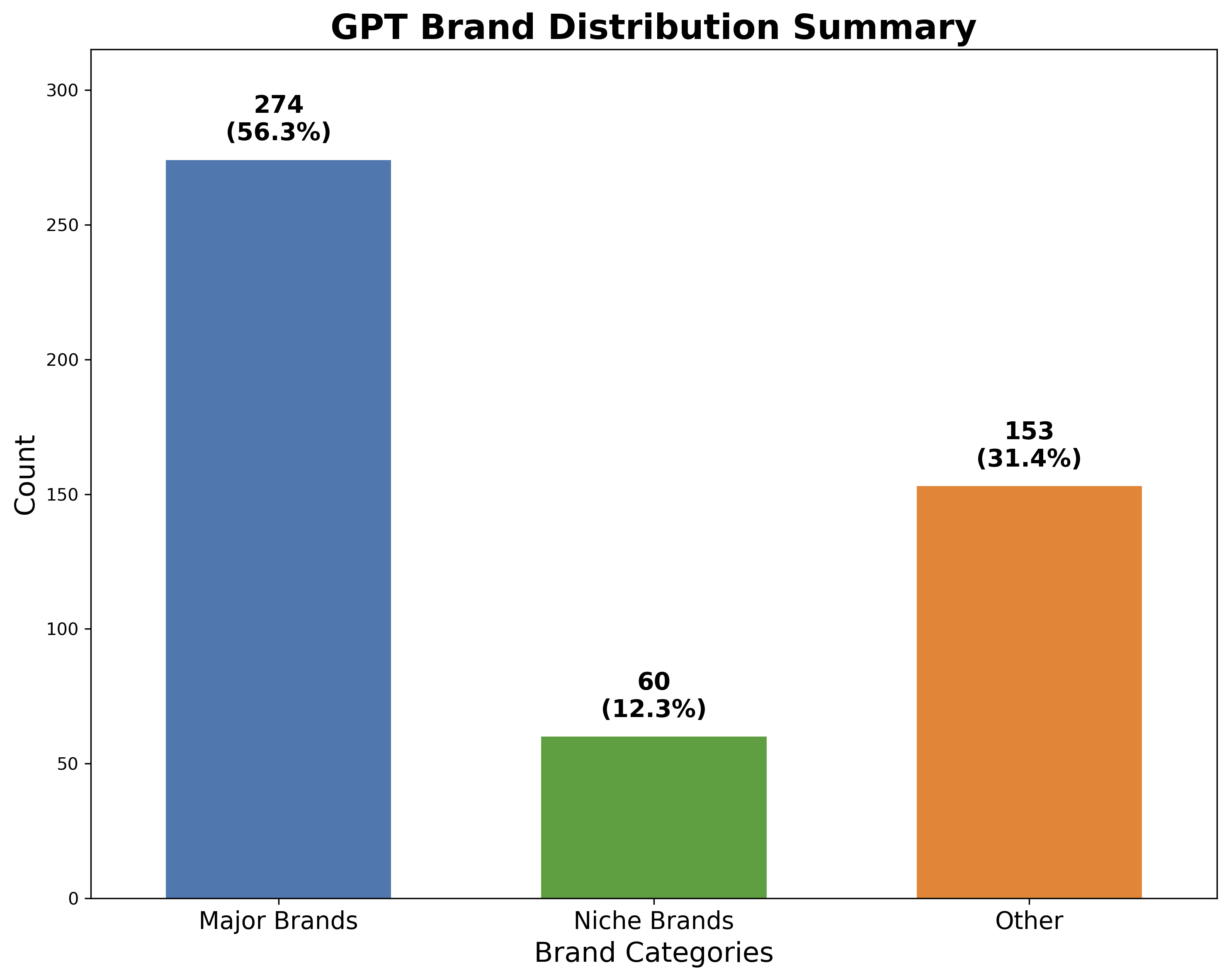}
    \caption{Big Brand Bias: Category summary (ChatGPT)}
    \label{fig:brand-summary-gpt}
\end{figure}

\begin{figure}[h]
    \centering
    \includegraphics[width=0.5\textwidth]{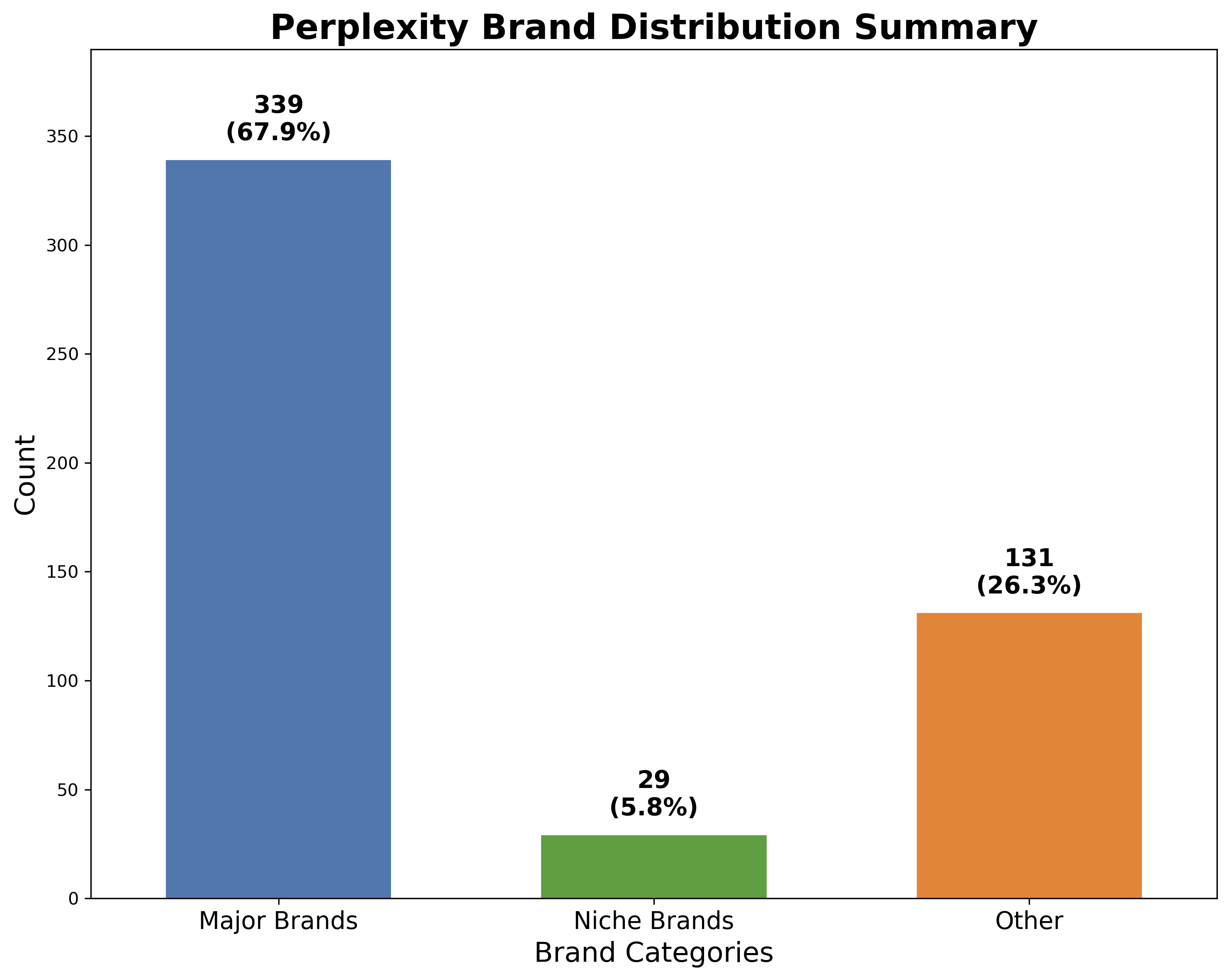}
    \caption{Big Brand Bias: Category summary (Perplexity)}
    \label{fig:brand-summary-perplexity}
\end{figure}

\paragraph{Results.}
Across both systems, answers skewed toward major brands. For ChatGPT, major brands account for 56.3\% of all identified brand mentions (274 of 487), niche brands for 12.3\% (60), and other brands for 31.4\% (153) (Figs.~\ref{fig:brand-distribution-gpt}, \ref{fig:brand-summary-gpt}). Perplexity exhibits an even stronger tilt toward major names: 67.9\% major (339 of 499), 5.8\% niche (29), and 26.3\% other (131) (Figs.~\ref{fig:brand-distribution-perplexity}, \ref{fig:brand-summary-perplexity}). Pooling the two systems yields a combined distribution of 62.2\% major, 9.0\% niche, and 28.8\% other (Figs.~\ref{fig:brand-distribution-combined}, \ref{fig:brand-summary-combined}). Taken together, these shares indicate that both models favor major brands, with Perplexity's preference visibly stronger (higher major share, lower niche share).

Brand-level concentration mirrors these aggregates. In both models, Coca-Cola and Pepsi dominate the head of the distribution (ChatGPT: 78 and 44 mentions; Perplexity: 107 and 58), with Dr Pepper consistently occupying the next tier (ChatGPT: 30; Perplexity: 34). Niche brands such as Jones Soda, Boylan, Faygo, and Sprecher do surface, but at much lower frequencies (Figs.~\ref{fig:brand-distribution-gpt}, \ref{fig:brand-distribution-perplexity}).

\begin{figure}[h]
    \centering
    \includegraphics[width=0.5\textwidth]{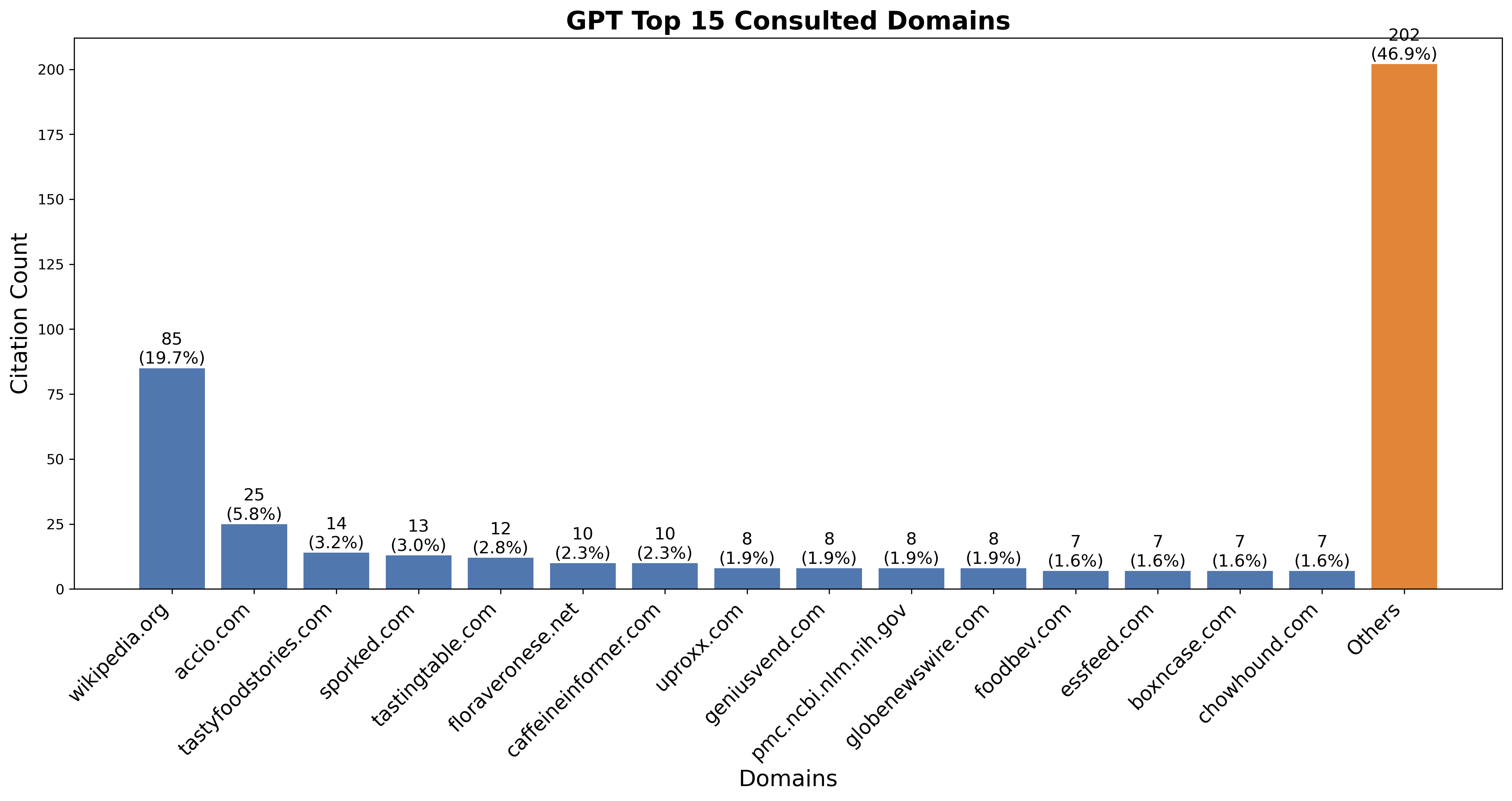}
    \caption{Big Brand Bias: Top consulted domains (ChatGPT)}
    \label{fig:domain-distribution-gpt}
\end{figure}

\begin{figure}[h]
    \centering
    \includegraphics[width=0.5\textwidth]{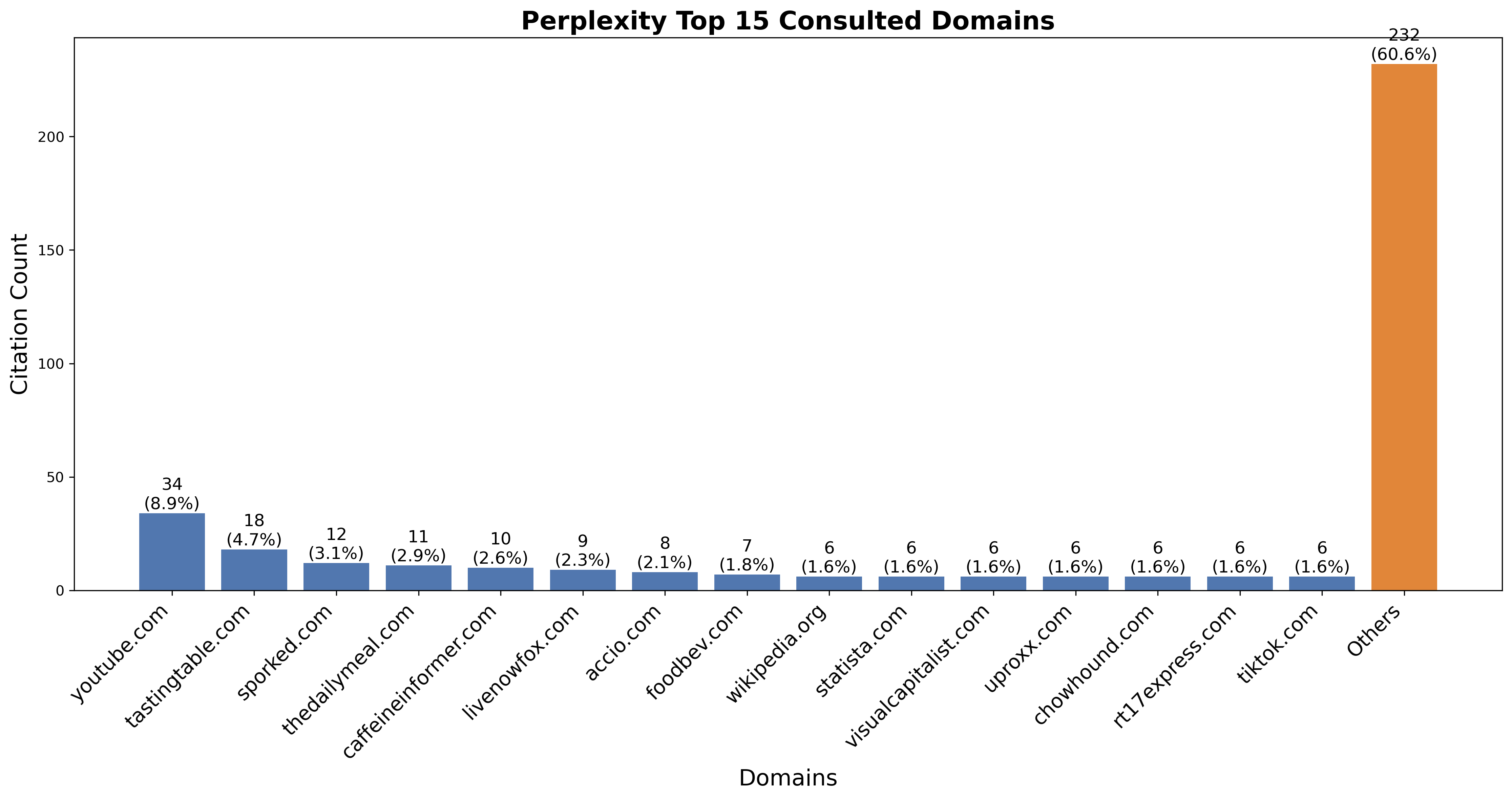}
    \caption{Big Brand Bias: Top consulted domains (Perplexity)}
    \label{fig:domain-distribution-perplexity}
\end{figure}

Cited-domain profiles also differ by system. ChatGPT shows a relatively concentrated pattern, relying heavily on Wikipedia ($\sim$19.7\% of citations) and a small set of consumer-review and food sites (e.g., accio.com, Tasting Table, Sporked) (Fig.~\ref{fig:domain-distribution-gpt}). Perplexity distributes citations across a broader long tail (the ``Others'' bucket $\approx$ 60.6\%), with YouTube as the single most frequent domain ($\sim$8.9\%) and additional coverage from Tasting Table, The Daily Meal, Caffeine Informer, Statista, Visual Capitalist, TikTok, and others (Fig.~\ref{fig:domain-distribution-perplexity}). Despite these different evidence profiles, Perplexity still returns more major brands overall than ChatGPT.

\paragraph{Interpretation.}
The two systems marshal evidence differently, one more encyclopedic and concentrated, the other more multimedia and dispersed, yet under unbranded prompts both default to market-leading brands. The relative magnitudes of the category shares suggest that, absent explicit constraints, source prominence and model priors jointly pull recommendations toward major labels.

\subsubsection{Bank Queries Across Personas}

\paragraph{Objective.}
We compare how four AI systems, Gemini, Perplexity, ChatGPT, Claude, rank banks when queries are phrased in a ranking format and conditioned on three personas: (i) young professionals entering the workforce, (ii) mid-career individuals with families, and (iii) retirees. The goal is to characterize differences in the domains of sources each model cites across personas, as well as the types of these sources (brand, social, earned).

\paragraph{Experimental Design.}
We adapted a comprehensive set of banking questions for each persona into bank-ranking prompts (e.g., fees and waivers, ATM networks, mobile features, APY/ CDs, mortgages/ HELOCs, fraud protection, retirement services). For each persona $\times$ model combination, we collected the full responses and citations, extracted domains, and classified them as brand (bank-owned), social (forums/communities), or earned (editorial/third-party). The extraction, normalization, and classification steps follow the common pipeline in \S5.1. We report (a) the distribution of domain types per persona $\times$ model and overall, and (b) the most frequent domains for each persona $\times$ model and in aggregate.

\paragraph{Results.}

\begin{figure}[h]
    \centering
    \includegraphics[width=0.5\textwidth]{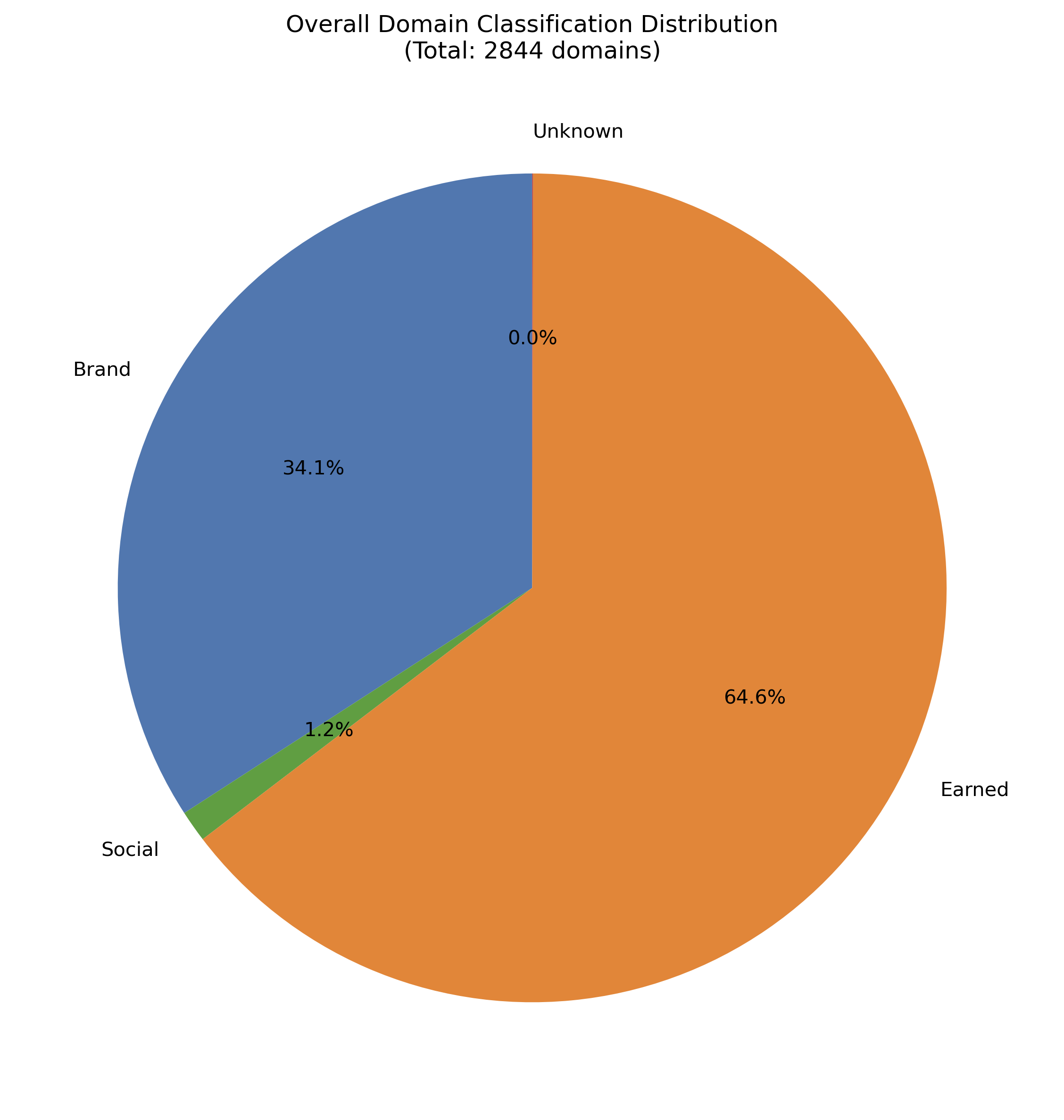}
    \caption{Bank Queries Across Personas: Overall domain-type distribution (all models/personas)}
    \label{fig:overall-classification-distribution}
\end{figure}

\textbf{Overall mix.} Across all models and personas, earned sources dominate, with brand second and social minimal (overall: earned 64.6\%, brand 34.1\%, social 1.2\%; see Fig.~\ref{fig:overall-classification-distribution}). This pattern reflects extensive reliance on editorial reviews and financial explainers in ranking-style answers.

\textbf{Differences across models.} The engine identity explains more variation than persona.

\begin{figure}[h]
    \centering
    \includegraphics[width=0.5\textwidth]{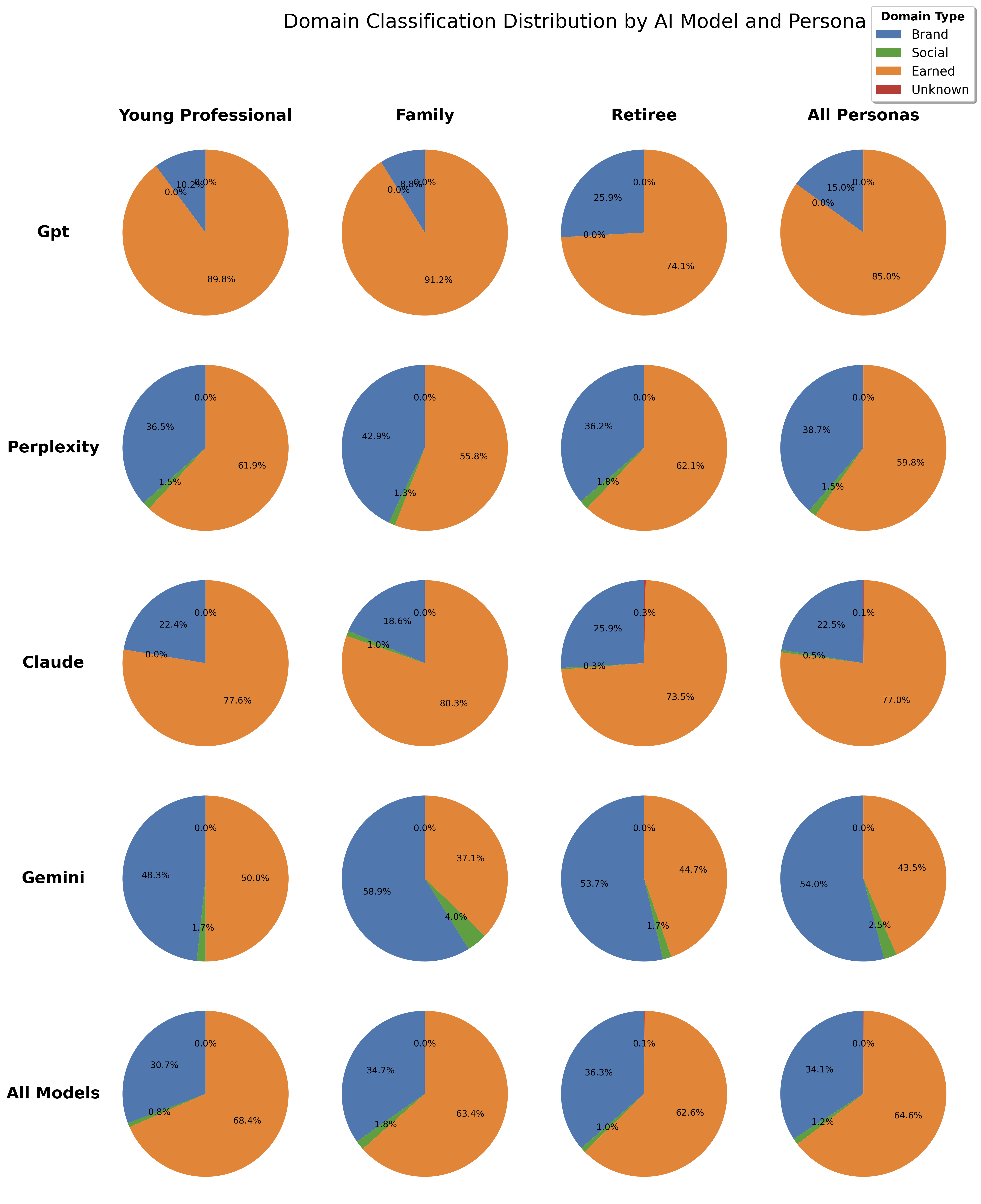}
    \caption{Bank Queries Across Personas: Domain-type shares by model $\times$ persona}
    \label{fig:classification-by-model-persona-grid}
\end{figure}

\begin{itemize}
    \item ChatGPT and Claude are the most \textit{earned-heavy}: for all three personas they concentrate citations on third-party editorial sites; brand shares remain comparatively low (Fig.~\ref{fig:classification-by-model-persona-grid}, first/third rows).
    \item Perplexity mixes earned and brand more evenly, but still leans earned across personas (Fig.~\ref{fig:classification-by-model-persona-grid}, second row).
    \item Gemini is the most \textit{brand-leaning}: brand sources are frequently at or above half of citations across personas, particularly for the family persona (Fig.~\ref{fig:classification-by-model-persona-grid}, fourth row).
\end{itemize}

\textbf{Persona effects.} Within-model shifts with persona are modest relative to cross-model gaps. A consistent nuance is that retirees tend to draw a slightly higher brand share than young professionals or families for the more earned-oriented models (e.g., ChatGPT, Claude), while Gemini remains brand-forward across all personas. In short, \textit{who the engine is} matters more than \textit{which persona} is queried.

\begin{figure}[h]
    \centering
    \includegraphics[width=0.5\textwidth]{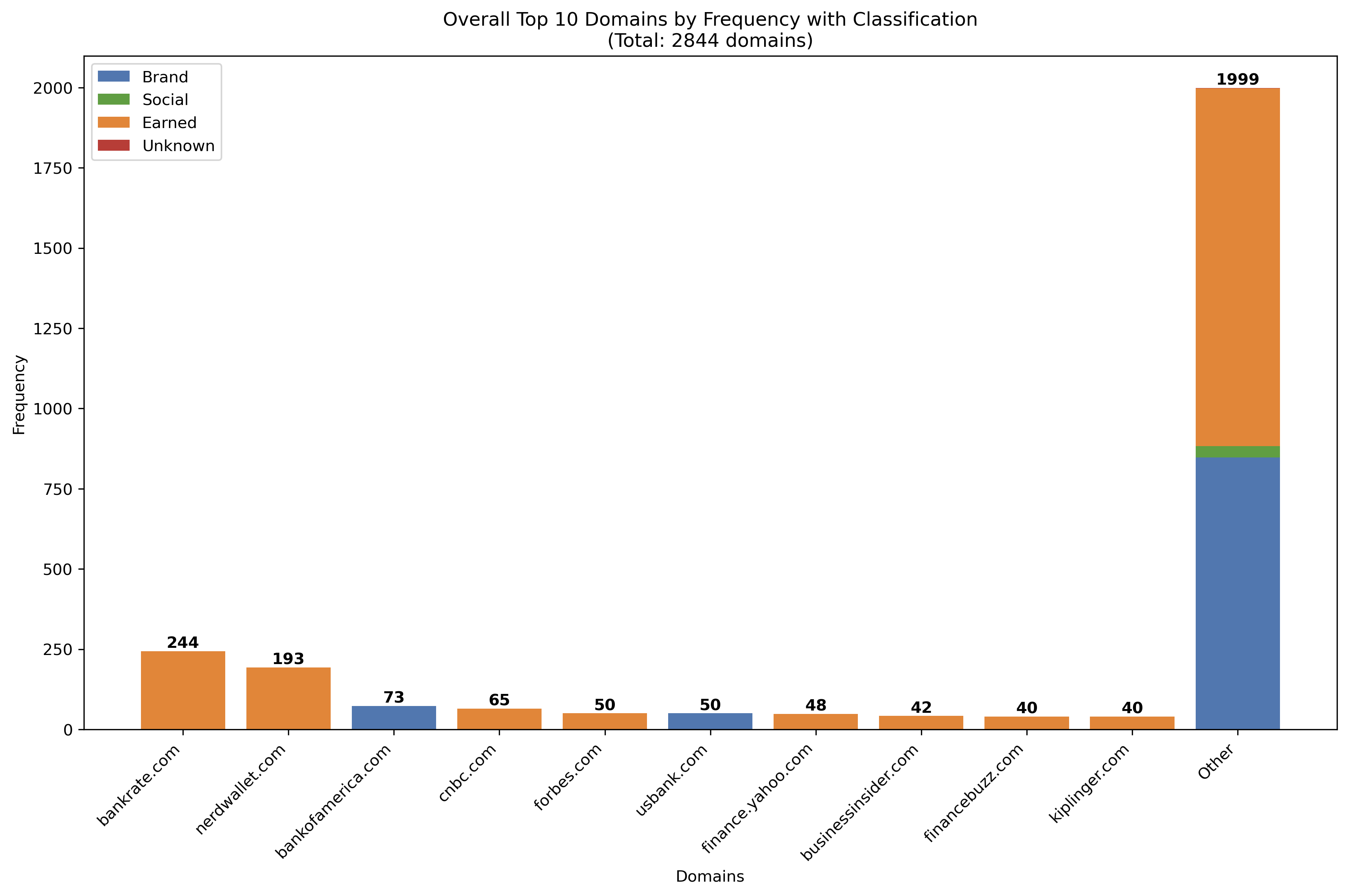}
    \caption{Bank Queries Across Personas: Overall top 10 domains with type breakdown}
    \label{fig:overall-top10-domains-classification}
\end{figure}

\textbf{Domain ecology (aggregate).} The aggregate top-domain view is led by Bankrate and NerdWallet, followed by mainstream business media (e.g., CNBC, Forbes, Yahoo Finance, Business Insider, Kiplinger). Bank of America and U.S. Bank are notable brand outliers within the top ten (Fig.~\ref{fig:overall-top10-domains-classification}). The long tail is substantial (``Other'' is the largest bar), indicating wide dispersion beyond a few anchors.

\textbf{Social signals.} Social sources contribute very little across all settings (near-zero slices in Fig.~\ref{fig:classification-by-model-persona-grid}, and $\sim$1\% overall in Fig.~\ref{fig:overall-classification-distribution}), suggesting that bank-ranking prompts elicit editorial/brand pages far more than community discussions.

\paragraph{Interpretation.}
Under ranking-style banking prompts, all systems converge on editorial financial media as primary evidence, but differ in how much they incorporate brand-owned pages. The ordering that emerges from these results is: Gemini (most brand-leaning) $\rightarrow$ Perplexity (balanced) $\rightarrow$ Claude / ChatGPT (most earned-heavy). Persona changes do modulate the mix, for instance, retirees occasionally tilt some engines toward brand resources, but these shifts are smaller than the between-engine differences. The dominance of Bankrate/NerdWallet and the breadth of the long tail point to a citation ecosystem where authoritative, evergreen review hubs anchor the answers, while brand sites are selectively injected depending on the engine.

\subsubsection{Language Sensitivity Experiment}

\paragraph{Objective.}
We examine how language affects AI search outputs across engines. Concretely, we ask whether the brands and domains surfaced for the same intents vary when prompts are translated into Chinese, Japanese, German, French, and Spanish (vs.\ English), and how the type and language of cited websites shift across languages.

\paragraph{Experimental Design.}
Design and pipeline exactly follow \S4.2.3 (the same language sensitivity experiment), except that we exclude Google and compare AI engines with one another (Gemini, Claude, ChatGPT, Perplexity). We use the same ten verticals and five target languages (Chinese, Japanese, German, French, Spanish vs.\ English). Apart from the additional brand-overlap heatmaps, all other metrics and reporting conventions are identical to \S4.2.3. Specifically:

\begin{itemize}
    \item \textbf{Domain-overlap heatmaps:} English--X overlaps by vertical, computed as the Jaccard index on cited-domain sets and averaged within each vertical (definition per \S4.2.3).
    \item \textbf{Brand-overlap heatmaps:} English--X overlaps by vertical, computed as the Jaccard index on extracted brand sets per query (from LLM answers), then averaged within each vertical (consistent with the domain-overlap calculation).
    \item \textbf{Aggregate pies:} overall \textbf{domain-type} mix (brand / social / earned) and \textbf{website-language} mix (English vs.\ non-English) across all verticals, same as in \S4.2.3.
\end{itemize}

Extraction, normalization, and classification follow the common pipeline in \S5.1.

\paragraph{Results.}

\textbf{Cross-language domain stability is limited and model-dependent.}  
Domain overlap varies sharply by engine. Claude shows the highest cross-language stability (frequent high overlaps across verticals; Fig.~\ref{fig:language-domain-overlap-claude}). Perplexity and Gemini tend to have much lower cross-language domain overlap overall (a few low to moderate cells, with the majority remaining very low or near zero; Figs.~\ref{fig:language-domain-overlap-perplexity}, \ref{fig:language-domain-overlap-gemini}). GPT shows the lowest overlap: in our sample, domain sets across languages are consistently near-zero, i.e., it taps different site ecosystems by language (Fig.~\ref{fig:language-domain-overlap-gpt}).

\begin{figure}[h]
    \centering
    \includegraphics[width=0.5\textwidth]{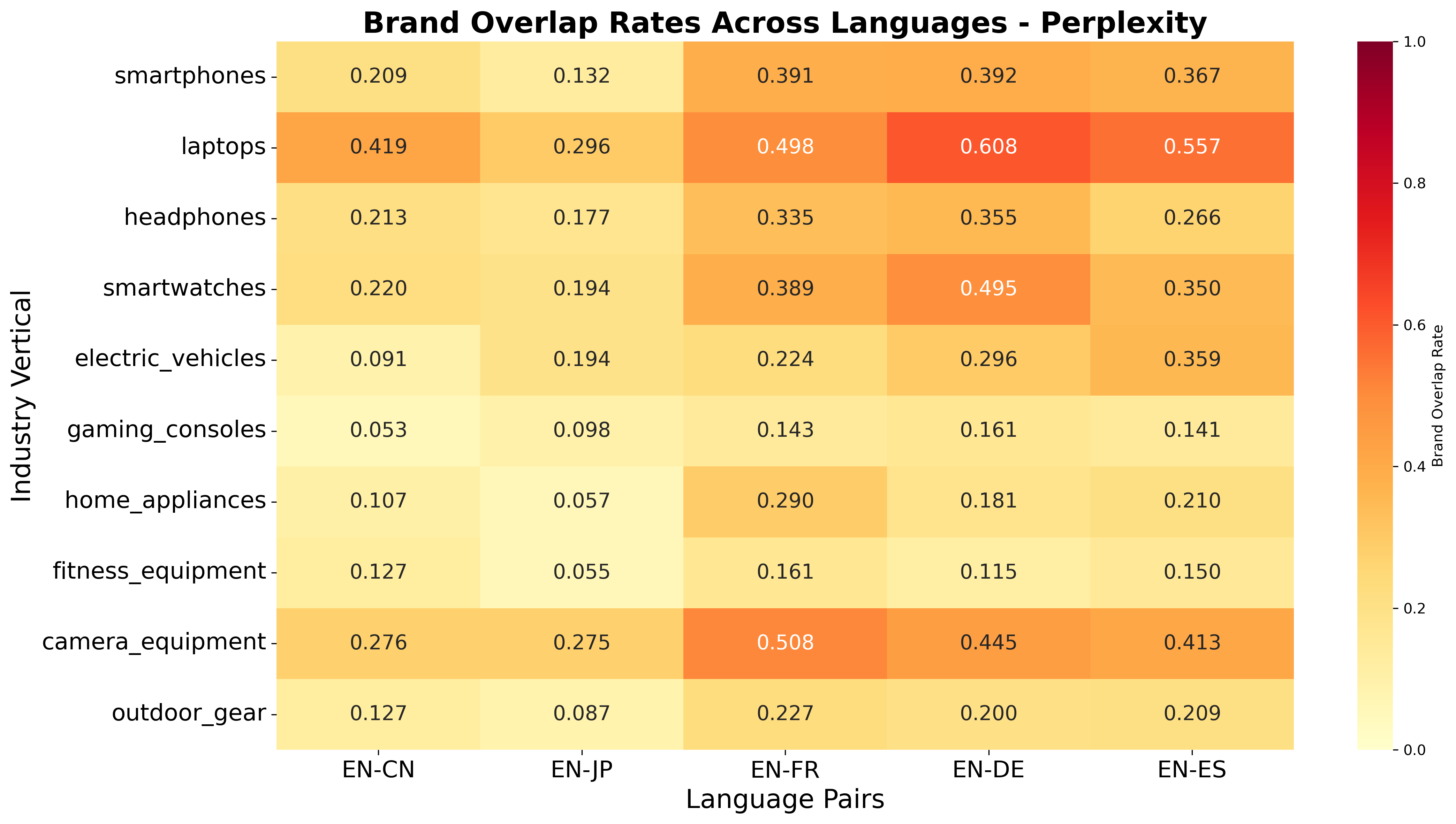}
    \caption{Language Sensitivity: Brand overlap heatmap, Perplexity}
    \label{fig:language-brand-overlap-perplexity}
\end{figure}

\begin{figure}[h]
    \centering
    \includegraphics[width=0.5\textwidth]{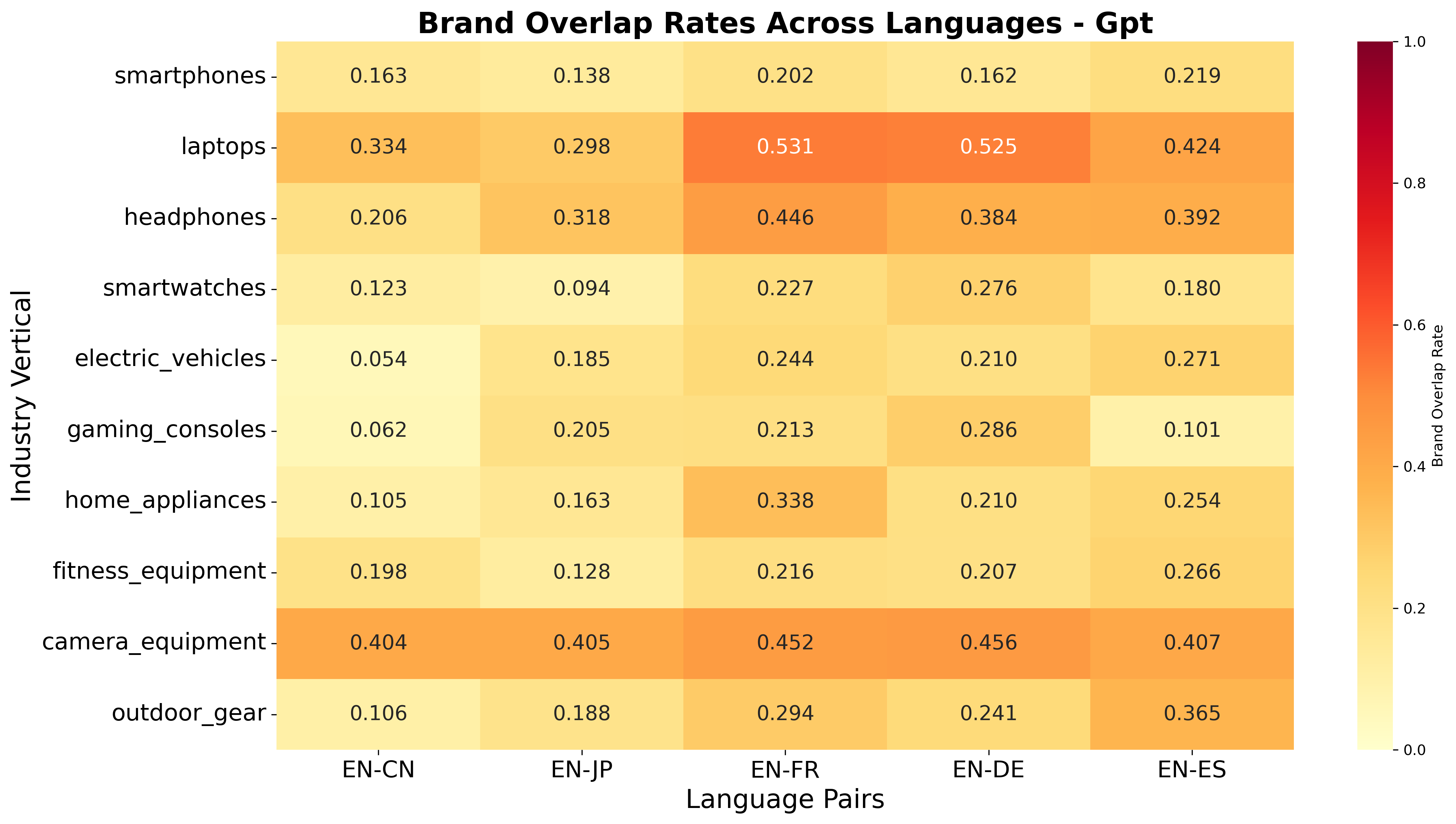}
    \caption{Language Sensitivity: Brand overlap heatmap, GPT}
    \label{fig:language-brand-overlap-gpt}
\end{figure}

\begin{figure}[h]
    \centering
    \includegraphics[width=0.5\textwidth]{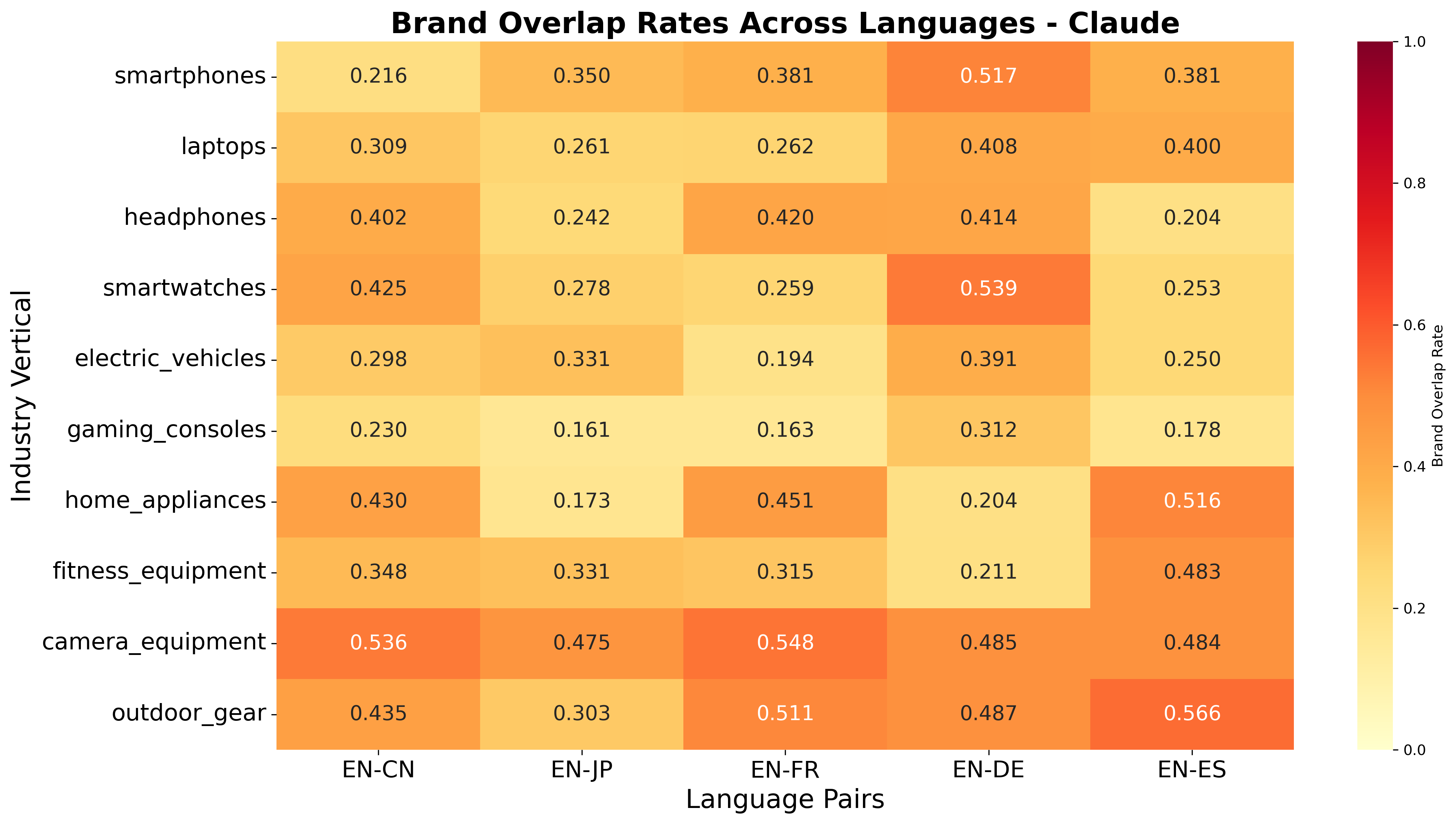}
    \caption{Language Sensitivity: Brand overlap heatmap, Claude}
    \label{fig:language-brand-overlap-claude}
\end{figure}

\begin{figure}[h]
    \centering
    \includegraphics[width=0.5\textwidth]{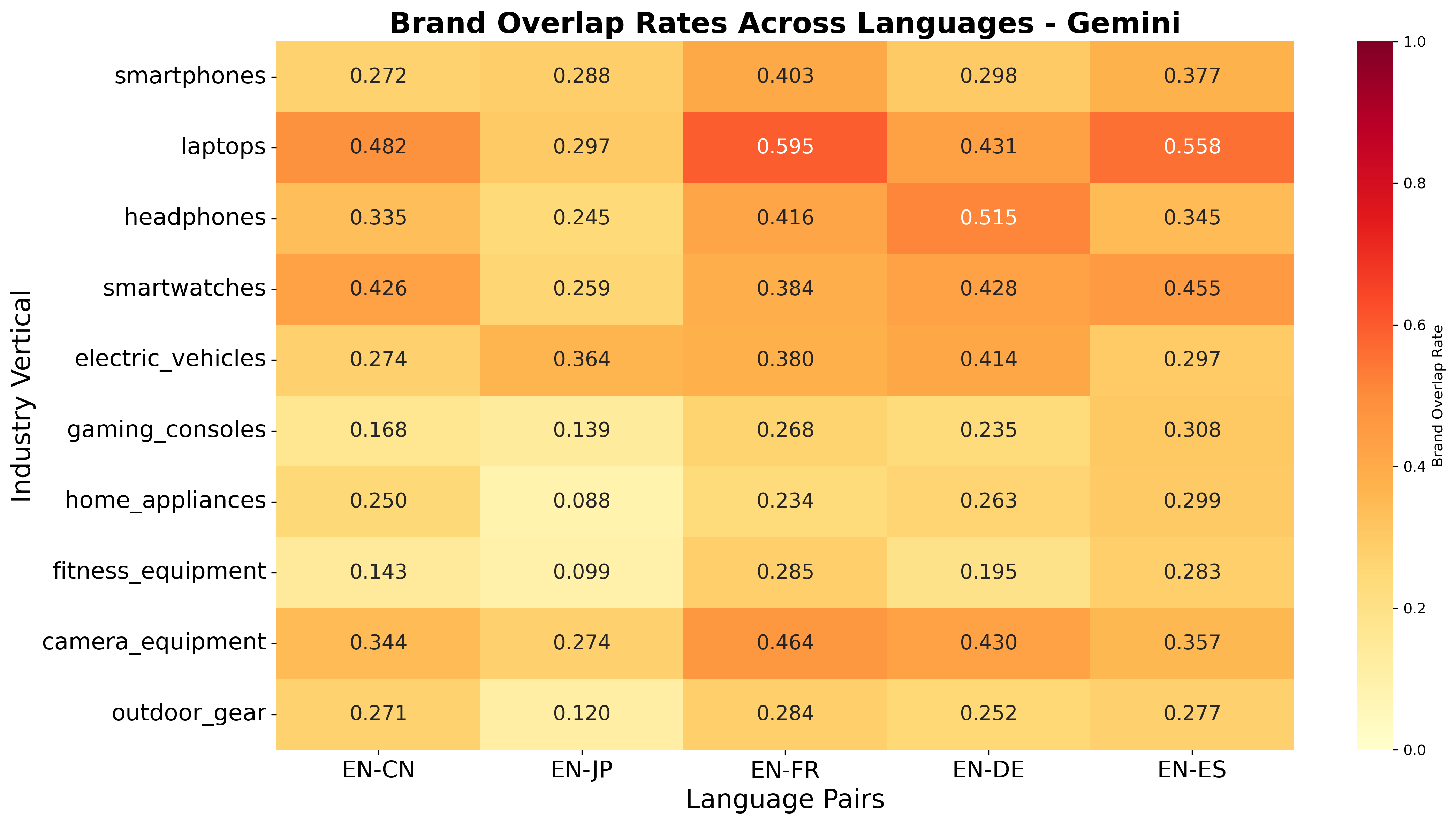}
    \caption{Language Sensitivity: Brand overlap heatmap, Gemini}
    \label{fig:language-brand-overlap-gemini}
\end{figure}

\textbf{Brand overlap also differs by engine.}  
Across the board, brand overlap is much higher than domain overlap. Patterns vary by model and vertical. For example, Perplexity and ChatGPT reach medium to high overlap in several high-concentration verticals (e.g., laptops, camera equipment; Figs.~\ref{fig:language-brand-overlap-perplexity}, \ref{fig:language-brand-overlap-gpt}). Claude is similarly strong across many categories (smartphones, camera, outdoor gear; Fig.~\ref{fig:language-brand-overlap-claude}), with pockets near or above 0.5. Gemini is comparable overall, with strong cells in laptops and several headphones/camera pairs while exhibiting more low values (<0.30) in home appliances and fitness equipment. Overall, no single engine dominates across all verticals; stability is vertical-dependent. (Fig.~\ref{fig:language-brand-overlap-gemini}).

\textbf{Source types remain earned-heavy across languages.}  
When aggregating all verticals, models, and languages, we can see that the engine identity explains more variation than languages used. The domain distributions follows a earned $\gg$ brand $\gg$ social pattern across AI engines (Fig.~\ref{fig:language-overall-domain-type-distribution}). This mirrors \S5.2.2: ranking-style prompts tend to elicit editorial evidence regardless of language. Regarding the differences across models, GPT and Claude are the most earned-heavy (Fig.~\ref{fig:language-overall-domain-type-distribution}, second and fourth column), while Perplexity and Gemini have a greater share of brand and social sources (but remains the earned $\gg$ brand $\gg$ social pattern; Fig.~\ref{fig:language-overall-domain-type-distribution}, third and fifth column), which is also similar to \S5.2.2.

\textbf{Website language depends on the engine.}  
Under non-English prompts, citations skew toward the target language, but to different degrees (Fig.~\ref{fig:language-website-language-distribution}): GPT and Perplexity are the most local-language heavy (Fig.~\ref{fig:language-website-language-distribution}, second and third column); Claude is the special case where English dominates and non-English is only a very small share (Fig.~\ref{fig:language-website-language-distribution}, fourth column); Gemini is more balanced, with the split fluctuating by language.

Across languages, a consistent gradient emerges (Fig.~\ref{fig:language-website-language-distribution}): Japanese and French prompts produce the strongest localization (target-language sites dominate across engines); Chinese is also highly localized but with a slightly larger English slices (especially clear for Gemini); German and Spanish are more mixed, often have larger English slices (especially Gemini where the English slices for these two languages are close to 50\%). This language gradient holds after excluding Claude's extreme English-heavy behavior and is most pronounced for GPT and Perplexity.

\paragraph{Interpretation.}
Language reshapes results at two levels: the evidence ecology (which sites are cited) and the brand lists. Claude reuses a relatively stable set of authority domains across languages; GPT effectively swaps in different site ecosystems by language, yielding very low cross-language domain overlap; Perplexity and Gemini sit between these poles. On the brand side, overlap is higher where global head brands dominate (e.g., cameras, laptops) and lower in more localized or long-tail categories (e.g. home appliances, fitness equipment). Implication: for consistent visibility in multilingual markets, brands need coverage in authoritative local-language media as well as English, and a multi-engine, multi-language distribution strategy is warranted.

\subsubsection{Paraphrase Sensitivity Experiment}

\paragraph{Objective.}
We test whether small, within-language changes in how a query is phrased alter what the systems surface. For the same intent, we compare the brands and domains returned when prompts are reworded and whether the mix of source types (brand / social / earned) shifts. This section mirrors the language experiment but replaces translation with controlled paraphrases.

\paragraph{Experimental Design.}
Design and pipeline exactly follow \S4.2.4 (the same paraphrase sensitivity experiment), except that we exclude Google and compare AI engines with one another (Gemini, ChatGPT, Perplexity). We use the same ten verticals and the same seven paraphrase templates: \textit{justification\_required}, \textit{source\_required}, \textit{quote\_required}, \textit{confidence\_score}, \textit{ranked\_order}, \textit{imperative\_list}, \textit{keyword\_only}. Apart from the additional brand-overlap heatmaps, all other metrics and reporting conventions are identical to \S4.2.4. Specifically:

\begin{itemize}
    \item \textbf{Domain-overlap heatmaps:} base--paraphrase overlaps by vertical, computed as the Jaccard index on cited-domain sets per query and averaged within each vertical (definition per \S4.2.4).
    \item \textbf{Brand-overlap heatmaps:} base--paraphrase overlaps by vertical, computed as the Jaccard index on extracted brand sets per query (from LLM answers), then averaged within each vertical (consistent with the domain-overlap calculation).
    \item \textbf{Aggregate pies:} overall \textbf{domain-type} mix (brand / social / earned) across all verticals, same as in \S4.2.4.
\end{itemize}

Extraction, normalization, and classification follow the common pipeline in \S5.1.

\paragraph{Results.}

\textbf{Paraphrases move results less than languages.}  
Relative to \S5.2.8 (language sensitivity test), paraphrasing induces smaller perturbations. Brand lists are generally robust to paraphrasing (high overlaps), while domain sets change more than brands but still remain much more stable than in the cross-language study.

\begin{figure}[h]
    \centering
    \includegraphics[width=0.5\textwidth]{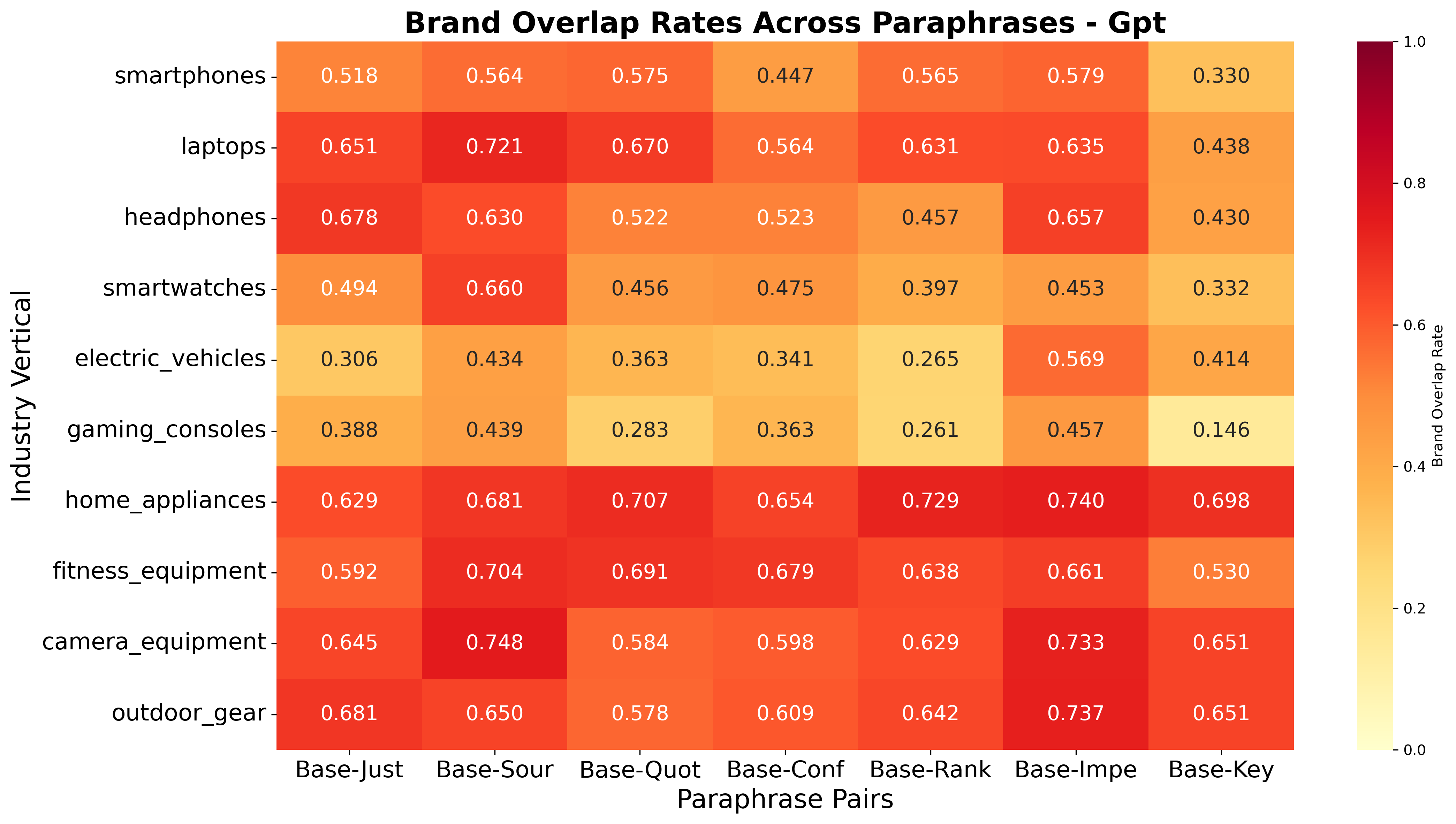}
    \caption{Paraphrase Sensitivity: Brand overlap heatmap, GPT}
    \label{fig:paraphrase-brand-overlap-gpt}
\end{figure}

\begin{figure}[h]
    \centering
    \includegraphics[width=0.5\textwidth]{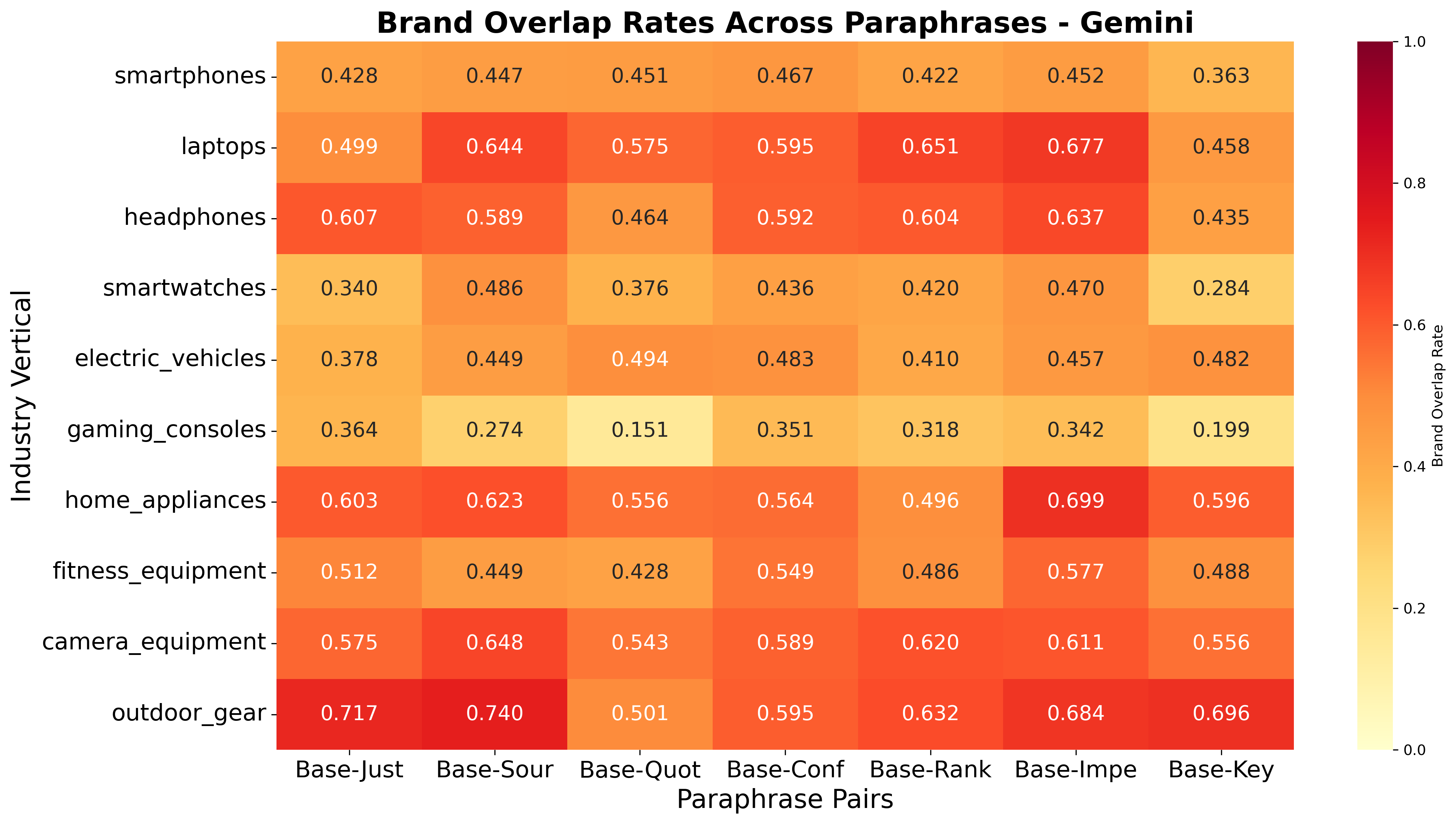}
    \caption{Paraphrase Sensitivity: Brand overlap heatmap, Gemini}
    \label{fig:paraphrase-brand-overlap-gemini}
\end{figure}

\begin{figure}[h]
    \centering
    \includegraphics[width=0.5\textwidth]{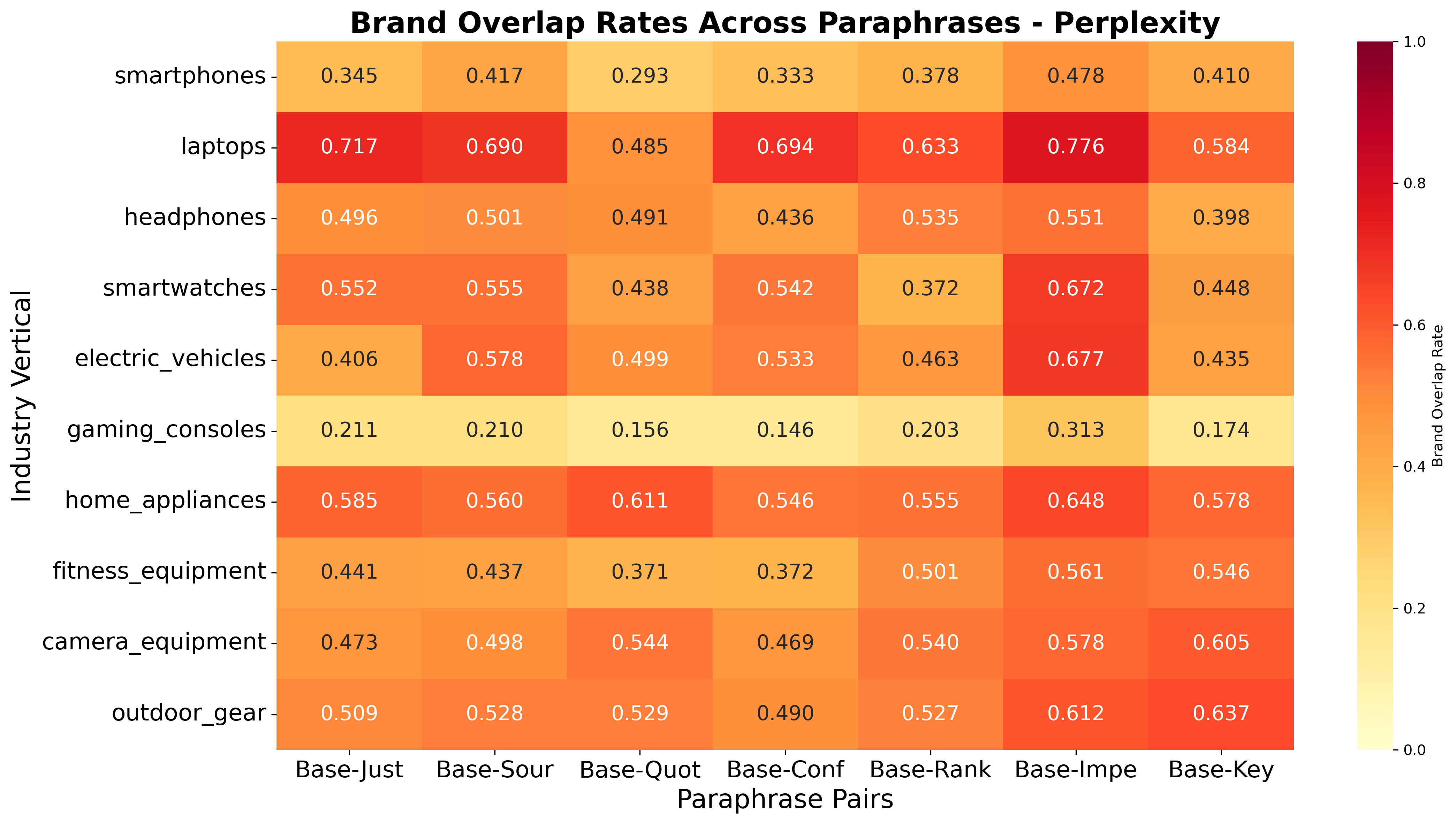}
    \caption{Paraphrase Sensitivity: Brand overlap heatmap, Perplexity}
    \label{fig:paraphrase-brand-overlap-perplexity}
\end{figure}

\textbf{Brand overlap (within engine).}  
Across verticals, GPT shows generally high cross-paraphrase stability (many cells exceed 0.6 or even 0.7), indicating that once intent is fixed, wording rarely overturns head brands (Fig.~\ref{fig:paraphrase-brand-overlap-gpt}). Gemini reaches moderately high to high overall (roughly 0.4–0.7 across most verticals), with a few low-overlap cells in verticals such as gaming consoles (Fig.~\ref{fig:paraphrase-brand-overlap-gemini}). Perplexity is comparable on average but displays the widest spread, gaming consoles notably lower, indicating stronger category-specific sensitivity (Fig.~\ref{fig:paraphrase-brand-overlap-perplexity}).

\textbf{Domain overlap (within engine).}  
Paraphrases affect which articles get cited more than which brands appear. Perplexity and Gemini show the highest domain stability across paraphrases (numerous mid-to-high overlaps; Figs.~\ref{fig:paraphrase-domain-overlap-perplexity}, \ref{fig:paraphrase-domain-overlap-gemini}). GPT has lower overlaps overall, i.e., it rotates supporting sites even when brand slates are similar (Fig.~\ref{fig:paraphrase-domain-overlap-gpt}).

\textbf{Source-type mix across paraphrase styles.}  
The earned-heavy pattern persists for all engines, and the distribution differences across AI engines also mirrors previous sections. Shares differ by engine, but the brand/social/earned distribution changes little across paraphrases (Fig.~\ref{fig:paraphrase-domain-type-distribution}).

\paragraph{Interpretation.}
Rewording primarily changes format and citations, not the core recommendations. Compared with translation, paraphrases seldom overturn brand lists, especially for GPT and Gemini, though Perplexity is more wording-sensitive in certain verticals. Domain selection is more pliable (notably for GPT), yet still more stable than in the language experiment. Practically, tuning content to satisfy sources/quotes/ranking instructions can influence which specific articles get cited, but language localization has a much larger impact on both the evidence ecology and brand visibility than prompt rephrasing alone.

\subsection{The Generative Engine Optimization Agenda}

\subsubsection{Engineer for Agency and Scannability}

A universal finding across all engines and verticals is the critical need for machine-readable, structured data. AI systems function as agents that must parse, interpret, and synthesize information to generate answers. Websites cluttered with marketing fluff and unstructured content will fail. The primary action is to treat your website as an API for AI. This requires rigorous implementation of technical SEO fundamentals combined with detailed schema markup (Schema.org) for all entities—products, specifications, prices, reviews, warranty details, and availability. This technical foundation is non-negotiable for being "do business with" by AI agents.

\subsubsection{Dominate Earned Media Across All Engines}
The most consistent and significant finding is the overwhelming bias of AI engines toward Earned media. Claude and ChatGPT are extremely earned-heavy, while Perplexity and Gemini incorporate more brand and social content but still prioritize earned sources. To win in AI search, a brand must shift its focus from creating owned content to systematically earning third-party validation. This means investing heavily in public relations, media outreach, and expert collaborations to secure features, reviews, and mentions in the authoritative publications and review sites that these engines favor. Building a portfolio of backlinks from these high-authority domains is not a secondary tactic but the core GEO strategy, as it directly feeds the AI's perception of your brand's expertise, authoritativeness, and trustworthiness (E-E-A-T).

\subsubsection{Engine-Specific Tactics for Brand Visibility}
The analysis reveals that engine choice fundamentally alters the information ecosystem a user encounters. Therefore, a one-size-fits-all approach is ineffective. For Claude and ChatGPT, which exhibit high cross-language brand stability and a strong earned bias, the strategy is to secure a position within the core set of globally recognized, authoritative domains in your vertical. For Perplexity, which incorporates more diverse sources including YouTube and retail sites, the strategy expands to include creating video content and ensuring product information is accurately listed on major retailer domains. Gemini, which shows a greater propensity to cite brand-owned properties, allows for a slightly more balanced approach that leverages both earned media and well-structured, deep content on your own domain.

\subsubsection{Multilingual Strategy: Localize Authority, Not Just Content}
The language sensitivity experiment demonstrates that success in non-English markets requires more than simple translation. GPT and Perplexity heavily localize, sourcing almost entirely from the target language's ecosystem. Claude, by contrast, reuses English-language authority domains across languages. This demands a dual-path strategy. To win on GPT and Perplexity in a specific region, you must build relationships with and earn coverage from the most authoritative local-language publishers and review sites. Simultaneously, strengthening your brand's presence in top-tier English-language earned media can improve visibility on Claude across many languages. A brand must audit the authority landscape in each target region and engine to allocate resources effectively.

\subsubsection{Content Strategy: Justify and Compare for the Shortlist}
AI search is not about generating ten blue links but about justifying a placement on a synthesized shortlist. The low domain overlap between engines indicates that each AI is synthesizing answers from a different set of sources, but they are all seeking clear, unambiguous justification. Website content must be explicitly engineered to answer comparison questions. This involves creating scannable, justification-rich content such as detailed comparison tables against competitors, bulleted pros and cons lists, and clear, bolded statements of value proposition (e.g., "longest battery life," "best for small families"). The brand that makes it easiest for the AI to extract reasons for its superiority will win the recommendation.

\subsubsection{Niche Brand Strategy: Overcome the Big Brand Bias}
The observed big brand bias presents a significant challenge for niche and indie brands. Unbranded queries default to market leaders. To break through, niche brands must over-invest in building tangible, verifiable authority. This can be achieved by dominating a specific, narrow niche through deep expert content and targeted earned media campaigns in specialty publications. They should also leverage strategies that work on Perplexity, such as creating high-quality YouTube review content and engaging with community discussions, to build a grassroots authority that can eventually be recognized by the more conservative engines like Claude and GPT.

\section{The Imperative for Principled GEO Methodologies and Services}

The experimental data reveals a critical truth: the AI search landscape is not a monolith but a fragmented, dynamic, and highly competitive ecosystem. Success is not achieved through isolated tactics but requires a continuous, principled, and strategic discipline. The old paradigm of periodic SEO audits is obsolete. The new paradigm demands a comprehensive GEO operating system—a suite of methodologies and managed services designed to achieve and defend dominance across all major AI engines. This system must be built on five core pillars.

First, a principled methodology requires continuous, engine-specific competitive intelligence. The low domain overlap and differing source biases mean that a strategy that works on Perplexity may fail on Claude. Brands cannot afford to guess. They need a service that continuously audits the digital ecosystem for their core topics, identifying exactly which sources each AI engine (Claude, GPT, Perplexity, Gemini) privileges. This involves mapping the "citation network" for your vertical: which earned media outlets, YouTube creators, and retail domains are consistently cited for key queries. This intelligence forms the absolute foundation of strategy, answering the critical questions of which sources are most important and what content to create by reverse-engineering the evidence base of the engines themselves.

Second, strategy must be guided by a structured content and justification framework. Knowing the sources is not enough; a methodology is needed to out-compete them on their own terms. This involves a deliberate shift from creating general top-of-funnel content to engineering "justification assets." A principled service will deploy a framework to audit existing content and plan new content against a strict checklist: Does it contain explicit, machine-scannable comparison data? Does it feature a clear, bolded value proposition that an AI can extract as a "justification attribute"? Is it structured with schema to be effortlessly parsed? This framework turns content creation from an art into a science, ensuring every asset is built to win a place in the AI's shortlist.

Third, brands must implement a systematic authority-building and earned media pipeline. The overwhelming bias toward earned media makes this the most critical investment. A principled methodology moves beyond sporadic PR to a managed service focused on GEO outcomes. This pipeline proactively identifies the top-tier authoritative sources in the mapped citation network and executes a continuous campaign to secure features, reviews, and expert commentary within them. The goal is not just a press mention but the deliberate cultivation of a backlink profile and online authority that every AI engine is trained to recognize and trust. This is how a brand ensures it is perceived as an authoritative source by the AI, making its inclusion in recommendations non-negotiable.

Fourth, a defensive and offensive ranking defense system is non-negotiable. The landscape is not static; competitors are continuously executing these same strategies. A principled methodology must therefore include continuous monitoring and defense. This involves tracking your ranking for core "shortlist" queries across all major AI engines and rapidly identifying any erosion in position. When a competitor gains ground, the system triggers a response: creating a more comprehensive justification asset, targeting a new earned media placement, or strengthening schema on a key page. This transforms GEO from a passive optimization task into an active, ongoing battle for visibility where speed and strategic response are paramount.

Finally, success requires an integrated, metrics-driven execution platform. These pillars cannot operate in silos. A true GEO service integrates them into a single dashboard and workflow. It connects the competitive intelligence to the content framework, directing which justification assets to build. It ties the authority pipeline directly to the target source list. It uses the ranking defense system to measure the ROI of every placed article and created asset. This closed-loop system allows for the strategic prioritization of engine-related tactics, answering how to prioritize based on where the greatest opportunity or threat exists, and ensuring resources are allocated to the efforts with the highest impact on AI visibility.

In conclusion, the complexity and competitiveness of the generative search environment have rendered ad-hoc SEO tactics obsolete. The only path to sustainable dominance is through a principled, disciplined, and continuous GEO methodology. This is not a one-time project but an essential managed service—an ongoing arms race where victory belongs to those with the best intelligence, the most impactful content, the strongest authority, and the fastest reaction time.

\section{Assumptions and Limitations of the Study}

This study provides a detailed snapshot of Generative AI search engine behaviors, but its findings are necessarily bounded by specific constraints in time, methodology, and data access. These limitations must be acknowledged to properly contextualize the results and their applicability.

First and foremost, the temporal nature of this analysis is a primary limitation. The data was collected in August 2025, and the behaviors, algorithms, and user interfaces of the services studied—Google, ChatGPT, Claude, Perplexity, and Gemini—are inherently dynamic. The competitive landscape of AI search is evolving at an accelerated pace. It is highly probable that these services will continually adjust their sourcing methodologies, ranking algorithms, and presentation of results (including how and if they cite sources) in response to competitive pressures, user feedback, and technological advancements. Therefore, the specific quantitative distributions of source types (Brand, Earned, Social) and the observed levels of inter-engine overlap are a product of this specific moment in time. Consequently, treating these findings as permanent truths is not advised. To remain relevant, the periodic evaluation and repetition of this study is imperative to track trends, identify shifts in strategy, and update optimization practices accordingly.

A second significant limitation lies in the source classification system. The study employed a proprietary three-tiered classification system (Brand, Earned, Social) to categorize domains. While every effort was made to ensure consistency and objectivity through a combination of predefined rules and AI-assisted labeling, this framework is ultimately a constructed model of the digital information ecosystem. The definitions of these categories, while logical, are subjective. For instance, the line between an "Earned" media outlet and a "Brand" blog can sometimes be blurry. A different research team utilizing an alternative classification system—for example, one with more granular categories like "News Media," "Professional Review," "User Review," "E-commerce," and "Corporate"—would inevitably produce different quantitative results. Therefore, the absolute percentages reported should be interpreted as illustrative of strong, relative trends (e.g., the dominance of Earned media) rather than as immutable facts. The comparative findings between engines are more robust than the absolute figures themselves.

Finally, a fundamental constraint of this external analysis is the lack of access to internal data. The study was conducted from the outside looking in, analyzing the outputs of these AI systems without access to their internal query logs, user data, ranking models, or training data intricacies. This "black box" nature means that while the study can accurately describe what is happening, the definitive why remains inferred. Different research threads around mechanistic interpretability will assist in that regard.

In summary, this study offers a valuable and rigorous comparative analysis of AI search engines at a critical point in their development. However, its conclusions are constrained by their temporal specificity, the subjective nature of its classification schema, and the inherent opacity of the systems being studied. These limitations do not invalidate the findings but rather define the scope within which they should be applied—as a powerful guide for current strategy that must be continually validated against the evolving reality of the search landscape.

\section{Acknowledgments}

We wish to thank the team at ktau.ai (www.ktau.ai) the leader in Generative Engine Optimization for their support.

\bibliographystyle{ACM-Reference-Format}
\bibliography{main}


\begin{thebibliography}{10}


\ifx \showCODEN    \undefined \def \showCODEN     #1{\unskip}     \fi
\ifx \showISBNx    \undefined \def \showISBNx     #1{\unskip}     \fi
\ifx \showISBNxiii \undefined \def \showISBNxiii  #1{\unskip}     \fi
\ifx \showISSN     \undefined \def \showISSN      #1{\unskip}     \fi
\ifx \showLCCN     \undefined \def \showLCCN      #1{\unskip}     \fi
\ifx \shownote     \undefined \def \shownote      #1{#1}          \fi
\ifx \showarticletitle \undefined \def \showarticletitle #1{#1}   \fi
\ifx \showURL      \undefined \def \showURL       {\relax}        \fi
\providecommand\bibfield[2]{#2}
\providecommand\bibinfo[2]{#2}
\providecommand\natexlab[1]{#1}
\providecommand\showeprint[2][]{arXiv:#2}

\bibitem[Aggarwal et~al\mbox{.}(2024)]%
        {aggarwal2024geogenerativeengineoptimization}
\bibfield{author}{\bibinfo{person}{Pranjal Aggarwal}, \bibinfo{person}{Vishvak Murahari}, \bibinfo{person}{Tanmay Rajpurohit}, \bibinfo{person}{Ashwin Kalyan}, \bibinfo{person}{Karthik Narasimhan}, {and} \bibinfo{person}{Ameet Deshpande}.} \bibinfo{year}{2024}\natexlab{}.
\newblock \bibinfo{title}{GEO: Generative Engine Optimization}.
\newblock
\showeprint[arxiv]{2311.09735}~[cs.LG]
\urldef\tempurl%
\url{https://arxiv.org/abs/2311.09735}
\showURL{%
\tempurl}


\bibitem[AI(2025)]%
        {ref5}
\bibfield{author}{\bibinfo{person}{Perplexity AI}.} \bibinfo{year}{2025}\natexlab{}.
\newblock \bibinfo{booktitle}{\emph{CEO says Perplexity hit 780M queries in May 2025}}.
\newblock
\urldef\tempurl%
\url{https://www.perplexity.ai/page/ceo-says-perplexity-hit-780m-q-dENgiYOuTfaMEpxLQc2bIQ}
\showURL{%
\tempurl}
\newblock
\shownote{[Online]. Available}.


\bibitem[Center(2025a)]%
        {ref2}
\bibfield{author}{\bibinfo{person}{Pew~Research Center}.} \bibinfo{year}{2025}\natexlab{a}.
\newblock \bibinfo{booktitle}{\emph{About a third of U.S. adults have used ChatGPT; usage has increased over the past year}}.
\newblock
\urldef\tempurl%
\url{https://www.pewresearch.org/short-reads/2025/06/25/34-of-us-adults-have-used-chatgpt-about-double-the-share-in-2023/}
\showURL{%
\tempurl}
\newblock
\shownote{[Online]. Available}.


\bibitem[Center(2025b)]%
        {ref6}
\bibfield{author}{\bibinfo{person}{Pew~Research Center}.} \bibinfo{year}{2025}\natexlab{b}.
\newblock \bibinfo{booktitle}{\emph{Google users are less likely to click on links when an AI summary appears in the results}}.
\newblock
\urldef\tempurl%
\url{https://www.pewresearch.org/short-reads/2025/07/22/google-users-are-less-likely-to-click-on-links-when-an-ai-summary-appears-in-the-results/}
\showURL{%
\tempurl}
\newblock
\shownote{[Online]. Available}.


\bibitem[Kumar and Lakkaraju(2024)]%
        {kumar2024manipulatinglargelanguagemodels}
\bibfield{author}{\bibinfo{person}{Aounon Kumar} {and} \bibinfo{person}{Himabindu Lakkaraju}.} \bibinfo{year}{2024}\natexlab{}.
\newblock \bibinfo{title}{Manipulating Large Language Models to Increase Product Visibility}.
\newblock
\showeprint[arxiv]{2404.07981}~[cs.IR]
\urldef\tempurl%
\url{https://arxiv.org/abs/2404.07981}
\showURL{%
\tempurl}


\bibitem[Similarweb(2024)]%
        {ref3}
\bibfield{author}{\bibinfo{person}{Similarweb}.} \bibinfo{year}{2024}\natexlab{}.
\newblock \bibinfo{booktitle}{\emph{ChatGPT Surges to 3.1B Visits in September 2024}}.
\newblock
\urldef\tempurl%
\url{https://www.similarweb.com/blog/insights/ai-news/chatgpt-topped-3-billion-visits-in-september/}
\showURL{%
\tempurl}
\newblock
\shownote{[Online]. Available}.


\bibitem[Stats(2025a)]%
        {ref4}
\bibfield{author}{\bibinfo{person}{StatCounter~Global Stats}.} \bibinfo{year}{2025}\natexlab{a}.
\newblock \bibinfo{booktitle}{\emph{AI Chatbot Market Share Worldwide}}.
\newblock
\urldef\tempurl%
\url{https://gs.statcounter.com/ai-chatbot-market-share}
\showURL{%
\tempurl}
\newblock
\shownote{[Online]. Available}.


\bibitem[Stats(2025b)]%
        {ref1}
\bibfield{author}{\bibinfo{person}{StatCounter~Global Stats}.} \bibinfo{year}{2025}\natexlab{b}.
\newblock \bibinfo{booktitle}{\emph{Search Engine Market Share Worldwide}}.
\newblock
\urldef\tempurl%
\url{https://gs.statcounter.com/search-engine-market-share}
\showURL{%
\tempurl}
\newblock
\shownote{[Online]. Available}.


\bibitem[Wan et~al\mbox{.}(2024)]%
        {wan2024evidencelanguagemodelsconvincing}
\bibfield{author}{\bibinfo{person}{Alexander Wan}, \bibinfo{person}{Eric Wallace}, {and} \bibinfo{person}{Dan Klein}.} \bibinfo{year}{2024}\natexlab{}.
\newblock \bibinfo{title}{What Evidence Do Language Models Find Convincing?}
\newblock
\showeprint[arxiv]{2402.11782}~[cs.CL]
\urldef\tempurl%
\url{https://arxiv.org/abs/2402.11782}
\showURL{%
\tempurl}


\bibitem[WIRED(2024)]%
        {ref7}
\bibfield{author}{\bibinfo{person}{WIRED}.} \bibinfo{year}{2024}\natexlab{}.
\newblock \bibinfo{booktitle}{\emph{Google Cut Back AI Overviews in Search Even Before Its 'Pizza Glue' Fiasco}}.
\newblock
\urldef\tempurl%
\url{https://searchengineland.com/google-ai-overviews-visibility-drops-15-percent-queries-442850?utm_source=chatgpt.com}
\showURL{%
\tempurl}
\newblock
\shownote{[Online]. Available}.


\end{thebibliography}

\end{document}